\documentclass[letterpaper,12pt]{article}
\usepackage{times}
\usepackage{empheq}
\usepackage{amsmath, amssymb}
\usepackage{amsthm}
\usepackage{bbm}
\usepackage[T1]{fontenc}
\usepackage[english]{babel}
\usepackage[utf8]{inputenc}
\usepackage{hyperref}
\usepackage{mathrsfs}
\usepackage{mathabx}
\usepackage{color}
\usepackage{anysize}
\usepackage{apacite}
\usepackage{natbib}

\marginsize{2cm}{2cm}{2cm}{2cm}
\usepackage[toc,page]{appendix}
\usepackage{url}
\usepackage{bm}
\usepackage{mathtools}
\usepackage[font={small,sc}]{caption}
\usepackage{ragged2e}
\usepackage[shortlabels]{enumitem}
\usepackage{tikz-cd}
\usepackage{algorithm}
\usepackage{algpseudocode}

\usepackage{arydshln}

\newtheorem{theorem}{Theorem}[section]
\newtheorem{definition}{Definition}[section]
\newtheorem{proposition}{Proposition}[section]

\newtheorem{lemma}{Lemma}[section]
\newtheorem{example}{Example}[section]
\newtheorem{remark}{Remark}[section]
\newtheorem{assumption}{Assumption}[section]

\AtBeginDocument{}

\newcommand{\Rd}{\mathbb{R}^{d}}

\newcommand{\Borel}[1]{\mathcal{B}(#1)}

\newcommand{\BorelRd}{\mathcal{B}(\mathbb{R}^{d})}

\newcommand{\Cov}{\mathbb{C}ov}
\newcommand{\Var}{\mathbb{V}ar}

\newcommand{\R}{\mathbb{R}}

\theoremstyle{remark}
\newtheorem{axiom}{Axiom}
\theoremstyle{remark}
\newtheorem{Rule}{Rule}[section]
\theoremstyle{remark}
\newtheorem{Step}{Step}[subsection]
\theoremstyle{remark}
\newtheorem{condition}{Condition}[subsection]

\title{Movement-based models for abundance data}

\author
{\\ \ \\ 
	 Ricardo Carrizo Vergara$^1$, Marc Kéry$^1$, Trevor Hefley$^2$\\
     \normalsize{$^{1}$Population Biology Group, Swiss Ornithological Institute, 6204 Sempach, Switzerland.}\\
	\normalsize{$^{2}$Department of Statistics,
Kansas State University, Manhattan, KS 66506, USA.}\\
}

\date{September 2024}

\begin{document}

\maketitle

\noindent {\bf Abstract} \quad We develop two statistical models for space-time abundance data based on a stochastic underlying continuous individual movement. In contrast to current models for abundance in statistical ecology, our models exploit the explicit connection between movement and counts, including the induced space-time auto-correlation. Our first model, called Snapshot, describes the counts of free moving individuals with a false-negative detection error. Our second model, called Capture, describes the capture and retention of moving individuals, and it follows an axiomatic approach based on three simple principles from which it is deduced that the density of the capture time is the solution of a Volterra integral equation of the second kind. Mild conditions are imposed to the underlying stochastic movement model, which is free to choose. We develop simulation methods for both models. The joint distribution of the space-time counts provides an example of a new multivariate distribution, here named the evolving categories multinomial distribution, for which we establish key properties. Since the general likelihood is intractable, we propose a pseudo-likelihood fitting method assuming multivariate Gaussianity respecting mean and covariance structures, justified by the central limit theorem. We conduct simulation studies to validate the method, and we fit our models to experimental data of a spreading population. We estimate movement parameters and compare our models to a basic ecological diffusion model. Movement parameters can be estimated using abundance data, but one must be aware of the necessary conditions to avoid underestimation of spread parameters.

\bigskip


\newpage

\section{Introduction}
\label{Sec:Intro}

The study of spatial population dynamics is a central endeavour in ecology and related sciences \citep{may1999unanswered}. It consists in the quantitative description of movement, birth and death processes in natural populations and the environmental factors influencing them. The complexities of ecological processes, ubiquitous measurement errors, and multiple spatial and temporal scales involved make this quantitative accounting challenging \citep{williams2002analysis,seber1986review}. Many different types of data contain information for this aim: presence/absence data, capture-recapture of marked individuals, survey counts at some space-time locations (abundance data), and individual tracking (movement data). Our work focuses on models for abundance data, but based on an underlying model of individual movement.

Models for abundance data are ubiquitous in statistical ecology \citep{andrewartha1954distribution,seber1982estimation} and have evolved in many directions involving various forms of detection errors \citep{krebs2009ecology,seber2023estimating}. For our work, the most important abundance models are ecological diffusion models (hereafter EcoDiff), used mainly to study the spread of populations and which consider the counts as statistical errors around a deterministic mean field solution to an advection-diffusion-reaction partial differential equation (PDE). Since the work of \citet{hotelling1927differential}, statisticians have focused on modelling errors around physically inspired differential equations, with diffusion PDEs being studied for ecological purposes by \citet{skellam1951random}. EcoDiff models became more popular at the beginning of the millennium \citep{okubo2001diffusion,wikle2003hierarchical}, and are currently a standard method for inference about spreading processes \citep{hooten2013computationally, roques2013modeles, soubeyrand2014parameter, hefley2017mechanism, louvrier2020mechanistic,lu2020nonlinear,eisaguirre2023informing}. Other important models for abundance are point process models such as Poisson and Cox process \citep{schoenberg2002point}, in which the mean intensity field can be chosen using data-fitting considerations \citep{diggle2010partial,gelfand2018bayesian} or following a mechanistic intuition, with intensities satisfying PDEs such as in EcoDiff models \citep{hefley2016hierarchical,zamberletti2022understanding}. In practice, spatio-temporal count data are relatively easy to collect and very widely available, especially owing to major citizen-science data collection initiatives such as eBird \citep{sullivan2009ebird}.

On the other hand, movement trajectory modelling has generated considerable interest in the recent statistical ecology community \citep{turchin1998quantitative,hooten2017animal}. Various kinds of stochastic processes are used to model animal trajectories, ranging from Brownian motion \citep{horne2007analyzing,kranstauber2019modelling} to sophisticated versions of Ornstein-Uhlenbeck processes and Itô diffusions \citep{fleming2014fine,gurarie2017correlated,michelot2019langevin}. While modern tracking techniques such as satellite transmitters have enabled progress in the collection of such data, logistical costs and the need to capture, handle, and place a device on an individual will always limit the cases where we can perform unbiased data analysis. Individual movement data is thus, as a general rule, more expensive and more intricate to analyse than abundance data.

The main issue with the aforementioned abundance models is that they do not use the conceptual and physical link between movement and counts. Therefore, even for models using mechanistic PDEs, there is no physical sense, and integrating both abundance and telemetry data in a single analysis is not immediate. More importantly, the space-time auto-correlation of the abundance field is not related to the movement: EcoDiff and Poisson point process models assume independence at different space-time locations, while the versions of Cox process already used in ecology \citep{diggle2010partial,gelfand2018bayesian} do not consider the movement of individuals. Covariance modelling is central in spatial statistics, where independence at different locations is often considered as insufficient to explain spatialized data and the choice of adequate covariance models is always challenging \citep{chiles1999geostatistics,wackernagel2003multivariate}.

In this context, we propose models for abundance which explicitly consider an underlying continuous individual movement modelled with stochastic processes. We focus on two situations: the Snapshot model for counts of free moving individuals with a false-negative detection error, and the Capture model for counts of individuals being captured in a capture domain with possible multiple releases. The main novelty is the establishment of the Snapshot and Capture modelling principles, which can be used with arbitrary trajectory models, and the study and use in practice of the mean and auto-correlation structures of the space-time abundance field induced by the movement. The distribution of a vector of counts generated by any of our models is an instance of the same multivariate distribution, new to our knowledge, which is here named the evolving categories multinomial distribution. It is described in section \ref{Sec:ECMDistribution}. The likelihood of this distribution remains intractable, therefore ad-hoc pseudo-likelihood methods are required for inference.

While the Snapshot model considers a simple binomial detection error, the Capture model is constructed with an axiomatic approach, where mechanistic and epistemological principles on the capture procedure are established, the distribution of capture times and positions being thereby deduced. In particular, we obtain the remarkable result on the capture time density, which must be the solution to a Volterra integral equation of the second kind. This result is used to provide formulae for the mean and covariance structures of the counts field of captured individuals.

For model fitting, we propose a frequentist pseudo-likelihood method using a Gaussian multivariate distribution respecting the first two moments, justified by the central limit theorem for large number of individuals. We explore the validity of this method through a simulation study, for which we developed simulation algorithms for our models. We apply our methods to an experimental data set consisting of counts of fruit flies being released in a meadow and captured repeatedly in traps \citep{edelsparre2021scaling}. We compare the performance of our models with a basic EcoDiff model, both for real data fitting and the simulation study. To our knowledge, this is the first published simulation study on EcoDiff models. We concluded that in appropriate settings it is indeed possible to infer movement parameters from broad-scale abundance data.

While the spirit of this modelling approach is not new since statistical physics has worked with similar notions for a long time \citep{doi1976stochastic}, this approach has not been explored in statistical ecology with the rigour we do. However, similar methods have been used in slightly different settings. \citet{glennie2021incorporating} use an underlying animal Brownian movement for distance sampling: it is a Snapshot model with a sophisticated detection probability. The authors fit the model to abundance and telemetry data jointly, but the fitting method requires a pseudo-likelihood for which it is not clear if the auto-correlation structure is respected.\citet{malchow2024fitting} modelled individuals moving, dying, growing and reproducing in discrete time, and then fit the model to broader scale data. However, the fitting method uses a (stochastic) pseudo-likelihood assuming independent errors around a simulated mean field, losing the auto-correlation. \citet{potts2023scale} provide a general outline for connecting individual and collective movement based on step selection analysis, to the large-scale animal space use, but this work does not aim to be usable for abundance data. We also mention the remarkable theoretical work of \citet{roques2022spatial}, where a connection between PDEs, stochastic PDEs, and particle movement is explored. However, the fitting method uses a Poisson process pseudo-likelihood, where the auto-correlation is again lost.

Our methodology seems therefore to be the only one currently available in statistical ecology with the versatility to use any desired continuous-time movement model and that respects the auto-correlation structure in the fitting method. Our work provides thus a prototypical and flexible movement-based modelling methodology of abundance, which is a first step towards a more comprehensive and mechanistic description of spatial population dynamics.

\textit{Notation.} For a random variable $X$ we denote $\mu_{X}$ its probability distribution (law measure). We may write $\mu_{X_{1} , \ldots , X_{n}}$ to be more explicit for a vector-valued random variable $X = (X_{1} , \ldots , X_{n})$. We sometimes abuse notation as follows: if $Y$ is a continuous random variable (no atoms), we may write $\mathbb{P}( X \in A \mid Y = y  )$ to denote $\mu_{X \mid Y = y}(A)$, where $\mu_{X \mid Y = y}$ is the conditional distribution of $X$ given $Y = y$. Conversely, we may write $\mu_{X \mid Y \in B}(A)$ to denote $\mathbb{P}( X \in A \mid Y \in B)$. $\mathbbm{1}$ is used for indicator functions. $\Borel{D}$ denotes the collection of Borel subsets of $D$.

\section{The evolving categories multinomial distribution}
\label{Sec:ECMDistribution}

Here we develop the main abstract multivariate distribution needed in our work. This distribution arises from the count of independent moving individuals. The main idea is that each individual belongs to a category at each time, and the possible categories also evolve with time.

Let $N$ be the number of individuals. Let $n > 0$ be the number of time steps. For each $k = 1 , \ldots , n$ we consider $m_{k} \geq 1$ mutually exclusive and exhaustive categories into which an individual may belong at time $k$. We denote them $C_{k,l}$, with $k = 1 , \ldots , n$ and $l = 1 , \ldots , m_{k}$. The categories are abstract and could represent whatever is needed in an application scenario: the places where moving individuals are counted at some times, the presidential candidates voters select from in an election with multiple rounds, the products available at different seasons that consumers may chose and buy, to mention but a few examples.

Next, we assume that each individual follows independently the same stochastic dynamics across the categories, that is, each individual will exhibit a unique sequence of categories across time, the sequence itself being random. We call path any sequence of categories across some times, and we call full path a path that considers the $n$ times. A full path is represented by an index vector $(l_{1} , \ldots , l_{n}) \in \lbrace 1 , \ldots , m_{1} \rbrace \times  \cdots \times \lbrace 1 , \ldots , m_{n} \rbrace$, where $l_{k}$ is the index of the category considered at time $k$. The key quantities of our construction are the full path probabilities, which give the probability that an individual follows a certain path:
\begin{equation}
\label{Eq:full pathProbs}
\begin{aligned}
    p_{l_{1} , \ldots , l_{n}}^{(1,\ldots,n)} = \hbox{probability of belonging} &\hbox{ to category } C_{1,l_{1}} \hbox{ at time } 1, \\ &\hbox{ to category } C_{2,l_{2}} \hbox{ at time } 2, \\
& \cdots \ , \hbox{ AND to the category } C_{n,l_{n}} \hbox{ at time } n.
\end{aligned} 
\end{equation}
These probabilities fully describe the stochastic dynamics of the ``movement'' of the individual among the categories. To fully specify the dynamics, $(\prod_{k=1}^{n}m_{k}) - 1$ of these probabilities have to be made precise, since they must sum to one:
\begin{equation}
\label{Eq:ProbSum1}
\sum_{l_{1}=1}^{m_{1}} \cdots \sum_{l_{n}=1}^{m_{n}} p_{l_{1} , \ldots , l_{n}}^{(1,\ldots,n)} = 1.
\end{equation}
Let us introduce notation for not necessarily full path probabilities. If $k_{1},  \ldots , k_{\tilde{n}}$ are $\tilde{n}$ (different) time indices in $\lbrace 1 , \ldots , n \rbrace$, for each $(l_{1}, \ldots , l_{\tilde{n}}) \in \lbrace 1 , \ldots , m_{k_{1}} \rbrace \times \cdots \times \lbrace 1 , \ldots , m_{k_{\tilde{n}}} \rbrace $ we denote
\begin{equation}
\label{Eq:SubPathProbs}
\begin{aligned}
p^{(k_{1} , \ldots , k_{\tilde{n}})}_{l_{1} , \ldots , l_{\tilde{n}}} = \hbox{probability of belonging} &\hbox{ to category } C_{k_{1},l_{1}} \hbox{ at time } k_{1}, \\ &\hbox{ to category } C_{k_{2},l_{2}} \hbox{ at time } k_{2}, \\
& \cdots \ , \hbox{ AND to category } C_{k_{\tilde{n}},l_{\tilde{n}}} \hbox{ at time } k_{\tilde{n}}.
\end{aligned}
\end{equation}
A projectivity consistency condition on the probabilities $p^{(k_{1} , \ldots , k_{\tilde{n}})}_{l_{1} , \ldots , l_{\tilde{n}}}$ must hold when changing $\tilde{n}$ and the indices $k_{1} , \ldots , k_{\tilde{n}}$ and $l_{1} , \ldots , l_{\tilde{n}}$: if $\tilde{n} \geq 2$, and $s \in \lbrace 1 , \ldots , \tilde{n} \rbrace $, then
\begin{equation}
\label{Eq:ProjSubPathProbs}
p_{l_{1} , \ldots , l_{s-1} ,  l_{s+1} , \ldots , l_{\tilde{n}}}^{(k_{1} , \ldots , k_{s-1} , k_{s+1} , \ldots , k_{\tilde{n}})} = \sum_{l_{s} = 1}^{m_{s}} p_{l_{1} , \ldots , l_{\tilde{n}}}^{(k_{1} , \ldots , k_{\tilde{n}})}. 
\end{equation}

Our focus is on the counts of individuals in each category at each time. We define thus
\begin{equation}
\label{Eq:DefQComponent}
    Q_{k,l} = ``\hbox{number of individuals belonging to category } C_{k,l} \hbox{ at time } k".
\end{equation}
We can assemble the variables in \eqref{Eq:DefQComponent} into a single \textit{random arrangement}
\begin{equation}
\label{Eq:DefQArrangment}
    \mathscr{Q} = \big(  Q_{k,l} \big)_{k \in \lbrace 1 , \ldots , n \rbrace , l \in \lbrace 1 , \ldots , m_{k} \rbrace}. 
\end{equation}

\begin{definition}
\label{Def:ECMDistribution}
    Under the previously defined conditions, we say that the random arrangement $\mathscr{Q}$ follows an evolving categories multinomial distribution (abbreviated ECM distribution).
\end{definition}

$\mathscr{Q}$ contains thus the information at the category level, losing the information of the paths effectively followed by each individual. $\mathscr{Q}$ can also be seen as a list of random vectors, $\mathscr{Q} = ( \vec{Q}_{k}  )_{k \in \lbrace 1 , \ldots , n \rbrace}$, with
\begin{equation}
\label{Eq:DefVectorQk}
    \vec{Q}_{k} = ( Q_{k,1} , \ldots , Q_{k,m_{k}} ), \quad \hbox{for all } k \in \lbrace 1 , \ldots , n \rbrace.
\end{equation}
These vectors have components that sum to $N$. If $m_{1} = \cdots = m_{n}$, $\mathscr{Q}$ can be seen as a random matrix. The arrangement $\mathscr{Q}$ could be wrapped into a random vector with $\sum_{k=1}^{n}m_{k}$ components, but for ease of presentation we maintain the indexing scheme proposed in \eqref{Eq:DefQArrangment}. The ECM distribution can be fully described through its characteristic function.

\begin{proposition}
\label{Prop:CharFunctionQ}
The characteristic function $\varphi_{\mathscr{Q}} : \mathbb{R}^{m_{1} + \cdots + m_{n}} \to \mathbb{C}$  of $\mathscr{Q}$ is
    \begin{equation}
    \label{Eq:CharFunctionQ}
        \varphi_{\mathscr{Q}}(\mathscr{\xi} ) = \left\{  \sum_{l_{1}=1}^{m_{1}} \cdots \sum_{l_{n}=1}^{m_{n}} e^{i( \xi_{1,l_{1}} + \cdots + \xi_{n,l_{n}} )} p_{l_{1}, \ldots , l_{n}}^{(1 , \ldots , n)} \right\}^{N}, 
    \end{equation}
    for all $ \mathscr{\xi} = \left( \xi_{k,l} \right)_{k \in \lbrace 1 , \ldots , n \rbrace, l \in \lbrace 1 , \ldots , m_{k} \rbrace} \in \mathbb{R}^{m_{1} + \cdots + m_{n}}$.\footnote{For ease of exposition, we consider the arrangements of the form $(\xi_{k,l})_{k \in \lbrace 1 , \ldots , n \rbrace, k \in \lbrace 1 , \ldots , m_{k} \rbrace}$ as members of space $\R^{m_{1} + \cdots + m_{n}}$, although the indexing differs from the traditionally used for vectors.}
\end{proposition}

In our work we do not provide an explicit formula for the probability mass function of the ECM distribution, that is, we do not compute $\mathbb{P}(\mathscr{Q} = \mathscr{q})$ for every possible consistent arrangement $\mathscr{q}$. Nonetheless, we provide the one-time and two-times marginal distributions, used for computing the mean and covariance of $\mathscr{Q}$. The following result is the reason the word ``multinomial'' is in the name of the ECM distribution.

\begin{proposition}
\label{Prop:OneTimeDistQ}
    Let $k \in \lbrace 1 , \ldots , n \rbrace$. Then,  $\vec{Q}_{k}$ is a multinomial random vector of size $N$ and vector of probabilities $(p_{1}^{(k)} , \ldots , p_{m_{k}}^{(k)})$.
\end{proposition}

This result can be seen intuitively: for a given time $k$ we wonder about how many independent individuals belong to each category $C_{k,1} , \ldots , C_{k,m_{k}}$, so it has indeed the form of a multinomial random vector. The two-times distribution is also related to the multinomial distribution.
\begin{proposition}
\label{Prop:TwoTimesDistQ}
    Let $k,k' \in \lbrace 1 , \ldots , n \rbrace$ with $k \neq k'$. Denote by $p_{l' \mid l}^{(k' \mid k)}$ the conditional probabilities $ p_{l' \mid l}^{(k' \mid k)} = p_{l,l'}^{(k , k')} / p_{l}^{k} $ for every $l = 1 , \ldots , m_{k}$, $l' = 1 , \ldots , m_{k'}$. Then, the random vector $\vec{Q}_{k'}$ conditioned on the random vector $\vec{Q}_{k}$ has the distribution of a sum of $m_{k}$ independent $m_{k'}$-dimensional multinomial random vectors, say $\vec{Y}_{1} , \ldots , \vec{Y}_{m_{k}}$, each vector $\vec{Y}_{l}$ having size $Q_{k,l}$ and vector of probabilities $( p_{1\mid l}^{(k'\mid k)} \ , \ \ldots \ , \ p_{m_{k'}\mid l}^{(k'\mid k)}) $.  
\end{proposition}

The structure of a sum of independent multinomial random vectors is a particular example of the Poisson-multinomial distribution \citep{daskalakis2015structure,lin2022poisson}. A related structure is obtained by \citet{roques2022spatial} in a context where $N$ is random and Poisson, resulting in a sum of independent random vectors with independent Poisson components. Propositions  \ref{Prop:OneTimeDistQ} and \ref{Prop:TwoTimesDistQ} allow us to compute the first and second moment structures of $\mathscr{Q}$.

\begin{proposition}
\label{Prop:MomentsECM}
    The mean and the covariance structures of $\mathscr{Q}$ are given respectively by
    \begin{equation}
    \label{Eq:MeanECM}
        E\left( Q_{k,l}  \right) = Np_{l}^{(k)},
    \end{equation}
\begin{equation}
\label{Eq:CovQ}
    \Cov\left( Q_{k,l} \ , \  Q_{k',l'}  \right)  = \begin{cases}
        N\left( p_{l,l'}^{(k,k')} - p_{l}^{(k)}p_{l'}^{(k')}  \right) \quad & \hbox{ if } k \neq k', \\
        N\left(   \delta_{l,l'}p_{l}^{(k)}    - p_{l}^{(k)}p_{l'}^{(k')}   \right) \quad & \hbox{ if } k = k'.
    \end{cases} 
\end{equation}
    for every $k,k' \in \lbrace 1 , \ldots , n \rbrace$ and $(l,l') \in \lbrace 1 , \ldots , m_{k} \rbrace \times \lbrace 1 , \ldots , m_{k'} \rbrace.$ Here $\delta_{l,l'}$ is the Kronecker delta notation: $\delta_{l,l'} = 1$ if $l=l'$, and $\delta_{l,l'} = 0$ if $l \neq l'$.
\end{proposition}

\section{Counting with error: the Snapshot model}
\label{Sec:Snapshot}

In almost every real abundance data set one does not have perfect abundance information, owing to measurement errors such as imperfect detection or species mis-identification \citep{williams2002analysis,kery2016modeling}. Such measurement errors are often conceptualized in hierarchical models \citep{royle2008hierarchical,newman2014modelling,cressie2015statistics}, where a true but latent state, the ``state process'', is distinguished from error-prone raw data, the ``data process''. Here we follow this approach by considering a false negative detection error.

We first describe the state process, that is, the  space-time random field containing the information of the true number of individuals at each space-time location. Let $N$ be the total number of individuals, which we suppose known and fixed. Let $t_{0}$ be an initial time. We denote $(X_{j}(t))_{t \geq t_{0}}$ the $\Rd$-valued stochastic process which describes the trajectory of individual $j$, for $j=1,\ldots,N$. The \textit{abundance random measure} $\Phi$ is defined as the space-time (generalized) random field identified with the family of random variables $\{ \Phi_{t}(A) \}_{A \in \BorelRd, t \geq t_{0} }$ given by
\begin{equation}
\label{Eq:DefPhit(A)}
\Phi_{t}(A) = \sum_{j = 1}^{N} \delta_{X_{j}(t)}(A) = \sum_{j=1}^{N} \mathbbm{1}_{A}\left\{ X_{j}(t) \right\},
\end{equation}
where we use both the Dirac-delta measure and the indicator function notations. That is, $\Phi_{t}(A)$ is simply the number of individuals in region $A$ at time $t$. We make the following assumption.

\begin{assumption}
\label{Ass:IIDTrajectories}
The processes $X_{j}$, for $j = 1 , \ldots , N $ are mutually independent and follow the same distribution as a reference process $X = \{X(t)\}_{t \geq t_{0}}$.
\end{assumption}

The properties of $\Phi$ can be studied in function of 
 $X$, without specifying which kind of process $X$ is exactly: we are free to choose it as deemed most appropriate. What is important to retain is that the individual movement induces mean and auto-correlation structures in a specific manner.

\begin{proposition}
\label{Prop:MomentsPhi}
The mean and covariance structures of the abundance random measure $\Phi$ are given respectively by
\begin{equation}
\label{Eq:MeanPHI}
E\left\{ \Phi_{t}(A) \right\} = N \mu_{X(t)}(A),
\end{equation}
\begin{equation}
\label{Eq:CovPHI}
\Cov\left\{ \Phi_{t}(A) , \Phi_{s}(B) \right\} = N \left\{ \mu_{X(t),X(s)}(A \times B ) - \mu_{X(t)}(A)\mu_{X(s)}(B)  \right\},
\end{equation}
for every $t,s \geq t_{0}$ and every $A,B \in \BorelRd$.
\end{proposition}

Now we define hierarchically the data process. We assume the count of individuals does not affect their movement, which occurs for example when taking a non-invasive photograph over a given space region (see e.g. \citep{williams2017estimating}). This is why we call this model the Snapshot model. We assume a false-negative detection error in each taken snapshot with a detection probability $p \in [0,1]$. The count random variables are defined as
\begin{equation}
\label{Eq:DefQtASnapshot}
    Q_{t}(A) = \hbox{``number of individuals detected in the snapshot taken over region $A$ at time $t$''}.
\end{equation}
 Our probabilistic model for the random field $Q = \left\{Q_{t}(A)\right\}_{t \geq t_{0}, A \in \BorelRd}$ obeys the following rules:
\begin{Rule}
\label{It:SnapshotDiffTimesIndepCondPHI} For any set of distinct times $t_{1} , \ldots , t_{n}$, the spatial stochastic processes $Q_{t_{1}}, \ldots , Q_{t_{n}}$ are mutually independent conditional on $\Phi$.
\end{Rule} 
\begin{Rule}
\label{It:SnapshotDisjSetsIndepCondPHIBinomial} If $A_{1} , \ldots , A_{n}$ are mutually disjoint regions, then for any $t \geq t_{0}$, the random variables $Q_{t}(A_{1}) , \ldots , Q_{t}(A_{n})$ are independent conditional on $\Phi$, with $Q_{t}(A_{j})$ following a binomial distribution of size $\Phi_{t}(A_{j})$ and success  probability $p$.
\end{Rule}

The distribution of $Q$ is thus obtained hierarchically through these conditional relations with respect to  $\Phi$. The random arrangements we can construct with $Q$ are actually sub-arrangements of ECM-distributed random arrangements. Explicitly, consider $t_{1} , \ldots , t_{n} \geq t_{0}$ time points. For each time $t_{k}$, consider $m_{k}$ snapshots taken over mutually disjoint regions of the space $A_{k,1}, \ldots , A_{k,m_{k}}$. We define an auxiliary stochastic process $(R_{t})_{t \geq t_{0}}$ as
\begin{equation}
\label{Eq:DefRtMath}
    R_{t_{k}} = N - \sum_{l = 1}^{m_{k}} Q_{t_{k}}(A_{k,l}), \quad \hbox{for all } k \in \lbrace 1 , \ldots , n \rbrace.
\end{equation}
$R_{t}$ is the number of non-detected individuals in the snapshots taken at time $t$. Consider the random arrangement $\mathscr{Q}$, interpreted as a list of random vectors $\mathscr{Q} = (\vec{Q}_{k})_{k=1 , \ldots , n}$, each $\vec{Q}_{k}$ given by
\begin{equation}
\label{Eq:DefMargVectorsQSnapshot}
    \vec{Q}_{k} = ( Q_{t_{k}}(A_{k,1}) , \ldots , Q_{t_{k}}(A_{k,m_{k}}) , R_{t_{k}} ). 
\end{equation}
In this scenario we can identify $m_{k}+1$ exclusive and exhaustive categories at each time $t_{k}$: 
\small
\begin{equation}
\label{Eq:DefCategSnapshot}
\begin{aligned}
    &\hbox{``individual $j$ belongs to category $C_{k,l}$''} \\
    & \Leftrightarrow \begin{cases}
    \hbox{``individual $j$ is on the snapshot taken over set $A_{k,l}$ at time $t_{k}$''} & \hbox{ if } l \leq m_{k}, \\
    \hbox{``individual $j$ has not been detected in any snapshot taken during time $t_{k}$''} & \hbox{ if } l = m_{k}+1.
\end{cases}
\end{aligned}
\end{equation}
\normalsize
\begin{proposition}
\label{Prop:SnapshotIsECM}
    The random arrangement $\mathscr{Q}$ follows an ECM distribution.
\end{proposition}

Modelling Rules \ref{It:SnapshotDiffTimesIndepCondPHI} and \ref{It:SnapshotDisjSetsIndepCondPHIBinomial} concern directly the distribution of $Q$, thus the arrangement $\mathscr{Q}$ is not explicitly defined as the aggregated count of independent individuals following a specified dynamics, as it is needed for the ECM distribution. Therefore, Proposition \ref{Prop:SnapshotIsECM} needs a proper mathematical justification which is given in supplementary material, section \ref{Proof:SnapshotIsECM}, where we also provide a method for computing the corresponding full path probabilities. The mean and covariance structures of the field $Q$ can be expressed through a direct link with the mean and covariance structures of $\Phi$.
\begin{proposition}
\label{Prop:MomentsSnapshot}
     The mean and covariance structures of the field $Q = (Q_{t}(A))_{t \geq t_{0}, A \in \BorelRd}$ are, as functions of the moments of $\Phi$, 
    \begin{equation}
    \label{Eq:MeanSnapshot}
        E\left\{ Q_{t}(A) \right\} = p E\left\{ \Phi_{t}(A)  \right\}, \quad \hbox{for all } A \in \BorelRd, t \geq t_{0},
    \end{equation}
    \begin{equation}
    \label{Eq:CovSnapshot}
        \Cov\left\{ Q_{t}(A) , Q_{s}(B) \right\} = \begin{cases}
            p^{2}\Cov\left\{ \Phi_{t}(A) , \Phi_{s}(B) \right\} & \hbox{ if } t \neq s \hbox{ or } A \cap B = \emptyset, \\
            p(1-p)E\left\{ \Phi_{t}(A) \right\} + p^{2}\Var\left\{ \Phi_{t}(A) \right\} & \hbox{ if } t = s \hbox{ and } A = B. 
        \end{cases}
    \end{equation}
\end{proposition}

\begin{remark}
\label{Rem:SnapshotNotTotallyDefined}
Technically we have not made precise the behaviour of $Q$ over non-disjoint spatial sets. For instance, if $B \subset A$, Rules \ref{It:SnapshotDiffTimesIndepCondPHI} and \ref{It:SnapshotDisjSetsIndepCondPHIBinomial} do not determine the distribution of the random vector $(Q_{t}(A),Q_{t}(B))$ for a given $t$. There are different modelling options for solving this issue. One is to assume that $Q_{t}$ must be additive in space. Another is to assume $Q_{t}(A)$ and $Q_{t}(B)$ always independent conditional on $\Phi$, as long as $A \neq B$. This last option follows a particular modelling intuition: if at time $t$ we take two different snapshots, one over $A$ and other over $B$, then both must be taken with different cameras, for which we may assume independent performance. In this work we shall only consider disjoint spatial regions for each time.
\end{remark}

\begin{remark}
The distribution of the abundance random measure $\Phi$ itself is also related to the ECM distribution. See Lemma \ref{Lemma:PHIisECM} in supplementary material, section \ref{Proof:MomentsPhi}.
\end{remark}

\section{Capture model}
\label{Sec:Capture}

\subsection{Generalities}
\label{Sec:GeneralitiesCapture}

The Capture model aims to describe the following situation. Individuals move in $\Rd$ from an initial time $t_{0}$. In a subdomain $D_{c} \subset \Rd$, called the capture domain, traps are set to capture and retain some of those. We aim for a stochastic model of the capture procedure leading to a model for counts of captured individuals at different locations. We continue to use the iid Assumption \ref{Ass:IIDTrajectories}, so general conclusions can be obtained from the analysis of a single reference individual. The individual trajectory $X = \left\{X(t)\right\}_{t \geq t_{0}}$ must consider that at a time $T_{c}$ the individual is captured and no longer moves. $T_{c}$ is a new continuous random variable, and the distribution of the pair $(X,T_{c})$ must be made precise.

\subsection{Distribution of the capture time}
\label{Sec:DistributionCaptureTime}

We consider a special auxiliary process $\tilde{X} = \left\{\tilde{X}(t) \right\}_{t \geq t_{0}}$ called the \textit{free trajectory}. This process represents \textit{the hypothetical movement the individual would exhibit in the absence of any capture procedure}. $\tilde{X}$ is purely auxiliary and speculative (there will never be any data for it), for which we can select a trajectory model with considerable freedom. The model for $\tilde{X}$ will then be used to construct reasonable models for the \textit{concrete} variables $X$ and $T_{c}$. The process $X$ will be now called the \textit{effective trajectory}, since it represents the trajectory the individual physically takes.

We begin with the assumption (which could be considered as a $0$th axiom) that before being captured, the individual follows the free trajectory. This imposes a deterministic expression for $X$ as a function of $\tilde{X}$ and $T_{c}$:
\begin{equation}
\label{Eq:EffTrajFunctionFreeTrajTc}
X(t) = \begin{cases}
\tilde{X}(t) & \hbox{if } t < T_{c} \\
\tilde{X}(T_{c}) & \hbox{if } t \geq T_{c}.
\end{cases}
\end{equation}
Thus, to fully specify the stochastic model it is sufficient to focus on the pair $(\tilde{X},T_{c})$. We propose the following axioms which are inspired from mechanistic and epistemological intuitions.

\begin{axiom}[No-capture possibility]
\label{Axiom:TcExtReal}
The individual may be captured at any time $t > t_{0}$ or it may be not captured at all: 
\begin{equation}
\label{Eq:SupportTc}
T_{c} \in  (t_{0},\infty)\cup \lbrace \infty \rbrace,
\end{equation}
the event $\lbrace T_{c} = \infty \rbrace$ meaning that the individual is not captured. $T_{c}$ is an extended real random variable.
\end{axiom}
\begin{axiom}[Stationary rate of growth of capture probability]
\label{Axiom:StatRateGrowthCaptProb}
The instantaneous rate of growth of the probability of capture of the individual, given that it is free and within the capture domain, equals a strictly positive constant $\alpha > 0$:
\begin{equation}
\label{Eq:RateGrowthCaptProb}
\lim_{\Delta t \to 0} \frac{\mathbb{P}\left\{  T_{c} \in [t , t+ \Delta t] \mid  T_{c} \geq t \ , \ \tilde{X}(t) \in D_{c}   \right\}}{\Delta t} = \alpha , \quad \hbox{for all } t \in (t_{0},\infty). 
\end{equation}
\end{axiom}
\begin{axiom}[Capture-position information]
\label{Axiom:CaptPosInfo} Regarding the free trajectory, knowing that the individual has been captured at a given time is equivalent to know that the individual is in the capture domain at that time. Mathematically,
\begin{equation}
\label{Eq:CaptPosInfo}
\mathbb{P}\left\{  \tilde{X}(t_{1}) \in A_{1} , \ldots , \tilde{X}(t_{n}) \in A_{n}  \mid T_{c} = t  \right\} = \mathbb{P}\left\{  \tilde{X}(t_{1}) \in A_{1} , \ldots , \tilde{X}(t_{n}) \in A_{n}  \mid \tilde{X}(t) \in D_{c}  \right\},
\end{equation} 
for every $t \in (t_{0} , \infty)$, every $t_{1} , \ldots , t_{n} \in  (t_{0} , \infty)  $, and every $A_{1}, \ldots , A_{n} \in \Borel{\Rd}$. 
\end{axiom}

Axiom \ref{Axiom:TcExtReal} is a minimal realistic condition regarding the values of $T_{c}$. Axiom \ref{Axiom:StatRateGrowthCaptProb} is a modelling assumption concerning the dynamics of the capture process, and it ensures that the longer the individual visits the capture domain, the more likely it is to be captured. The value $\alpha$ is called the \textit{rate of growth of capture probability}, and it is the only parameter for the capture procedure. We add the adjective \textit{stationary} since $\alpha$ does not change with $t$. The capture-position information Axiom \ref{Axiom:CaptPosInfo} is not an assumption about the physical reality, but rather about the information we can have when knowing a particular part of it: it is an epistemological axiom.

The following Theorem shows that under these axioms the distribution of $T_{c}$ cannot be chosen arbitrarily: there is a specific relation between the distribution of $T_{c}$ and that of $\tilde{X}$. 

\begin{theorem}
\label{Theo:DistributionTc}
Suppose the free trajectory $\tilde{X}$ satisfies the two following conditions:
\begin{condition}
\label{It:FreeTrajProbDcNotNull} $\mu_{\tilde{X}(t)}(D_{c})  > 0$ for every $t > t_{0}$.
\end{condition}
\begin{condition}
\label{It:ContinuityFreeTraj} The function $(t,u) \mapsto \mu_{\tilde{X}(t),\tilde{X}(u)}(D_{c}\times D_{c})$ is continuous over $(t_{0}, \infty)\times (t_{0}, \infty)$.
\end{condition}
Suppose there exists an extended real random variable $T_{c}$ satisfying Axioms \ref{Axiom:TcExtReal}, \ref{Axiom:StatRateGrowthCaptProb}, \ref{Axiom:CaptPosInfo}. Then, a unique probability distribution is possible for $T_{c}$, which must have a continuous density over $(t_{0},\infty)$ which coincides with the solution to the Volterra integral equation of the second kind
\begin{equation}
\label{Eq:fTcVolterra}
f_{T_{c}}(t) = \alpha \mu_{\tilde{X}(t)}(D_{c}) - \alpha \int_{t_{0}}^{t} \mu_{\tilde{X}(t) \mid \tilde{X}(u) \in D_{c}}(D_{c}) f_{T_{c}}(u) du.
\end{equation}
\end{theorem}

We make the following remarks about Theorem \ref{Theo:DistributionTc}. 

\begin{remark}
\label{Rem:fTcNotProbDensity}
Since $T_{c}$ is extended real, the density  $f_{T_{c}}$ is not exactly a probability density function because in general $\int_{t_{0}}^{\infty} f_{T_{c}}(t)dt \leq 1$. However, since $\int_{t_{0}}^{\infty} f_{T_{c}}(t)dt + \mathbb{P}( T_{c} = \infty ) = 1$,  $f_{T_{c}}$ is sufficient to fully characterize the distribution of $T_{c}$.
\end{remark}

\begin{remark}
\label{Rem:DoesTcExist}
Theorem \ref{Theo:DistributionTc} does not assert the existence of the random variable $T_{c}$: it only shows the unique possible distribution for it. Proving the existence of $T_{c}$ technically means constructing a probability space supporting both $\tilde{X}$ and $T_{c}$. In this aim, one could proceed as follows: assuming $\tilde{X}$ satisfies Conditions \ref{It:FreeTrajProbDcNotNull} and \ref{It:ContinuityFreeTraj}, take the solution to the Volterra equation \eqref{Eq:fTcVolterra} and verify if $f_{T_{c}} \geq 0 $ and $\int_{t_{0}}^{\infty}f_{T_{c}}(u)du \leq 1$. Construct then the finite-dimensional distributions of $(\tilde{X}, T_{c})$ with Axiom \ref{Axiom:CaptPosInfo}, and verify the Kolmogorov consistency conditions. Proposition \ref{Prop:AlphaSmallTcPosInt} provides a partial result in this respect for a finite time horizon, which is sufficient for our applications.
\end{remark}

\begin{proposition}
\label{Prop:AlphaSmallTcPosInt}
Let $t_{H} > t_{0}$. Suppose $\alpha$ satisfies $\alpha < \frac{1}{t_{H}-t_{0}}$. Then, the unique solution to the Volterra equation \eqref{Eq:fTcVolterra} is positive over the interval $[t_{0},t_{H}]$ and $\int_{t_{0}}^{t_{H}} f_{T_{c}}(u)du < 1$.
\end{proposition}

To help the understanding of Theorem \ref{Theo:DistributionTc}, we propose some examples where the density $f_{T_{c}}$ can be obtained explicitly.

\begin{example}
\label{Ex:TcExponential}
Suppose $D_{c} = \Rd$, that is, the whole space is the capture domain. Then, $T_{c}$ is an exponential random variable with rate $\alpha$:
\begin{equation}
\label{Eq:TcExponential}
    f_{T_{c}}(t) = \alpha e^{-\alpha (t-t_{0})}, \quad t \geq t_{0}.
\end{equation}
In this case $\mathbb{P}(T_{c} = \infty) = 0$.
\end{example}

\begin{example}
\label{Ex:XfreeEscaping}
Consider an individual escaping from the origin with a random velocity given by a standard Gaussian vector. Suppose $D_{c}$ is the closed ball of radius $1$ centred at the origin. Then,
\begin{equation}
\label{Eq:fTcXfreeEscaping}
    f_{T_{c}}(t) = \alpha e^{-\alpha(t-t_{0})}F_{\chi^{2}_{d}}\left\{ (t-t_{0})^{-2} \right\} , \quad t \geq t_{0}.
\end{equation}
where $F_{\chi^{2}_{d}}$ denotes the cumulative distribution function of a $\chi^{2}$ random variable with $d$ degrees of freedom. In this case $\mathbb{P}(T_{c} = \infty) > 0$.
\end{example}

\subsection{Release times}
\label{Sec:ReleaseTime}

Consider now a time $t_{L} > t_{0}$ at which a captured individual is released and can be then recaptured. Two capture times are involved, the time of first capture $T_{c}^{(1)}$ and the time of second capture $T_{c}^{(2)}$. One has $T_{c}^{(2)} > \max \lbrace T_{c}^{(1)} , t_{L} \rbrace$, and if $T_{c}^{(1)} > t_{L}$, then $T_{c}^{(2)} = \infty$.
$T_{c}^{(2)}$ may be $\infty$ even if $T_{c}^{(1)} \leq t_{L} $. We establish the following principle: if the individual has never been captured then it follows the free trajectory, and if it is captured and then released \textit{it continues the same trajectory it would have followed if it had never been captured}. The effective trajectory $X$ as a function of $\tilde{X}, T_{c}^{(1)},T_{c}^{(2)}$ is then
\begin{equation}
\label{Eq:EffectTrajFunctionTc1Tc2Xfree}
    X(t) = \begin{cases}
    \tilde{X}(t) & \hbox{ if } t \leq T_{c}^{(1)} \\
    \tilde{X}(T_{c}^{(1)}) & \hbox{ if } T_{c}^{(1)} < t \leq  t_{L} \hbox{ or } t_{L} < T_{c}^{(1)} < t   \\
    \tilde{X}( T_{c}^{(1)} + t - t_{L} ) & \hbox{ if } T_{c}^{(1)} \leq t_{L} < t \leq T_{c}^{(2)} \\
    \tilde{X}( T_{c}^{(1)} + T_{c}^{(2)} - t_{L} ) & \hbox{ if } T_{c}^{(1)} \leq t_{L} \ ( <  ) \ T_{c}^{(2)} < t.
    \end{cases}
\end{equation}
Let us focus on the following auxiliary extended real random variable 
\begin{equation}
\label{Eq:DefTc2tilde}
    \tilde{T}_{c}^{(2)} =  T_{c}^{(1)} + T_{c}^{(2)} - t_{L} = T_{c}^{(2)} - ( t_{L} - T_{c}^{(1)} ).
\end{equation}
$\tilde{T}_{c}^{(2)}$ can be interpreted as the \textit{hypothetical time of second capture if the individual were released immediately after being captured at first time before $t_{L}$}. If $T_{c}^{(1)} > t_{L}$ then $\tilde{T}_{c}^{(2)} = T_{c}^{(2)} = \infty $. The notation of the capture positions gets simplified:
\begin{equation}
\label{Eq:CapturePositionsWithHypotheticalTimes}
     X(T_{c}^{(1)}) =  \tilde{X}(T_{c}^{(1)})\quad , \quad 
 X(T_{c}^{(2)})  = \tilde{X}(\tilde{T}_{c}^{(2)}).
\end{equation}

We propose a model for the triplet $(\tilde{X}, T_{c}^{(1)} , \tilde{T}_{c}^{(2)})$, equations \eqref{Eq:EffectTrajFunctionTc1Tc2Xfree} \eqref{Eq:DefTc2tilde} implying deterministically a model for the triplet $(X, T_{c}^{(1)} , T_{c}^{(2)})$. Our model will respect the following rules.

\begin{Rule}
\label{Rule:DistTc1DistTc} $T_{c}^{(1)}$ has the same distribution as $T_{c}$, the capture time without release.
\end{Rule}
\begin{Rule}
\label{Rule:DistTc2tildeGivenTc1} The hypothetical second-capture time follows the ``as if nothing had happened'' intuition:  the  distribution of $\tilde{T}_{c}^{(2)}$ given that $T_{c}^{(1)} = u$ for some $u \in (t_{0},t_{L}]$ is the distribution of $T_{c}$ given that the individual has not been captured and is in the capture domain at $u$:
    \begin{equation}
    \label{Eq:DistTc2tildeGivenTc1}
         \mu_{\tilde{T}_{c}^{(2)} \mid T_{c}^{(1)} = u} = \mu_{T_{c} \mid T_{c} > u , \tilde{X}(u) \in D_{c}}, \quad  u \in (t_{0} , t_{L}].
    \end{equation}
\end{Rule}
\begin{Rule}
\label{Rule:CaptPosInfoWithLiberation} The capture-position information axiom holds in the following sense:
    \begin{equation}
    \label{Eq:CapturePositionFirstCapture}
        \mu_{\tilde{X}(t_{1}) , \ldots , \tilde{X}(t_{n}) \mid  T_{c}^{(1)} = u } = \mu_{\tilde{X}(t_{1}) , \ldots , \tilde{X}(t_{n}) \mid \tilde{X}(u) \in D_{c} }, \quad \hbox{ if }  u < \infty;
    \end{equation}
    \begin{equation}
    \label{Eq:CapturePositionFirst&SecondCapture}
        \mu_{\tilde{X}(t_{1}) , \ldots , \tilde{X}(t_{n}) \mid \tilde{T}_{c}^{
        (2)} = v , T_{c}^{(1)} = u } = \mu_{\tilde{X}(t_{1}) , \ldots , \tilde{X}(t_{n}) \mid \tilde{X}(u) \in D_{c} , \tilde{X}(v) \in D_{c} }, \quad \hbox{ if } t_{0} < u \leq t_{L}, u < v < \infty.
    \end{equation}
    for every $t_{1} , \ldots , t_{n} \in (t_{0} , \infty)$.
\end{Rule}

Rules \ref{Rule:DistTc1DistTc} and \ref{Rule:CaptPosInfoWithLiberation} imply that the distribution of $(\tilde{X},T_{c}^{(1)})$ is the same as the distribution of $(\tilde{X},T_{c})$ developed in section \ref{Sec:DistributionCaptureTime}. For the conditional distribution of $\tilde{T}_{c}^{(2)}$ given $T_{c}^{(1)}$ we will require the auxiliary function
\begin{equation}
\label{Eq:DefPhiTc}
    \phi_{T_{c}}(t) = \frac{f_{T_{c}}(t)}{\mu_{\tilde{X}(t)}(D_{c})}, \quad t > t_{0}.
\end{equation}

\begin{proposition}
\label{Prop:DistributionTripletTc2tildeTc1Xfree}
    Rules \ref{Rule:DistTc1DistTc}, \ref{Rule:DistTc2tildeGivenTc1}, \ref{Rule:CaptPosInfoWithLiberation} are consistent and sufficient to fully characterize the distribution of the triplet $(\tilde{X} , T_{c}^{(1)} , \tilde{T_{c}}^{(2)})$. If these hold, the conditional distribution $\mu_{\tilde{T}^{(2)}_{c} \mid T_{c}^{(1)} = u}$ for $u \in (t_{0},t_{L}]$ has the density over $(t_{0},\infty)$
    \begin{equation}
    \label{Eq:DensityTc2tildeGivenTc1}
        f_{\tilde{T}_{c}^{(2)} \mid T_{c}^{(1)} = u}(v) = \alpha \mu_{\tilde{X}(v) \mid \tilde{X}(u) \in D_{c} }(D_{c}) \frac{\phi_{T_{c}}(v)}{\phi_{T_{c}}(u)} \mathbbm{1}_{(u,\infty)}(v).
    \end{equation}
\end{proposition}

\begin{remark}
\label{Rem:DensityTc2tildeGivenTc1NotProba}
As in Remark \ref{Rem:fTcNotProbDensity}, the conditional density $ f_{\tilde{T}_{c}^{(2)} \mid T_{c}^{(1)} = u}$ is not necessarily a probability density, since we have $\int_{t_{0}}^{\infty} f_{\tilde{T}_{c}^{(2)} \mid T_{c}^{(1)} = u}(v)dv \leq 1 $.
\end{remark}

In this work we consider only one release time. For general scenarios with $m-1$ release times $t_{L}^{(1)} < \cdots < t_{L}^{(m-1)}$, $m \geq 2$, one can proceed analogously defining hypothetical capture times $T_{c}^{(1)} , \tilde{T}_{c}^{(2)} , \ldots , \tilde{T}_{c}^{(m)}$, respecting Rule \ref{Rule:DistTc1DistTc} and hierarchically defining
\begin{equation}
\label{Eq:CondDistManyCaptureTimes}
    \mu_{\tilde{T}_{c}^{(k)} \mid ( T_{c}^{(1)} , \tilde{T}_{c}^{(2)} , \ldots , \tilde{T}_{c}^{(k-1)} ) = (u_{1} , u_{2} , \ldots , u_{k-1} )  } = \mu_{T_{c} \mid T_{c} > u_{k-1} , \tilde{X}(u_{1}) \in D_{c} , \ldots , \tilde{X}(u_{k-1}) \in D_{c} },
\end{equation}
for every $k = 2 , \ldots , m$, and every $(u_{1} , \ldots , u_{k-1})$ such that $t_{0} < u_{1} \leq t_{L}^{(1)}$ and $ 0 < u_{j} - u_{j-1} \leq t_{L}^{(j)} - t_{L}^{(j-1)}$ for every $j = 2 , \ldots , k-1$. The dependence structure between $\tilde{X}$ and the hypothetical capture times can be given by a suitable adaptation of Rule  \ref{Rule:CaptPosInfoWithLiberation}.

\subsection{Distribution of the counts of captured individuals}
\label{Sec:DistCaptureCount}

Consider $N$ independent individuals following this capture procedure, with or without release. Consider the random field $Q = \left\{Q_{t}(A)\right\}_{t \geq t_{0}, A \in \BorelRd}$ having the abundance information of captured individuals:
\begin{equation}
\label{Eq:DefQtACapture}
    Q_{t}(A) = \hbox{``number of retained individuals in (some point of) the set $A$ at time $t$''. }
\end{equation}
``Retained'' means that the individual has been captured at some moment before $t$: it is a cumulative information. Individuals can only be captured in $D_{c}$, thus $Q_{t}(A) = 0$ if $A \subset D_{c}^{c}$.

Let us consider $t_{1} < \cdots < t_{n}$ time steps. For each $k \in \lbrace 1 , \ldots , n \rbrace$, we consider $m_{k} \geq 1 $ sets $A_{k,1} , \ldots , A_{k,m_{k}}$ forming a  partition of $D_{c}$. We define the stochastic process $(L_{t})_{t \geq t_{0}}$ as
\begin{equation}
\label{Eq:DefLtCaptureMath}
    L_{t} = N - Q_{t}(D_{c}).
\end{equation}
$L_{t}$ is the number of free individuals at time $t$. Let $\mathscr{Q} = (\vec{Q}_{k})_{k \in \lbrace 1 , \ldots , n \rbrace}$ be the random arrangement interpreted as a list of random vectors given by
\begin{equation}
\label{Def:MargVectorsCapture}
    \vec{Q}_{k} = \left( Q_{t_{k}}(A_{k,1}) , \ldots , Q_{t_{k}}(A_{k,m_{k}}) , L_{t_{k}} \right), \quad \hbox{for all } k \in \lbrace 1 , \ldots , n \rbrace.
\end{equation}
We identify the exclusive and exhaustive categories for each time $C_{k,l}$ as
\small
\begin{equation}
\label{Eq:DefCategCapture}
        \hbox{``individual $j$ belongs to category $C_{k,l}$''} \\
        \Leftrightarrow \begin{cases}
            \hbox{``individual $j$ is retained in set $A_{k,l}$ at time $t_{k}$''} & \hbox{ if } l \leq m_{k}, \\
            \hbox{``individual $j$ is free at time $t_{k}$''} & \hbox{ if } l = m_{k}+1.
        \end{cases}
\end{equation}
\normalsize

\begin{proposition}
\label{Prop:QCaptureIsECM}
    The random arrangement $\mathscr{Q}$ follows an ECM distribution.
\end{proposition}

Unlike in the Snapshot model, here $\mathscr{Q}$  is defined as the count of independent individuals following the same stochastic dynamics over categories. Thus, Proposition \ref{Prop:QCaptureIsECM} follows by construction. 

\begin{proposition}
\label{Prop:MomentsCapture}
    The mean and covariance structures of the field $Q$  considering one release time $t_{L}$, are respectively (assuming $A,B \subset D_{c}$)
    \small
    \begin{equation}
    \label{Eq:MeanQCapture}
        E\left\{  Q_{t}(A) \right\} = N\begin{cases}
            \int_{t_{0}}^{t} \mu_{\tilde{X}(u)}(A)\phi_{T_{c}}(u)du & \hbox{ if } t \leq t_{L}, \\
            \int_{t_{L}}^{t} \mu_{\tilde{X}(u)}(A)\phi_{T_{c}}(u)du + \alpha \int_{t_{0}}^{t_{L}} \int_{u}^{u + t-t_{L}} \mu_{\tilde{X}(u),\tilde{X}(v)}(D_{c} \times A)\phi_{T_{c}}(v)dvdu &\hbox{ if } t > t_{L},
        \end{cases}
    \end{equation}
    \begin{equation}
    \label{Eq:CovQCapture}
\begin{aligned}
    &\Cov\left\{ Q_{t}(A) , Q_{s}(B) \right\}  =-\frac{1}{N}E\left\{Q_{t}(A)\right\}E\left\{Q_{s}(B)\right\} \\
    &+N\begin{cases}
             \int_{t_{0}}^{t \wedge s} \mu_{\tilde{X}(u)}( A \cap B ) \phi_{T_{c}}(u)du & \hbox{ if } t,s \leq t_{L}, \\
            \alpha\int_{t_{0}}^{t}\int_{u}^{u + s-t_{L}}\mu_{\tilde{X}(u),\tilde{X}(v)}( A \times B )\phi_{T_{c}}(v)dvdu  & \hbox{ if } t \leq t_{L} < s, \\
            \alpha\int_{t_{0}}^{t_{L}} \int_{u}^{u + t\wedge s  - t_{L}} \mu_{\tilde{X}(u),\tilde{X}(v)}\{D_{c} \times (A\cap B) \}   \phi_{T_{c}}(v)dvdu + \int_{t_{L}}^{t\wedge s}\mu_{\tilde{X}(u)}( A \cap B ) \phi_{T_{c}}(u)du  & \hbox{ if } t_{L} < t,s.
        \end{cases}
\end{aligned}
\end{equation}
\normalsize
\end{proposition}

\section{Simulation methods}
\label{Sec:Simulation}

\subsection{Simulation of the Snapshot model}
\label{Sec:SimSnapshot}

In the Snapshot model the properties of the dynamics are entirely determined by the trajectory process $X$. We assume the model for $X$ is fixed and determined by a parameter $\theta_{X}$, and that we can simulate trajectories under this model for any finite set of time points. Then, we count the points that move considering some detection error. We outline the algorithm as follows: 
\begin{Step}
Simulate $N$ independent trajectories according to $X$ over the snapshot times.
\end{Step}
\begin{Step}
Count the number of individuals at each snapshot region-time.
\end{Step}
\begin{Step}
Obtain the number of detected individuals through the binomial Rule \ref{It:SnapshotDisjSetsIndepCondPHIBinomial} at each snapshot.
\end{Step}

\subsection{Simulation of the Capture model}
\label{Sec:SimCapture}

We aim to simulate the abundance information over a capture domain $D_{c}$ within a finite-horizon time interval $(t_{0} , t_{H}]$. Consider a partition of the capture domain, $D_{c} = \bigcup_{l=1}^{m}A_{l}$. The information to be simulated is the number of captured individuals at every $A_{l}$ and every time. 

Our method follows a capture-time-to-capture-position logic, first simulating the capture times, and then the capture positions following suitable conditional distributions described below. The method requires $f_{T_{c}}$, solution to the Volterra equation \eqref{Eq:fTcVolterra}. We use a simple numerical method with Riemann sums for solving the Volterra equation, with same-length sub-intervals and right tag-points. We construct thus a regular time-grid $t_{0} < t_{1} <  \cdots  < t_{n} = t_{H}$, with $\Delta t = t_{k}-t_{k-1}$ reasonably small. For convenience, we propose a discrete-time simulation, where the capture time belongs to the same time-grid used for the discretization of the Volterra equation. 

\begin{proposition}[Capture-time to capture-position conditional distributions]
\label{Prop:CaptTimetoCaptPosCond}
    Consider the case with one release time $t_{L} > t_{0}$. Consider a time horizon $t_{H} \in (t_{L} , \infty)$. Then, 
    \begin{enumerate}[(i)]
        \item \label{It:CondDistCaptPosCapturedOnce} The conditional distribution of the first capture position given $T_{c}^{(1)}$ and $T_{c}^{(2)} > t_{H}$ (or equivalently, given $T_{c}^{(1)}$ and $\tilde{T}_{c}^{(2)} > T_{c}^{(1)} + t_{H}-t_{L}$)  is
        \small
        \begin{equation}
        \label{Eq:CondDistCaptPosCapturedOnce}
        \mu_{X(T_{c}^{(1)}) \mid T_{c}^{(1)} = u , T_{c}^{(2)} > t_{H}}(A) = \begin{cases}
            \frac{\mu_{\tilde{X}(u)}(A) }{ \mu_{\tilde{X}(u)}(D_{c}) } & \hbox{ if } u > t_{L},  \\
            \frac{  \mu_{\tilde{X}(u)}(A) - \frac{\alpha}{\phi_{T_{c}}(u)}\int_{u}^{u+t_{H}-t_{L}} \mu_{\tilde{X}(u),\tilde{X}(v)}(A \times D_{c})\phi_{T_{c}}(v)dv }{ \mu_{\tilde{X}(u)}(D_{c}) \left\{  1 - \int_{u}^{u+t_{H}-t_{L}} f_{\tilde{T}_{c}^{(2)} \mid T_{c}^{(1)} = u }(v)dv \right\}     } & \hbox{ if } u \leq t_{L},
        \end{cases}
    \end{equation}
        \normalsize
    for all $A \in \Borel{D_{c}}$.
    \item \label{It:CondDistCaptPosCapturedTwice} The conditional distribution of the capture positions given $T_{c}^{(1)}, \tilde{T}_{c}^{(2)}$ when $ T_{c}^{(1)} \leq t_{L} < T_{c}^{(2)} \leq t_{H}$ (or equivalently, when $ T_{c}^{(1)} \leq t_{L} $ and $ \tilde{T}_{c}^{(2)} < T_{c}^{(1)} + t_{H} - t_{L} $)  is
    \begin{equation}
    \label{Eq:CondDistCaptPosCapturedTwice}
        \mu_{X(T_{c}^{(1)}) , X(T_{c}^{(2)}) \mid \tilde{T}_{c}^{(2)} = v , T_{c}^{(1)} = u }(A \times B ) = \frac{\mu_{\tilde{X}(u),\tilde{X}(v)}(A\times B) }{ \mu_{\tilde{X}(u),\tilde{X}(v)}(D_{c}\times D_{c})},
    \end{equation}
    for all $A,B \in \Borel{D_{c}}$ and all $u,v$ such that $t_{0} < u \leq t_{L}$ and $u < v \leq u + t_{H} - t_{L}$.
    \end{enumerate}
\end{proposition}

 We assume the distribution of the free trajectory $\tilde{X}$ is fixed and determined by a parameter $\theta_{\tilde{X}}$, and that it can be simulated over any finite quantity of time points. The function  $(t,s) \mapsto \mu_{\tilde{X}(t),\tilde{X}(s) \mid \theta_{\tilde{X}}}(D_{c}\times D_{c}) $ plays a key role. We assume the probabilities of the form $\mu_{\tilde{X}(t_{k}),\tilde{X}(t_{k'})\mid \theta_{\tilde{X}}}(A_{l}\times A_{l'})$ for $k'\leq k$ can be computed. For instance, if the $A_{l}$'s are rectangles and $\tilde{X}$ is a Gaussian process, then libraries in many programming languages allow the computation of these probabilities with controllable efficiency and precision \citep{cao2022tlrmvnmvt}. The general outline the method is the following.

\begin{Step}
Solve numerically the Volterra equation \eqref{Eq:fTcVolterra} to obtain a discretized version of $f_{T_{c}}$ over the regular grid $t_{0} < t_{1} < \cdots < t_{n} = t_{H}$ (ensure the release time $t_{L}$ is contained in the grid). For this, use the partition $A_{1} , \ldots , A_{m}$ and obtain discretized versions of the functions $\mu_{\tilde{X}(\cdot) \mid \theta_{\tilde{X}}}(D_{c})$ and $\mu_{\tilde{X}(\cdot),\tilde{X}(\cdot) \mid \theta_{\tilde{X}}}(D_{c}\times D_{c}) $ and solve the associated linear system.
\end{Step}
\begin{Step}
Simulate independently the first capture times for $N$ individuals according to $f_{T_{c}}$.
\end{Step}
\begin{Step}
For individuals captured for the first time before $t_{L}$, simulate their second capture time using a discretized version of conditional density $f_{\tilde{T}_{c}^{(2)} \mid T_{c}^{(1)} = u}$ in Proposition \ref{Prop:DistributionTripletTc2tildeTc1Xfree}.
\end{Step}
\begin{Step}
For each captured individual, simulate the capture position(s) using Proposition \ref{Prop:CaptTimetoCaptPosCond} applied to the sets  $A_{1} , \ldots , A_{m}$.
\end{Step}
\begin{Step}
Aggregate the results on an arrangement $\mathscr{Q}$ considering when and where the individuals are captured. Keep the cumulative information over time up to $t_{L}$ and then after it.
\end{Step}

\section{Application}
\label{Sec:Application}

\subsection{Maximum Gaussian Likelihood Estimation}
\label{Sec:MGLE}

Since we lack a tractable likelihood for the ECM distribution, we cannot use a traditional maximum likelihood estimation (MLE). For a two-times setting one can rely on Propositions \ref{Prop:OneTimeDistQ} and \ref{Prop:TwoTimesDistQ} and hierarchically construct the likelihood, but even then, computing the Poisson-multinomial part can be expensive \citep{lin2022poisson}. Therefore, we  propose a maximum Gaussian likelihood estimation (MGLE) method for settings where the number of individuals is large.

Let $\mathscr{Q}^{(N)}$ be an ECM random arrangement of $N$ individuals. By definition, $\mathscr{Q}^{(N)}$ is the addition of $N$ iid random arrangements, therefore $\mathscr{Q}^{(N)}$ is asymptotically Gaussian as $N \to \infty$. More precisely, let $m_{\mathscr{Q}^{(N)}}$ and $\Sigma_{\mathscr{Q}^{(N)}}$ be respectively the mean and the covariance of $\mathscr{Q}^{(N)}$. Then, when $N \to \infty$ 
\begin{equation}
    \label{Eq:QasymptGaussian}
    \frac{\mathscr{Q}^{(N)} - N m_{ \mathscr{Q}^{(1)} } }{\sqrt{N}} \stackrel{l}{\to} \mathcal{N}(  0 , \Sigma_{\mathscr{Q}^{(1)}} ), 
\end{equation}
where $\stackrel{l}{\to}$ denotes the convergence in distribution. Thus, when $N$ is large enough, a Gaussian likelihood can be used as a pseudo-likelihood for fitting the ECM distribution. This method has the particularity of respecting both mean and covariance structures. 

Replacing an exact likelihood with a multivariate Gaussian approximation using central limit theorems is a common practice in statistics. For the Poisson-multinomial distribution, which is here involved, we refer to \citep{daskalakis2015structure,lin2022poisson}. For validating the method empirically, in section \ref{Sec:MGLEsimstudy} we do simulation studies, varying $N$ and movement parameters.

\subsection{A simple EcoDiff model as a reference}
\label{Sec:EcoDiff}

EcoDiff models treat the space-time count data as independent. In their simplest form, the marginal distribution of each count is Poisson, with rates that differ by space-time location. The intensity function is proportional to the solution of a diffusion PDE. In this work, we consider a simple advection-diffusion PDE of the form
\begin{equation}
\label{Eq:EcoDiffPDE}
\begin{cases}
\frac{\partial u}{\partial t} + v^{T}\nabla u - \frac{\sigma^{2}}{2} \Delta u = 0 & \hbox{ over } \Rd \times (t_{0} , \infty ) \\
u( \cdot , t_{0} ) = \delta_{0} & \hbox{ over } \Rd,
\end{cases}
\end{equation}
where $\sigma >0$ and $v \in \Rd$ is a velocity (advection) vector. $\nabla$ denotes the spatial gradient and $\Delta$ the spatial Laplacian. The solution $u$ is known: for every $t > t_{0}$, $u( \cdot , t )$ is a Gaussian density over $\Rd$ with mean vector $v(t-t_{0})$ and covariance matrix $\sigma^{2}(t-t_{0}) I_{d}$, $I_{d}$ being the identity matrix of size $d$. In other words, $u(\cdot , t )$ is the density of $v(t-t_{0}) + \sigma B(t-t_{0})$, where $B$ is a $d$-dimensional standard Brownian motion.

 If $Q_{t}(A)$ denotes the count of individuals in region $A$ at time $t$, the model proposes
\begin{equation}
\label{Eq:EcoDiffSpatialIntensity}
    E\left\{Q_{t}(A)\right\} = Np\int_{A} u(x,t)dx,
\end{equation}
where $N$ is the total number of individuals and $p \in [0,1]$ is a detection probability parameter. An EcoDiff model is then equivalent to a time sequence of independent spatially inhomogeneous Poisson processes, with spatial intensity \eqref{Eq:EcoDiffSpatialIntensity}. Note in addition that the expectation \eqref{Eq:EcoDiffSpatialIntensity} coincides with the expectation $E\left\{Q_{t}(A)\right\}$ when $Q$ follows a Snapshot model with same $N$ and $p$ and with individual trajectory $X(t) = v(t-t_{0}) + \sigma B(t-t_{0})$, see section \ref{Sec:Snapshot} and Eq. \eqref{Eq:MeanSnapshot}. As a final remark, the MGLE fitting method also makes sense for EcoDiff, since the Gaussian distribution can be used for approximating the Poisson distribution when rates are large $(N \to \infty)$.

\subsection{The case study data set}

As a case study, we used a subset of the data from a field experiment conducted by \citet{edelsparre2021scaling}. They released a known number of fruit flies (\textit{Drosophila melanogaster}) of three genetic strains in the middle of a meadow of $1$ ha covered with an irregular grid of $227$ traps that could retain flies without killing them. The traps were visited at irregular times after initial release for the next $3$ days, and trapped flies were counted and released. Here we restrain ourselves to the data of the Rover strain, where $N=5644$ individuals were initially released, and we consider the trap counts for the first two checks at $0.5$ and $1.5$h after first release. \citet{edelsparre2021scaling} suggest that the process of spread contains both a diffusion and an advection component, the latter likely induced by prevailing winds with a north-westerly direction. 
\begin{figure}[h]
    \centering   \includegraphics[width=0.8\textwidth]{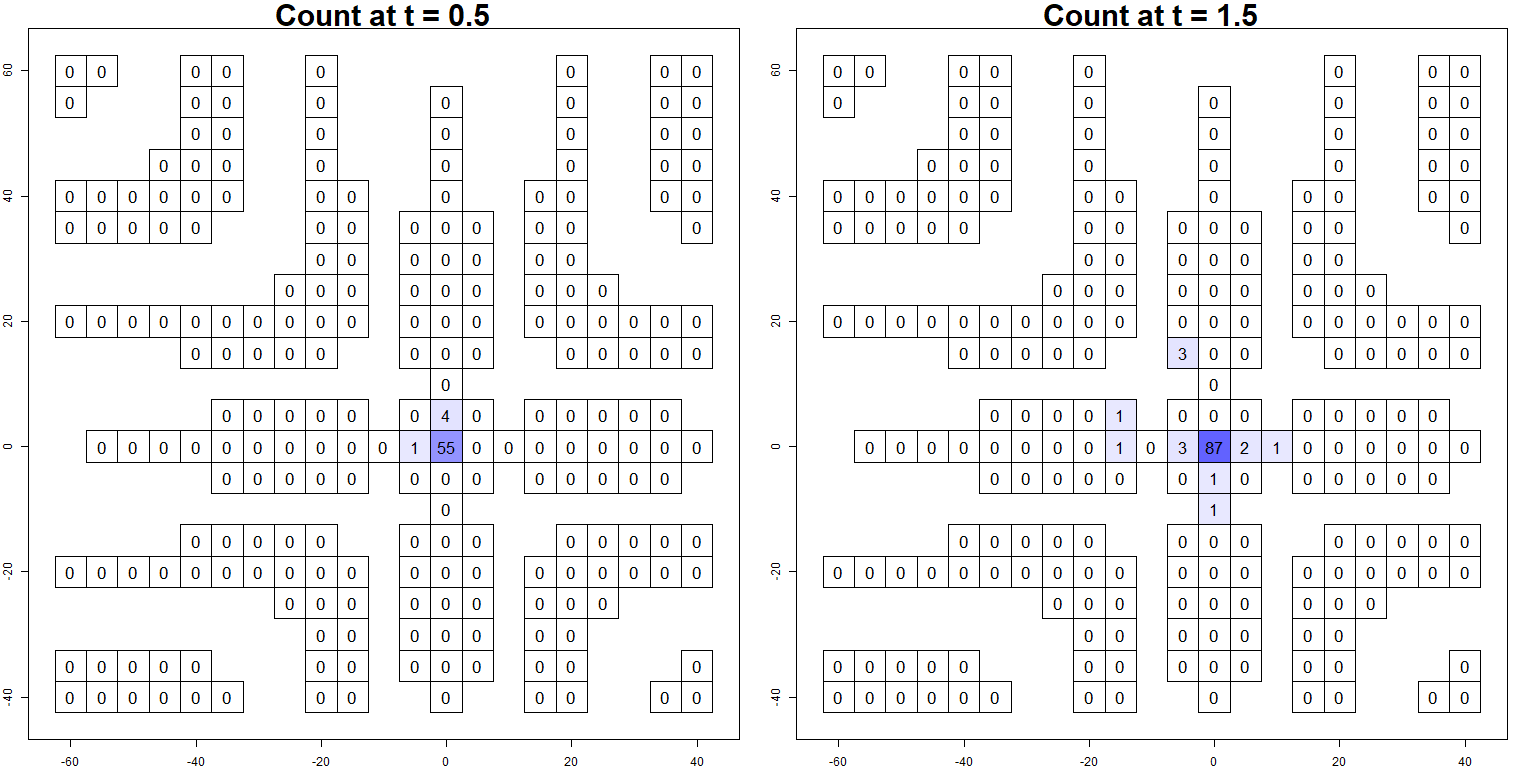}
    \caption{Experimental data of Rover flies and trap positions for $t= 0.5, 1.5$ h. Each square is a $5$m $\times 5$m region with a trap at the center, the associated number being the count of flies at that trap. The origin $(0,0)$ is the position of the trap from where the flies were released at $t = 0$.}
    \label{Fig:RoverFlyData}
\end{figure}

To each trap location we associated a square surrounding region with the trap at the center. The size of each square is the maximum so the squares do not overlap ($5$m side length). The so-recorded abundance field and the spatial trap configuration are presented in Figure \ref{Fig:RoverFlyData}. For the Snapshot model, each square is a snapshot region. For the Capture model, the union of the squares is the capture domain, and we retain this partition for the simulation algorithm.

\subsection{Simulation-based validation of the MGLE method}
\label{Sec:MGLEsimstudy}

In order to validate the MGLE method we conduct simulation studies for the three models here presented: Snapshot, Capture and EcoDiff. The space-time configuration is the one corresponding to the fly data set. We use a Brownian motion with advection as underlying trajectory in the Snapshot model and as free trajectory in the Capture model: 
\begin{equation}
    X(t) = v(t-t_{0}) + \sigma B(t-t_{0}),
\end{equation}
where $B$ is a standard $2$-dimensional Brownian motion, $v = (v_{x} , v_{y}) \in \R^{2}$ is the advection vector, and $\sigma > 0 $ is the spread parameter which controls the diffusivity of the population. The advection $v$ describes a net average population movement towards some direction. Four scalar parameters must be estimated for each model: $\sigma, v_{x}, v_{y}$ regarding the movement, $p$ for Snapshot and EcoDiff, and $\alpha$ for Capture. We gauged the quality of the estimates for $N = 10^{2}, 10^{3}, 5644, 10^{4}, 10^{5}$ and for different values of spread $\sigma = 2, 5 , 10$, representing slow, medium and fast spread. Inspired by the fly data set, we fixed a small advection $v = (-1,1)$, a small detection probability $p = 0.1$, and a small rate of growth of capture probability $\alpha = 0.1$. We simulated $300$ samples for each configuration and model and used MGLE for model fitting. For comparison, we also did exact MLE for the EcoDiff model. Here we present a synthetic description of the results, the details can be found in the supplementary material, section \ref{Sec:ResultsSimuStudy}.

Our study shows that MGLE respects the theory since better estimates are obtained as $N$ grows, suggesting asymptotically consistent estimation. However, the spread parameter $\sigma$ is always underestimated and more so for smaller $N$. For $N = 10^2$, the estimate of $\sigma$ is in average nearly $40\%$ smaller for Snapshot and EcoDiff, and more than $50\%$ for Capture. For EcoDiff with exact MLE, the bias in the estimation of $\sigma$ is negligible for $N \geq 10^3$. For all the models and fitting methods, from $N \geq 10^{5}$ the average underestimation is less than $3\%$. Estimates of the advection parameter $v$ were nearly unbiased for $N \geq 10^{3}$ in the low and middle spread scenarios, but were more biased with higher spread, becoming acceptable for $N \geq 10^{5}$. Estimates of $p$ and $\alpha$ were approximately unbiased for  $N \geq 5644$, with some positive bias for smaller $N$.

With MGLE, the negative bias in the estimation of $\sigma$ was smaller for slow spread scenarios than for medium-to-high spread scenarios. For instance, for  $N = 5644$ and $\sigma = 2$, we found a negative bias of around $5\%$ for all models, while for $\sigma = 5 , 10$, the negative bias was around $10\%$ for the EcoDiff and Snapshot models and $15\%$ for the Capture model. Similarly, the estimates of $v$, $p$ and $\alpha$ were all better for small or medium spread, and tended to require higher $N$ for high spread scenarios. This general behaviour is partially explained by the spatial sampling configuration (Figure \ref{Fig:RoverFlyData}), for which in small-spread scenarios more individuals are likely to be counted, providing more information.

While the bias of MGLE was reduced with greater $N$ in all scenarios, the coverage probabilities of the 95\%  Gauss confidence intervals and ellipsoids were quite poor and grew only very slowly with $N$, without reaching the nominal 95\%. This was more important for mid-to-high levels of spread. Clearly, MGLE yields too small variances for the estimates and thus too narrow confidence intervals. This pattern was more pronounced in the estimation of $\sigma$, where the bias partially explains the poor coverage probabilities of the overall confidence ellipsoids. In contrast, the coverage probabilities when using the exact MLE method to the EcoDiff model were near $95\%$ already for $N \geq 10^{3}$. This suggests that the Gaussian approximation is the main reason of the poor coverage probabilities with the MGLE method, which can be explained by the high contrast between the symmetric behaviour of the Gaussian distribution and the asymmetry of the Poisson and multinomial distributions for low rates.

\subsection{Fitting to real data}
\label{Sec:MGLEfitData}

Results of fitting our models to the fly data with MGLE are presented in Table \ref{tab:MGLEdata}, containing point estimates, the maximal Gaussian log-likelihood and its continuity-corrected version (see section \ref{Sec:ImplementationDetails} in supplementary material for more information on the continuity correction method). For sake of completeness, we add the $95\%$ Gauss confidence intervals, even if our simulation study shows that their coverage may be poor. Results of MLE for the EcoDiff model are also presented, for which we emphasize that the maximal log-likelihood must be compared with the continuity-corrected log-likelihoods obtained with MGLE. Since all models have $4$ parameters to be fitted, there is no need of computing Akaike information indexes for model comparison: comparing the likelihoods is enough.

\begin{table}[h]
    \caption{Fitting results for the rover fly data}
    \resizebox{\textwidth}{!}{%
   \begin{tabular}{ |c||c|c| 
 c |c|c|c|c|  }
 \hline
 \multicolumn{8}{|c|}{MGLE} \\
 \hline
 Model & $\hat{\sigma}$ & $\hat{v}_{x}$ &  $\hat{v}_{y}$ & $\hat{p}$ & $\hat{\alpha}$ & $\log(\mathcal{L})$ &  c.c. $\log(\mathcal{L})$ \\
 \hline
 EcoDiff & \begin{tabular}{@{}c@{}}$2.849$\\ $[2.758 ,  2.940]$\end{tabular}  &  \begin{tabular}{@{}c@{}}$-1.280$ \\  $[-1.557 , -1.003 ]$ \end{tabular} &  \begin{tabular}{@{}c@{}} $1.805$  \\  $[1.578 , 2.032 ]$ \end{tabular}& \begin{tabular}{@{}c@{}} $0.0294$ \\  $[0.0262 , 0.0326]$ \end{tabular} & $ - $ & $2274.47$  & $ -210.714$ \\
 \hdashline
 Snapshot   & \begin{tabular}{@{}c@{}}$2.852$ \\  $[2.761 , 2.943 ]$ \end{tabular} & \begin{tabular}{@{}c@{}}$-1.277$ \\  $ [-1.552 , -1.001 ]$ \end{tabular} & \begin{tabular}{@{}c@{}}$1.802$ \\  $[1.575 ,  2.028]$ \end{tabular} & \begin{tabular}{@{}c@{}}$0.0290$ \\ $[0.0259 , 0.0321]$  \end{tabular} &  $-$ & $2273.362$  & $-211.569$ \\
 \hdashline
 Capture   & \begin{tabular}{@{}c@{}} $3.140$\\   $[3.043 , 3.237]$ \end{tabular} & \begin{tabular}{@{}c@{}} $-1.897$\\ $[ -2.249 , -1.546 ]$  \end{tabular} & \begin{tabular}{@{}c@{}} $2.499$\\  $[2.219 ,    2.779 ]$ \end{tabular} & $-$ & \begin{tabular}{@{}c@{}} $0.03241$ \\  $[ 0.0285 ,  0.0363 ]$ \end{tabular} & $2334.782$ & $-148.619$  \\
 \hline
 \multicolumn{8}{|c|}{MLE} \\
 \hline
 EcoDiff & \begin{tabular}{@{}c@{}} $2.075$  \\ $[1.889 , 2.262 ]$ \end{tabular}& \begin{tabular}{@{}c@{}} $-0.302$ \\ $[-0.675  , 0.071 ]$\end{tabular}  &  \begin{tabular}{@{}c@{}} $0.361$ \\ $[ -0.012 ,  0.733 ]$\end{tabular}  &  \begin{tabular}{@{}c@{}} $0.0142$ \\ $[  0.0120  , 0.0164 ]$\end{tabular} & $-$  & $-$ & $-138.733$
 \\
 \hline
\end{tabular}
}
\label{tab:MGLEdata}
\caption*{MGLE: maximum Gaussian likelihood estimation method; MLE: maximum likelihood estimation method; $\log(\mathcal{L})$: maximum log-likelihood obtained ; c.c. $\log(\mathcal{L})$: continuity corrected log-likelihood.}
\end{table}

According to the MGLE, the Capture model fits the data better than the other two models. This is no surprise, since the Capture model is conceptually a more loyal representation of the studied phenomenon, considering in particular the cumulative in time information. Under the Capture model the estimates for movement parameters are higher than in both Snapshot and EcoDiff: more than $10\%$ higher for $\sigma$ and  more than $40\%$ higher for the magnitude of the components of $v$. This is also expected, since if one fits a model which assumes free individuals to a data set obtained from retained individuals, then one will underestimate the movement potential of the population. We remark that all the models yielded a non-null advection in the north-west direction, which coincides with the overall wind direction during the experiment \citep{edelsparre2021scaling}.

Estimates and maximum likelihoods under EcoDiff and Snapshot models with MGLE hardly differ. As an explanation, with the MGLE method both models assume Gaussian data with same mean vector (section \ref{Sec:EcoDiff}) but different covariance matrix, which is diagonal for EcoDiff. When using parameter values similar to the MGLE estimates, the theoretical covariances for Snapshot are actually small, making no big difference with respect to EcoDiff. We verified that for other parameters values the Snapshot model yields greater auto-correlation magnitudes, implying a larger discrepancy with EcoDiff. This is shown in Table \ref{tab:SnapshotEcoDiffComparison}, where three different parameter settings are proposed, the associated maximum absolute space-time auto-correlation for the Snapshot model is shown, and the Gaussian log-likelihoods under both models for the fly data are compared.

\begin{table}[h]
      \caption{EcoDiff v/s Snapshot models comparison}
   \resizebox{\textwidth}{!}{%
   \begin{tabular}{ |c||c|c| 
 c | }
 \hline
   Values of $(\sigma, v_{x} , v_{y} , p)$ & Snapshot max. abs. auto-corr. &  EcoDiff $\log(\mathcal{L})$   &  Snapshot $\log(\mathcal{L})$ \\
 \hline
 $(2.85 , -1.28 , 1.8 , 0.03)$ & $0.0092$ &  $2133.61$ &  $2132.37$  \\
 $(10 , -5 , 5 , 0.9)$ & $0.0649$ & $-2914.79$ &   $-8062.40$ \\
 $(5 , 0 , 0 , 0.5)$ & $0.0982$ & $-557.80$ &  $-2421.74$ \\
 \hline
\end{tabular}
}
\label{tab:SnapshotEcoDiffComparison}
\caption*{Snapshot max. abs. auto-corr.: maximal auto-correlation at different space-time locations for the Snapshot model; EcoDiff $\log(\mathcal{L})$: Gaussian continuous likelihood approximation for EcoDiff model over fly data set; Snapshot $\log(\mathcal{L})$: Gaussian continuous likelihood approximation for Snapshot model over fly data set.}
\end{table}

The comparison between MGLE and exact MLE for the EcoDiff model provides insights on the limitations of our methods and the relevance of the estimates for interpreting the fly data. There is a huge difference between MLE and MGLE estimates. MLE provides substantially smaller estimates than the MGLE: the spread parameter is $35\%$ smaller, the magnitudes of the two components of the advection vector are between $20$ and $25\%$ of the MGLE estimates, and the detection probability is reduced by $50\%$. Confidence intervals do not overlap for any of the parameters. In contrast, our simulation study shows that on average the MGLE underestimates the spread parameter more than the MLE. This indicates that the data are erratic with respect to a typical simulation of the EcoDiff model. Interestingly enough, EcoDiff with MLE provided the best model in terms of maximum obtained likelihood. It is not clear whether using a Snapshot or Capture model with exact MLE will provide higher likelihoods.

\section{Discussion}
\label{Sec:Discussion}

We conclude that, with the right models, fitting methods and space-time data configuration, it is possible to learn about features of the movement of individuals using only ``cheap'' abundance data. This comes from our simulation study, were we could estimate movement parameters with controllable bias for a scenario with only two time steps. However, one must be aware of the conditions in which our methods are more effectively applicable. MGLE requires large sample sizes in terms of number of individuals. A negative bias for spread parameters is present even for very large sample sizes, and there is a need of ad-hoc uncertainty quantification. We suggest to conduct tailor-made simulation studies to validate our methods when applied to any new data set, for one must evaluate their performance for given space-time designs and sample sizes. This also permits to judge whether the data are enough similar to typical simulated scenarios of a model. We have shown that this is not the case for the fly data and the studied EcoDiff model.

The lack of a tractable likelihood for the ECM distribution remains a challenge for data fitting. Our MGLE pseudo-likelihood method seems to be the only one available which explicitly respects the covariance structure, but has considerable limitations, so other option must be explored. Pairwise composite likelihood approaches \citep{lindsay1988composite,padoan2010likelihood,bevilacqua2012estimating} seem adequate here since we know the two-times marginals of the ECM distribution. Since we can simulate our models, likelihood-free methods can also be explored. Approximate Bayesian computation methods \citep{sisson2018handbook} are a simple option, but sufficient summary statistics and discrepancy functions have to be chosen and calibrated, which can be challenging in the context of a spreading population. Neural network-based inference \citep{lenzi2023neural,zammit2024neural} is another option, where the estimators are computed by a neural network trained over simulated data.

What is important to retain, more than the Snapshot and Capture models themselves, are the Snapshot and Capture modelling principles: the notions of counting moving individuals with detection error and capturing individuals with an axiomatically specified capture procedure. These principles can be applied beyond the scenarios here explored. For instance, in almost every ecological application $N$ is not known. In such case, one can use a prior distribution on $N$ and then use our models hierarchically. The joint estimation of $N$, movement and detection parameters requires special attention and will be possible only with highly informative data. Our work can also be extended by using various trajectory models. Ornstein-Uhlenbeck processes can be used for modelling territoriality \citep{fleming2014fine,eisaguirre2021multistate} and Itô diffusions are a rich source of models for which the mean fields are solutions to more general PDEs known as Fokker-Planck equations \citep{gardiner2009stochastic}, providing a connection with more sophisticated EcoDiff models where covariates can be considered \citep{michelot2019langevin}. Our modelling approach also allows to integrate theoretically telemetry, abundance and presence/absence data in a single analysis, and the continuous space-time formulation allows the use of irregularly sampled data. This is an important open research road for which many practical aspects such as likelihood specifications and space-time scale issues must be considered. To finish, the inclusion of non-independent trajectories, as well as death and reproduction processes is yet to be explored, for which the ECM distribution structure is no longer present but the Snapshot and Capture principles can in any case be applied.

\section*{Acknowledgement}

We warmly thank Denis Allard for his insights and suggestions on the format and content of this article, which greatly improved its quality with respect to the preprint version. We also thank Allan Edelsparre and his collaborators for providing us with an inspirational case study data.

This research has been founded by Swiss National Science Foundation grant number $310030\_207603$.

\begin{appendices}

This supplementary material contains technical details of the implementation of our methods, detailed results of the simulation study presented in section \ref{Sec:MGLEsimstudy}, and proofs of the mathematical claims here presented. Reproducible \texttt{R} routines  are openly available in the Github project \url{https://github.com/Vogelwarte/Movement-based-models-for-abundance-data}, where instructions of use and pseudo-codes are provided.


\section{Implementation details}
\label{Sec:ImplementationDetails}

Concerning the application section \ref{Sec:Application}, some technical aspects of the there used methods are here specified.

For the MGLE technique, in order to avoid singular or near-singular covariance matrices, we use a regularisation technique by adding  $\epsilon = 10^{-6}$ to the variance of each component of the Gaussian vector. This multivariate regularized Gaussian likelihood is the one to be maximized in the fitting procedure. Once the maximal continuous likelihood is obtained, a continuity-corrected version of the log-likelihood is computed in order to make this quantity comparable to likelihoods obtained from integer-valued probability distributions such as the EcoDiff model. This continuity correction is done computing $\mathbb{P}(  X \in [ x - 0.5 , x + 0.5 ] )$, where $X$ is the corresponding Gaussian vector and $x$ is the integer-valued data vector which the model must fit.

The estimations are obtained using the \texttt{optim} function of \texttt{R} with a \texttt{L-BFGS-B} algorithm. The Hessian at the optimal values, which is used to compute estimated $95\%$ Gaussian coverage intervals, is obtained using the \texttt{hessian} function from package \texttt{numDeriv}. We use the \texttt{dmnorm} function from package \texttt{mnormt} for computing the multivariate Gaussian densities, and the \texttt{pmvnorm} function from package \texttt{TruncatedNormal} for computing the corresponding continuity corrected log-likelihood, which is known to be one of the most precise options available in \texttt{R} to compute multivariate Gaussian probabilities \citep{cao2022tlrmvnmvt}. 

For the EcoDiff exact MLE, due to the high quantity of zero theoretical data (near $0$ Poisson rate for some space-time locations), we also used a regularisation technique for the Poisson distribution by adding $\epsilon = 10^{-24}$ to the rate of every Poisson component.

Concerning the implementation of the Capture model, we use a $1$-minute-resolution time grid $(\Delta t = 1/60)$ for the discrete-time grid in the simulation algorithm. The functions that simulate and compute the moments of the Capture model are ad-hoc for the Brownian motion case here presented. A method for approximating the bi-variate Gaussian probabilities required in the simulation method (section \ref{Sec:SimCapture}) is used. The method replaces conditional probabilities of the form $\mu_{\tilde{X}(t_{k}) \mid \tilde{X}(t_{k'}) \in A_{l'}  }(A_{l})$ by $\mu_{\tilde{X}(t_{k}) \mid \tilde{X}(t_{k'}) = z_{l'}  }(A_{l})$, where $z_{l'}$ is the center of the square $A_{l'}$, for $k \geq k'$. From the Markovian property of Brownian motion, such conditional probabilities do not depend upon $t_{k'}$ but only on the lag $t_{k}-t_{k'}$. This allows to compute approximate probabilities faster.

\section{Detailed results of the simulation study}
\label{Sec:ResultsSimuStudy}

The results of the simulation study are exposed here in four tables. Tables \ref{Tab:MGLESnapshot}, \ref{Tab:MGLECapture}, \ref{Tab:MGLEEcoDiff} contains the synthesized results for the MGLE method for the Snapshot, Capture and EcoDiff models respectively, and Table \ref{Tab:MLEEcoDiff} contains the results of the MLE method for the EcoDiff model. For each configuration, the average of the point estimates over the $300$ simulated data sets is presented, and the coverage probability in the $95\%$ Gaussian confidence interval is given in the following parentheses. In the final column we give the coverage probability of a joint $95\%$ Gaussian confidence ellipsoid.

\begin{table}[h]
    \caption{MGLE of Snapshot parameters on simulation study. Results over $300$ simulations.}
    \begin{center}
        \resizebox{60ex}{!}{%
    \begin{tabular}{ |c||c|c|c|c|c|  }
 \hline
   \multicolumn{6}{|c|}{MGLE for Snapshot, $(\sigma , v_{x} , v_{y} , p) = (2 , -1 , 1 , 0.1)$ } \\
 \hline
 $N$ & $\hat{\sigma}_{mean}$  & $\hat{v}_{x, mean}$  &  $\hat{v}_{y, mean}$  &  $\hat{p}_{mean}$  & ellipsoid $CP$
 \\
 \hline
$10^{2}$& 1.222 (0.06) & -1.137 (0.323) & 1.032 (0.227) & 0.133 (0.727) &  0.007 \\ 
  $10^{3}$& 1.753 (0.053) & -0.957 (0.737) & 0.961 (0.77) & 0.102 (0.94) & 0.027 \\ 
  5644 & 1.897 (0.137) & -0.995 (0.917) & 0.993 (0.863) & 0.1 (0.943) &  0.217 \\ 
  $10^{4}$& 1.934 (0.25) & -0.996 (0.923) & 0.999 (0.943) & 0.1 (0.937) &  0.37 \\ 
  $10^{5}$ & 1.989 (0.727) & -1.001 (0.957) & 1.001 (0.95) & 0.1 (0.937) &  0.837 \\ 
 \hline
 \multicolumn{6}{|c|}{MGLE for Snapshot, $(\sigma , v_{x} , v_{y} , p) = (5 , -1 , 1 , 0.1)$ } \\
 \hline
 $N$ & $\hat{\sigma}_{mean}$  & $\hat{v}_{x, mean}$  &  $\hat{v}_{y, mean}$  &  $\hat{p}_{mean}$  & ellipsoid $CP$
 \\
 \hline
$10^{2}$& 2.857 (0.003) & -0.898 (0.247) & 0.885 (0.157) & 0.189 (0.163) &  0 \\ 
  $10^{3}$& 3.964 (0.01) & -0.973 (0.3) & 0.983 (0.283) & 0.112 (0.633) & 0 \\ 
  5644 & 4.46 (0.02) & -0.99 (0.41) & 1.009 (0.403) & 0.099 (0.913) & 0 \\ 
  $10^{4}$& 4.593 (0.033) & -1.034 (0.513) & 1.032 (0.55) & 0.099 (0.86) &  0.007 \\ 
  $10^{5}$ & 4.902 (0.033) & -1.009 (0.853) & 1 (0.873) & 0.099 (0.787) & 0.03 \\ 
 \hline
 \multicolumn{6}{|c|}{MGLE for Snapshot, $(\sigma , v_{x} , v_{y} , p) = (10 , -1 , 1 , 0.1)$ } \\
 \hline
 $N$ & $\hat{\sigma}_{mean}$  & $\hat{v}_{x, mean}$  &  $\hat{v}_{y, mean}$  &  $\hat{p}_{mean}$  & ellipsoid $CP$
 \\
 \hline
$10^{2}$& 5.716 (0) & -0.617 (0.077) & 0.74 (0.1) & 0.318 (0.013) & 0 \\ 
  $10^{3}$& 8.221 (0.013) & -0.728 (0.2) & 0.761 (0.18) & 0.137 (0.027) &  0 \\ 
  5644 & 9.082 (0.013) & -0.802 (0.377) & 0.723 (0.31) & 0.104 (0.74) & 0 \\ 
  $10^{4}$& 9.272 (0.033) & -0.877 (0.36) & 0.788 (0.377) & 0.101 (0.9) & 0 \\ 
  $10^{5}$ & 9.88 (0.16) & -0.995 (0.76) & 0.998 (0.77) & 0.1 (0.837) & 0.127 \\
 \hline
\end{tabular}
}
\end{center}
\caption*{Ellipsoid CP: coverage probability of the $95\%$ Gaussian confidence ellipsoid for the four parameters estimates.}
    \label{Tab:MGLESnapshot}
\end{table}

\begin{table}[h]
    \caption{MGLE of Capture parameters on simulation study. Results over $300$ simulations.}
    \begin{center}
        \resizebox{60ex}{!}{%
    \begin{tabular}{ |c||c|c|c|c|c|  }
 \hline
   \multicolumn{6}{|c|}{MGLE for Capture, $(\sigma , v_{x} , v_{y} , \alpha) = (2 , -1 , 1 , 0.1)$ } \\
 \hline
 $N$ & $\hat{\sigma}_{mean}$  & $\hat{v}_{x, mean}$  &  $\hat{v}_{y, mean}$  &  $\hat{\alpha}_{mean}$  & ellipsoid $CP$
 \\
 \hline
$10^{2}$& 0.969 (0.051) & -1.301 (0.207) & 1.308 (0.191) & 0.13 (0.828) & 0.02 \\ 
  $10^{3}$& 1.697 (0.086) & -1.009 (0.699) & 1.058 (0.637) & 0.101 (0.93) & 0.047 \\ 
  5644 & 1.882 (0.172) & -0.981 (0.855) & 0.991 (0.867) & 0.1 (0.926) & 0.246 \\ 
  $10^{4}$& 1.916 (0.293) & -0.984 (0.879) & 0.984 (0.891) & 0.1 (0.938) & 0.363 \\ 
  $10^{5}$ & 1.986 (0.812) & -1.004 (0.934) & 1.001 (0.969) & 0.1 (0.953) &  0.891 \\ 
 \hline
 \multicolumn{6}{|c|}{MGLE for Capture, $(\sigma , v_{x} , v_{y} , \alpha) = (5 , -1 , 1 , 0.1)$ } \\
 \hline
 $N$ & $\hat{\sigma}_{mean}$  & $\hat{v}_{x, mean}$  &  $\hat{v}_{y, mean}$  &  $\hat{\alpha}_{mean}$  & ellipsoid $CP$
 \\
 \hline
$10^{2}$& 2.46 (0.01) & -0.808 (0.268) & 0.762 (0.224) & 0.204 (0.275) & 0 \\ 
  $10^{3}$& 3.756 (0.007) & -0.94 (0.295) & 0.86 (0.353) & 0.115 (0.647) & 0 \\ 
  5644 & 4.278 (0.007) & -0.966 (0.553) & 0.929 (0.485) & 0.1 (0.895) & 0.003 \\ 
  $10^{4}$& 4.423 (0.024) & -0.994 (0.539) & 0.948 (0.539) & 0.099 (0.868) & 0 \\ 
  $10^{5}$ & 4.853 (0.051) & -0.987 (0.851) & 0.995 (0.8) & 0.099 (0.817) & 0.044 \\ 
 \hline
 \multicolumn{6}{|c|}{MGLE for Capture, $(\sigma , v_{x} , v_{y} , \alpha) = (10 , -1 , 1 , 0.1)$ } \\
 \hline
 $N$ & $\hat{\sigma}_{mean}$  & $\hat{v}_{x, mean}$  &  $\hat{v}_{y, mean}$  &  $\hat{\alpha}_{mean}$  & ellipsoid $CP$
 \\
 \hline
$10^{2}$& 4.771 (0) & -0.747 (0.094) & 0.321 (0.109) & 0.369 (0.027) & 0 \\ 
  $10^{3}$& 7.786 (0.004) & -1.157 (0.156) & 1.021 (0.164) & 0.149 (0.012) & 0 \\ 
  5644 & 8.796 (0.031) & -0.651 (0.246) & 0.96 (0.266) & 0.106 (0.711) & 0 \\ 
  $10^{4}$& 9.016 (0.008) & -0.73 (0.387) & 0.82 (0.383) & 0.102 (0.906) &  0 \\ 
  $10^{5}$ & 9.704 (0.078) & -0.954 (0.715) & 0.958 (0.711) & 0.099 (0.77) & 0.027 \\ 
 \hline
\end{tabular}
}
\end{center}
\caption*{Ellipsoid CP: coverage probability of the $95\%$ Gaussian confidence ellipsoid for the four parameters estimates.}
    \label{Tab:MGLECapture}
\end{table}

\begin{table}[h]
    \caption{MGLE of EcoDiff parameters on simulation study. Results over $300$ simulations.}
    \begin{center}
        \resizebox{60ex}{!}{%
    \begin{tabular}{ |c||c|c|c|c|c|  }
 \hline
   \multicolumn{6}{|c|}{MGLE for EcoDiff, $(\sigma , v_{x} , v_{y} , p) = (2 , -1 , 1 , 0.1)$ } \\
 \hline
 $N$ & $\hat{\sigma}_{mean}$  & $\hat{v}_{x, mean}$  &  $\hat{v}_{y, mean}$  &  $\hat{p}_{mean}$  & ellipsoid $CP$
 \\
 \hline
$10^{2}$& 1.238 (0.087) & -1.208 (0.313) & 1.232 (0.297) & 0.139 (0.753) & 0 \\ 
  $10^{3}$& 1.755 (0.053) & -0.988 (0.72) & 0.967 (0.707) & 0.102 (0.93) &  0.027 \\ 
  5644 & 1.901 (0.163) & -0.996 (0.887) & 0.987 (0.893) & 0.1 (0.927) &  0.243 \\ 
  $10^{4}$& 1.934 (0.273) & -1.002 (0.933) & 0.996 (0.903) & 0.1 (0.943) &  0.427 \\ 
  $10^{5}$ & 1.991 (0.757) & -1.004 (0.923) & 1.002 (0.957) & 0.1 (0.933) &  0.843 \\ 
 \hline
 \multicolumn{6}{|c|}{MGLE for EcoDiff, $(\sigma , v_{x} , v_{y} , p) = (5 , -1 , 1 , 0.1)$ } \\
 \hline
 $N$ & $\hat{\sigma}_{mean} $  & $\hat{v}_{x, mean}$  &  $\hat{v}_{y, mean}$  &  $\hat{p}_{mean}$ & ellipsoid $CP$
 \\
 \hline
$10^{2}$& 2.812 (0) & -0.862 (0.257) & 1.027 (0.227) & 0.214 (0.18) &  0 \\ 
  $10^{3}$& 3.928 (0.003) & -0.956 (0.303) & 0.946 (0.313) & 0.115 (0.563) & 0 \\ 
  5644 & 4.505 (0.02) & -1.09 (0.39) & 1.021 (0.373) & 0.101 (0.927) & 0 \\ 
  $10^{4}$& 4.621 (0.047) & -0.982 (0.547) & 1.006 (0.573) & 0.099 (0.87) & 0 \\ 
  $10^{5}$ & 4.911 (0.05) & -1.008 (0.847) & 1 (0.857) & 0.099 (0.78) & 0.04 \\ 
 \hline
 \multicolumn{6}{|c|}{MGLE for EcoDiff, $(\sigma , v_{x} , v_{y} , p) = (10 , -1 , 1 , 0.1)$ } \\
 \hline
 $N$ & $\hat{\sigma}_{mean}$  & $\hat{v}_{x, mean}$  &  $\hat{v}_{y, mean}$  &  $\hat{p}_{mean}$  & ellipsoid $CP$
 \\
 \hline
$10^{2}$& 5.748 (0.003) & -1.089 (0.117) & 1.199 (0.123) & 0.445 (0.007) & 0 \\ 
  $10^{3}$& 8.211 (0.007) & -0.78 (0.213) & 0.769 (0.15) & 0.143 (0.01) & 0 \\ 
  5644 & 9.095 (0.04) & -0.873 (0.327) & 0.857 (0.33) & 0.105 (0.753) & 0 \\ 
  $10^{4}$& 9.316 (0.027) & -0.846 (0.357) & 0.798 (0.4) & 0.101 (0.89) & 0.003 \\ 
  $10^{5}$ & 9.863 (0.107) & -0.984 (0.79) & 0.978 (0.783) & 0.1 (0.89) & 0.07 \\ 
 \hline
\end{tabular}
}
\end{center}
\caption*{Ellipsoid CP: coverage probability of the $95\%$ Gaussian confidence ellipsoid for the four parameters estimates.}
    \label{Tab:MGLEEcoDiff}
\end{table}

\begin{table}[h]
    \caption{MLE of EcoDiff parameters on simulation study. Results over $\leq 300$ simulations.}
    \begin{center}
    \resizebox{60ex}{!}{%
    \begin{tabular}{ |c||c|c|c|c|c|  }
 \hline
   \multicolumn{6}{|c|}{MLE for EcoDiff, $(\sigma , v_{x} , v_{y} , p) = (2 , -1 , 1 , 0.1)$ } \\
 \hline
 $N$ & $\hat{\sigma}_{mean}$  & $\hat{v}_{x, mean}$  &  $\hat{v}_{y, mean}$  &  $\hat{p}_{mean}$  & ellipsoid $CP$
 \\
 \hline
$10^{2}$& 1.859 (0.967) & -1.067 (0.947) & 1.062 (0.92) & 0.101 (0.917) &  0.873 \\ 
  $10^{3}$& 1.99 (0.957) & -1.014 (0.957) & 1.004 (0.94) & 0.1 (0.957) &  0.957 \\ 
  5644 & 1.997 (0.933) & -1.007 (0.943) & 0.996 (0.957) & 0.1 (0.937) &  0.933 \\ 
  $10^{4}$& 1.999 (0.96) & -1.002 (0.963) & 0.998 (0.933) & 0.1 (0.953) &  0.95 \\ 
  $10^{5}$ & 2 (0.937) & -1.002 (0.95) & 1.001 (0.98) & 0.1 (0.943) & 0.967 \\ 
 \hline
 \multicolumn{6}{|c|}{MLE for EcoDiff, $(\sigma , v_{x} , v_{y} , p) = (5 , -1 , 1 , 0.1)$ } \\
 \hline
 $N$ & $\hat{\sigma}_{mean}$  & $\hat{v}_{x, mean}$  &  $\hat{v}_{y, mean}$  &  $\hat{p}_{mean}$  & ellipsoid $CP$
 \\
 \hline
$10^{2}$& 4.693 (0.887) & -0.959 (0.94) & 1.068 (0.927) & 0.101 (0.927) &  0.853 \\ 
  $10^{3}$& 4.957 (0.953) & -1.017 (0.933) & 0.999 (0.947) & 0.1 (0.963) &  0.947 \\ 
  5644 & 4.997 (0.967) & -1.01 (0.95) & 1.011 (0.943) & 0.1 (0.953) &  0.957 \\ 
  $10^{4}$& 4.998 (0.947) & -0.988 (0.933) & 0.999 (0.96) & 0.1 (0.967) &  0.953 \\ 
  $10^{5}$ & 5.001 (0.973) & -1.004 (0.96) & 1 (0.953) & 0.1 (0.943) &  0.95 \\ 
 \hline
 \multicolumn{6}{|c|}{MLE for EcoDiff, $(\sigma , v_{x} , v_{y} , p) = (10 , -1 , 1 , 0.1)$ } \\
 \hline
 $N$ & $\hat{\sigma}_{mean}$  & $\hat{v}_{x, mean}$  &  $\hat{v}_{y, mean}$  &  $\hat{p}_{mean}$  & ellipsoid $CP$
 \\
 \hline
$10^{2}$& 9.472 (0.883) & -1.023 (0.933) & 1.159 (0.923) & 0.103 (0.94) & 0.887 \\ 
  $10^{3}$& 9.926 (0.937) & -0.953 (0.94) & 0.991 (0.97) & 0.101 (0.96) & 0.957 \\ 
  5644 & 9.999 (0.947) & -0.987 (0.96) & 0.987 (0.967) & 0.1 (0.943) & 0.953 \\ 
  $10^{4}$& 10.001 (0.953) & -0.982 (0.933) & 1.003 (0.947) & 0.1 (0.947) & 0.95 \\ 
  $10^{5}$ & 9.996 (0.963) & -1.006 (0.987) & 0.996 (0.947) & 0.1 (0.953) & 0.977 \\ 
 \hline
\end{tabular}
}
\end{center}
\caption*{Ellipsoid CP: coverage probability of the $95\%$ Gaussian confidence ellipsoid for the four parameters estimates.}
    \label{Tab:MLEEcoDiff}
\end{table}

\section{Proofs}

\subsection{Proof of Proposition \ref{Prop:CharFunctionQ}}
\label{Proof:CharFunctionQ}

Let us begin by defining $\mathscr{X}$, an $n$-dimensional random array with dimensions $m_{1} \times \cdots \times m_{n}$ containing the counts of individuals adopting different paths. That is,
\begin{equation}
\label{Eq:DefMultiArrayX}
    \mathscr{X} = \big(  X_{l_{1} , \ldots , l_{n}} \big)_{(l_{1} , \ldots , l_{n}) \in \lbrace 1 , \ldots , m_{1} \rbrace \times \cdots \times \lbrace 1 , \ldots , m_{n} \rbrace },
\end{equation}
where each component $X_{l_{1} , \ldots , l_{n}}$ is intended to mean
\begin{equation}
\label{Eq:DefComponentMultiArrayX}
\begin{aligned}
X_{l_{1} , \ldots , l_{n}} = \hbox{``number of individuals belonging} &\hbox{ to category } C_{1,l_{1}} \hbox{ at time } 1, \\ &\hbox{ to category } C_{2,l_{2}} \hbox{ at time } 2, \\
& \cdots \ , \hbox{ AND to category } C_{n},l_{n} \hbox{ at time } n".
\end{aligned}
\end{equation}
Since each one of the $N$ individuals follows independently one and only one full path, with the probability of the full path $(l_{1} , \ldots , l_{n})$ being $p_{l_{1} , \ldots , l_{n}}^{(1,\ldots,n)}$ for every individual, then $\mathscr{X}$ is a multinomial random array, in the sense that it can be wrapped into a random vector of dimension $\prod_{k=1}^{n}m_{k}$ which follows a multinomial distribution of size $N$ and vector of probabilities given by the full path probabilities. It is concluded that the characteristic function of $\mathscr{X}$ is
\begin{equation}
    \label{Eq:CharFunctionX}
        \varphi_{\mathbf{X}}( \mathscr{\xi} ) = \left\{ \sum_{l_{1}=1}^{m_{1}}\sum_{l_{2}=1}^{m_{2}} \cdots \sum_{l_{n}=1}^{m_{n}} e^{i \xi_{l_{1}, \ldots , l_{n} }} p_{l_{1}, \ldots , l_{n}}^{(1,\ldots , n)}  \right\}^{N},
    \end{equation}
for every multi-dimensional array $ \mathscr{\xi} = \left( \xi_{l_{1} , \ldots , l_{n}}  \right)_{l_{1} , \ldots , l_{n}} \in \mathbb{R}^{m_{1}\times m_{2} \times \cdots \times m_{n}}$.

Let us now obtain the characteristic function of $\mathscr{Q}$. By definition, one can link mathematically $\mathscr{Q}$ to $\mathscr{X}$ through the sum over all-but-one margin of the random array $\mathbf{X}$, thus considering all possible paths which pass through a given category at a given time:
\begin{equation}
\label{Eq:DefQComponentMath}
    Q_{k,l} = \sum_{l_{1}=1}^{m_{1}} \cdots \sum_{l_{k-1}=1}^{m_{k-1}} \sum_{l_{k+1}=1}^{m_{k+1}} \cdots \sum_{l_{n} = 1}^{m_{n}} X_{l_{1} , \ldots , l_{k-1} , l , l_{k+1} , \ldots , l_{n}}.
\end{equation}
Using expression \eqref{Eq:DefQComponentMath}, we take $\mathscr{\xi} = \left( \xi_{k,l} \right)_{k,l}$ to be any arrangement in $\R^{m_{1} + \cdots + m_{n}}$ and we compute 
\begin{equation}
\begin{aligned}
    \varphi_{\mathscr{Q}}(\mathscr{\xi}) &= E\left( e^{i \sum_{k=1}^{n}\sum_{l_{k}=1}^{m_{k}} \xi_{k,l} Q_{k,l}} \right) \\
    &= E\left( e^{i\sum_{k=1}^{n}\sum_{l=1}^{m_{k}} \xi_{k,l} \sum_{l_{1}=1}^{m_{1}} \cdots \sum_{l_{k-1}=1}^{m_{k-1}}\sum_{l_{k+1}=1}^{m_{k+1}} \cdots \sum_{l_{n}=1}^{m_{n}} X_{l_{1} , \ldots , l_{k-1} , l , l_{k} , \ldots , l_{n}}  } \right).  \\
    &= E\left\{ e^{i\sum_{l_{1}=1}^{m_{1}} \cdots\sum_{l_{n}=1}^{m_{n}} \left( \sum_{k=1}^{n} \xi_{k,l_{k}} \right) X_{l_{1} , \ldots , l_{n}}  } \right\} \\
    &= \varphi_{\mathscr{X}}(\mathscr{\xi})\left( \xi_{1,l_{1}} + \cdots + \xi_{n,l_{n}} \right) \\
    &= \left\{ 
 \sum_{l_{1}=1}^{m_{1}} \cdots \sum_{l_{n}=1}^{m_{n}} e^{i(\xi_{1,l_{1}} + \cdots + \xi_{n,l_{n}} )} p_{l_{1} , \ldots , l_{n}}^{(1,\ldots,n)} \right\}^{N}. \quad \blacksquare
    \end{aligned}
\end{equation}

\subsection{Proof of Proposition \ref{Prop:OneTimeDistQ}}
\label{Proof:OneTimeDistQ}

Let $\vec{\xi} = (\xi_{1} , \ldots , \xi_{m_{k}} ) \in \R^{m_{k}}$. The characteristic function of the random vector $\vec{Q}_{k}$ evaluated at $\vec{\xi}$ can be simply obtained by evaluating the characteristic function of the full random arrangement $\mathscr{Q}$, in an arrangement $\tilde{\mathscr{\xi}} = (\tilde{\xi}_{k',l})_{k',l}$ defined such that $\tilde{\xi}_{k',l} = 0 $ if $k' \neq k$ and $\tilde{\xi}_{k,l} = \xi_{l}$ for every $l = 1 , \ldots , m_{k}$. Thus,
\begin{equation}
\label{Eq:CharacQkDevelopment}
\begin{aligned}
        \varphi_{\vec{Q}_{k}}(\vec{\xi}) &= \varphi_{\mathscr{Q}}(\tilde{\mathscr{\xi}}) \\
        &=\left\{ \sum_{l_{1}=1}^{m_{1}} \cdots \sum_{l_{n}=1}^{m_{n}} e^{i(\tilde{\xi}_{1,l_{1}} + \cdots + \tilde{\xi}_{n,l_{n}}   ) } p_{l_{1} , \ldots , l_{n}}^{(1,\ldots,n)} \right\}^{N} \\
        &= \left\{ \sum_{l_{1}=1}^{m_{1}} \cdots \sum_{l_{n}=1}^{m_{n}} e^{i\xi_{l_{k}} } p_{l_{1} , \ldots , l_{n}}^{(1,\ldots,n)} \right\}^{N} \\
        &=  \left\{ \sum_{l_{k}=1}^{m_{k}}e^{i\xi_{l_{k}} } \sum_{l_{1}=1}^{m_{1}} \cdots \sum_{l_{k-1}}^{m_{k-1}}
        \sum_{l_{k+1}=1}^{m_{k+1}} \cdots \sum_{l_{n}=1}^{m_{n}}p_{l_{1} , \ldots , l_{n}}^{(1,\ldots,n)} \right\}^{N}.
\end{aligned}
\end{equation}
Now, from the projectivity condition \eqref{Eq:ProjSubPathProbs} on the paths probabilities, one must have
\begin{equation}
\label{Eq:ProjectProbsTok}
    \sum_{l_{1}=1}^{m_{1}} \cdots \sum_{l_{k-1}}^{m_{k-1}}
        \sum_{l_{k+1}=1}^{m_{k+1}} \cdots \sum_{l_{n}}^{m_{n}}p_{l_{1} , \ldots , l_{n}}^{(1,\ldots,n)}  = p_{l_{k}}^{(k)}.
\end{equation}
Using \eqref{Eq:ProjectProbsTok} in \eqref{Eq:CharacQkDevelopment},we obtain
\begin{equation}
    \varphi_{\vec{Q}_{k}}(\vec{\xi}) = \left\{ \sum_{l_{k}=1}^{m_{k}}e^{i\xi_{l_{k}} } p_{l_{k}}^{(k)} \right\}^{N},
\end{equation}
which coincides with the characteristic function of a multinomial random vector of size $N$ and probabilities vector $(p_{1}^{(k)} , \ldots , p_{m_{k}}^{(k)})$. $\blacksquare$

\subsection{Proof of Proposition \ref{Prop:TwoTimesDistQ}}
\label{Proof:TwoTimesDistQ}

We remind that for two random vectors $\vec{X}$ and $\vec{Y}$ tanking values in $\R^{m}$ and $\R^{n}$ respectively, the conditioned characteristic function $\varphi_{\vec{X} \mid \vec{Y} = \vec{y}} $ satisfies
\begin{equation}
\label{Eq:ConnectionCharacPairConditionalDist}
\varphi_{\vec{X},\vec{Y}}(\vec{\xi},\vec{\eta}) = \int_{\R^{n}} \varphi_{\vec{X} \mid \vec{Y} = \vec{y}}(\vec{\xi})e^{i\vec{\eta}^{T}\vec{y}} d\mu_{\vec{Y}}(\vec{y}), \quad \vec{\xi} \in \R^{m}, \vec{\eta} \in \R^{n}. 
\end{equation}
In our case, consider $k' \neq k$. Using that $\vec{Q}_{k}$ follows a multinomial distribution (Proposition \ref{Prop:OneTimeDistQ}) we can use the principle \eqref{Eq:ConnectionCharacPairConditionalDist} and conclude that the joint characteristic function of $\vec{Q}_{k'}$ and $\vec{Q}_{k}$ satisfies
\begin{equation}
\label{Eq:JointQkQtilde{k}UsingQkMultinomial}
\begin{aligned}
    \varphi_{\vec{Q}_{k'} , \vec{Q}_{k}}(\vec{\xi} , \vec{\eta} ) &= \sum_{\substack{\vec{q} \in \lbrace 0 , \ldots , N \rbrace^{m_{k}} \\
    sum(\vec{q}) = N}} \varphi_{\vec{Q}_{k'} \mid \vec{Q}_{k} = \vec{q} }(\vec{\xi} ) e^{i\vec{\eta}^{T}\vec{q}} \mathbb{P}\left( \vec{Q}_{k} = \vec{q} \right) \\
    &= \sum_{\substack{\vec{q} \in \lbrace 0 , \ldots , N \rbrace^{m_{k}} \\
    sum(\vec{q}) = N}} \varphi_{\vec{Q}_{k'} \mid \vec{Q}_{k} = \vec{q} }(\vec{\xi} ) e^{i\vec{\eta}^{T}\vec{q}} \frac{N!}{q_{1}! \cdots q_{m_{k}}!} \left( p_{1}^{(k)} \right)^{q_{1}} \cdots \left( p_{l_{m_{k}}}^{(k)} \right)^{q_{m_{k}}}, 
    \end{aligned}
\end{equation}
for every $ \vec{\xi} = (\xi_{1} , \ldots , \xi_{m_{k'}} ) \in \R^{m_{k'}}$, and every $\vec{\eta} = (\eta_{1} , \ldots , \eta_{m_{k}} ) \in \R^{m_{k}}$. Here we denoted $sum(\vec{q}) = q_{1} + \cdots + q_{m_{k}}$. On the other hand, an explicit expression for $\varphi_{\vec{Q}_{k'} , \vec{Q}_{k}}$ can be obtained in a similar manner as the marginal characteristic function $\varphi_{\vec{Q}_{k}}$ obtained in the proof of Proposition \ref{Prop:OneTimeDistQ}, that is, by considering an arrangement with convenient null components for evaluating $\varphi_{\mathscr{Q}}$ and 
 using the projectivity condition \eqref{Eq:ProjSubPathProbs}. The final result (details are left to the reader) is
\begin{equation}
\label{Eq:JointCharacQkQtilde{k}}
    \varphi_{\vec{Q}_{k'} , \vec{Q}_{k}}(\vec{\xi} , \vec{\eta} ) = \left[ \sum_{l_{k'}=1}^{m_{k'}} \sum_{l_{k}=1}^{m_{k}} e^{i(\xi_{l_{k'}} + \eta_{l_{k}} )} p_{l_{k},l_{k'}}^{(k,k')} \right]^{N},
\end{equation}
 Using the factorisation $ p_{l_{k},l_{k'}}^{(k,k')}  =  p_{l_{k'} \mid l_{k}}^{(k' \mid k)} p_{l_{k}}^{(k)}$ and the multinomial theorem, we obtain
\small
\begin{equation}
\label{Eq:JointCharacQkQtilde{k}Development}
\begin{aligned}
    \varphi_{\vec{Q}_{k'} , \vec{Q}_{k}}(\vec{\xi} , \vec{\eta} ) &= \left\{ \sum_{l_{k}=1}^{m_{k}} e^{i\eta_{l_{k}}} p_{l_{k}}^{(k)} \sum_{l_{k'}=1}^{m_{k'}} e^{i\xi_{l_{k'}}} p_{l_{k'} \mid l_{k}}^{(k' \mid k)} \right\}^{N}  \\
    &= \sum_{\substack{\vec{q} \in \lbrace 0 , \ldots , N \rbrace^{m_{k}} \\
    sum(\vec{q}) = N}} \frac{N!}{q_{1}! \cdots q_{m_{k}}!} e^{i \vec{\eta}^{T}\vec{q} } \left\{ p_{1}^{(k)} \right\}^{q_{1}} \cdots \left\{ p_{l_{m_{k}}}^{(k)} \right\}^{q_{m_{k}}} \prod_{l_{k}=1}^{m_{k}} \left\{ \sum_{l_{k'}=1}^{m_{k'}} e^{i\xi_{l_{k'}}} p_{l_{k'} \mid l_{k}}^{(k' \mid k)}   \right\}^{q_{l_{k}}}.
\end{aligned}
\end{equation}
\normalsize
Comparing term-by-term expressions  \eqref{Eq:JointQkQtilde{k}UsingQkMultinomial} and \eqref{Eq:JointCharacQkQtilde{k}Development}, we conclude
\begin{equation}
\label{Eq:ConditionalCharFunctionQtildekQk}
    \varphi_{\vec{Q}_{k'} \mid \vec{Q}_{k} = \vec{q} }(\vec{\xi} ) = \prod_{l_{k}=1}^{m_{k}} \left\{ \sum_{l_{k'}=1}^{m_{k'}} e^{i\xi_{l_{k'}}} p_{l_{k'} \mid l_{k}}^{(k' \mid k)}   \right\}^{q_{l_{k}}}.
\end{equation}
The comparing term-by-term method is justified from the linear independence of the functions $\eta \in \R^{m_{k}} \mapsto e^{i\eta^{T}\vec{q}}$ for $\vec{q} \in \lbrace 0 , \ldots , N \rbrace^{m_{k}}$. Characteristic function \eqref{Eq:ConditionalCharFunctionQtildekQk} is thus  the product of $m_{k}$ multinomial characteristic functions, each one with the desired sizes and probabilities vectors. $\blacksquare$

\subsection{Proof of Proposition \ref{Prop:MomentsECM}}
\label{Proof:MomentsECM}

From Proposition \ref{Prop:OneTimeDistQ}, since $\vec{Q}_{k}$ is multinomial, one has immediately $E(Q_{k,l}) = Np_{l}^{(k)}$ for every $l \in \lbrace 1 , \ldots , m_{k} \rbrace$, which proves the mean structure \eqref{Eq:MeanECM}. For the same reason, we have
\begin{equation}
\label{Eq:CovQsamek}
\Cov\left( Q_{k,l} , Q_{k,l'} \right) = N \left\{\delta_{l,l'}p_{l}^{(k)} - p_{l}^{(k)}p_{l'}^{(k)} \right\},
\end{equation}
which covers the case $k = k'$. For the case $k \neq k'$, we use Proposition \ref{Prop:TwoTimesDistQ} and the properties of conditional expectations:
\small
\begin{equation}
    \begin{aligned}
        E(Q_{k,l}Q_{k',l'}) &= E\left\{ Q_{k,l}E\left( Q_{k',l'} \mid \vec{Q}_{k} \right)   \right\} \\
        &= E\left\{  Q_{k,l} \sum_{l''=1}^{m_{k}} Q_{k,l''} p_{l'\mid l''}^{(k'\mid k)}   \right\} && \substack{\hbox{(expectation of sum of indep.} \\ \hbox{multinomials)} } \\
        &= \sum_{l''=1}^{m_{k}} E\left( Q_{k,l} Q_{k,l''}\right) p_{l'\mid l''}^{(k'\mid k)} \\
        &= \sum_{l''=1}^{m_{k}} \left\{ \delta_{l,l''} N p_{l}^{(k)} + N(N-1)p_{l}^{(k)} p_{l''}^{(k)} \right\} p_{l'\mid l''}^{(k' \mid k)} && \hbox{(using \eqref{Eq:CovQsamek})} \\
        &= N p_{l}^{(k)}p_{l'\mid l}^{(k'\mid k)} + N(N-1) p_{l}^{(k)}\sum_{l''=1}^{m_{k}} p_{l'',l'}^{(k , k')} && \substack{\hbox{(split the sum and use conditional } \\ \hbox{probability products)} } \\
        &= N p_{l , l'}^{(k,k')} + N(N-1)p_{l}^{(k)} p_{l'}^{(k')}. && \substack{\hbox{(conditional probability products } \\ \hbox{and projectivity condition \eqref{Eq:ProjSubPathProbs} )}}
    \end{aligned}
\end{equation}
\normalsize
The covariance formula \eqref{Eq:CovQ} is then obtained  from $\Cov( Q_{k,l}  ,  Q_{k',l'} ) = E( Q_{k,l}Q_{k',l'} ) - E( Q_{k,l} ) E( Q_{k',l'})$. $\blacksquare$

\subsection{Proof of Proposition \ref{Prop:MomentsPhi}}
\label{Proof:MomentsPhi}

We are going to use a Lemma which will also be useful for further mathematical results.

\begin{lemma}
\label{Lemma:PHIisECM}
Let $\Phi$ be the abundance abundance random field \eqref{Eq:DefPhit(A)}. Let $t_{1} , \ldots , t_{n} \geq t_{0}$ be mutually different time steps. For each $k = 1 , \ldots , k$, let $A_{k,1} , \ldots , A_{k,m_{k}}$ be a partition of $\Rd$ made of $m_{k} \geq 1$ (Borel) subsets. Then, the random arrangement $\bm{\Phi} = \left\{ \Phi_{t_{k}}(A_{k,l}) \right\}_{k \in \lbrace 1 , \ldots , n \rbrace , l \in \lbrace 1 , \ldots , m_{k} \rbrace}$ follows an ECM distribution, with full path probabilities
\begin{equation}
\label{Eq:PathProbPHI}
p_{l_{1} , \ldots , l_{n}}^{(1, \ldots , n)} = \mu_{X(t_{1}) , \ldots , X(t_{n})}\left( A_{1,l_{1}} \times \cdots \times A_{n,l_{n}} \right).
\end{equation}
\end{lemma}

\textbf{Proof of Lemma \ref{Lemma:PHIisECM}:} If we consider the categories $C_{k,l}$ for $k = 1 , \ldots , n$ and $l = 1, \ldots , m_{k}$ defined as
\begin{equation}
\label{Eq:DefCategPHI}
\hbox{``individual $j$ belongs to category $C_{k,l}$ at time $t_{k}$''} \Leftrightarrow \hbox{``individual $j$ is in set $A_{k,l}$ at time $t_{k}$''},
\end{equation}
then by definition of the abundance random measure $\Phi$, $\bm{\Phi}$ is constructed as the aggregated count of individuals following independently the same stochastic dynamics belonging to mutually exclusive and exhaustive categories evolving across time. Therefore, $\bm{\Phi}$ follows an ECM distribution. In addition, since the individuals trajectories are iid following the same distribution as $X$, the full path probabilities are
\begin{equation}
\label{Eq:ProofFullPathProbPHI}
\begin{aligned}
p_{l_{1} , \ldots , l_{n}}^{(1, \ldots , n)} &= \mathbb{P}\left\{ X(t_{1}) \in A_{1,l_{1}} , \ldots , X(t_{n}) \in A_{n,l_{n}} \right\} \\
&= \mu_{X(t_{1}) , \ldots , X(t_{n})}\left( A_{1,l_{1}} \times \cdots \times A_{n,l_{n}} \right). \quad \blacksquare
\end{aligned}
\end{equation}

\textbf{Proof of Proposition \ref{Prop:MomentsPhi}:} 
Using Lemma \ref{Lemma:PHIisECM} together with Proposition \ref{Prop:MomentsECM}, using the partitions $A, A^{c}$ for the time $t$ and $B,B^{c}$ for the time $s$. $\blacksquare$

\subsection{Proof of Proposition \ref{Prop:SnapshotIsECM}}
\label{Proof:SnapshotIsECM}

We first use a Lemma concerning the probability generating function of an ECM random arrangement.

\begin{lemma}
\label{Lemma:GenFunctionQ}
    The probability generating function of an ECM random arrangement $\mathscr{Q}$, with number of individuals $N$, number of times $n$, with $m_{k}$ categories at each time $k$, and with full path probabilities $p_{l_{1} , \ldots , l_{n}}^{(1,..,n)}$ is
    \begin{equation}
    \label{Eq:GenFunctionQ}
        G_{\mathscr{Q}}(\mathscr{z}) = \left\{ \sum_{l_{1}=1}^{m_{1}} \cdots \sum_{l_{n}=1}^{m_{n}} p_{l_{1} , \ldots , l_{n}}^{(1,\ldots,n)} \prod_{k=1}^{n} z_{k,l_{k}} \right\}^{N},
    \end{equation}
    for every arrangement of complex numbers $\mathscr{z} = (z_{k,l})_{k,l}$ such that $|z_{k,l}| \leq 1$ for every $k,l$.
\end{lemma}

\textbf{Proof of Lemma \ref{Lemma:GenFunctionQ}:} It is enough to verify for arrangements $\mathscr{z} = (z_{k,l})_{k,l}$ such that $z_{k,l} \neq 0$. Express every complex number $z_{k,l}$ in the form $z_{k,l} = e^{\log(r_{k,l}) + i \xi_{k,l}}$, with $r_{k,l} = |z_{k,l}| > 0 $. Then, use the same arguments as in the proof of Proposition \ref{Prop:CharFunctionQ} presented in section \ref{Proof:CharFunctionQ} for the characteristic function $\varphi_{\mathscr{Q}}$, which is based on expressing $\mathscr{Q}$ in terms of the underlying multinomial random array $\mathscr{X}$. $\blacksquare$

\textbf{Proof of Proposition \ref{Prop:SnapshotIsECM}:} Let us compute the characteristic function of the random arrangement $\mathscr{Q}$. Let $\mathscr{\xi} = (\xi_{k,l})_{k,l}$ be an arrangement in $ \R^{(m_{1}+1) + \cdots + (m_{k} + 1)}$. Then,
\begin{equation}
\label{Eq:CharQtildeSnapshotDevelopment}
    \begin{aligned}
      \varphi_{\mathscr{Q}}(\mathscr{\xi} ) &= E\left(  e^{ i \sum_{k=1}^{n} \sum_{l = 1}^{m_{k}+1}  Q_{k,l} \xi_{k,l} } \right)  \\
      &= E\left(  E\left[  e^{ i \sum_{k=1}^{n} \left\{ R_{t_{k}}\xi_{k,m_{k}+1}  
 + \sum_{l = 1}^{m_{k}} Q_{t_{k}}(A_{k,l}) \xi_{k,l} \right\} } \mid \Phi \right] \right) \\
      &=  E\left( \prod_{k=1}^{n}  E\left[  e^{ i \left\{ R_{t_{k}}\xi_{k,m_{k}+1} + \sum_{l = 1}^{m_{k}} Q_{t_{k}}(A_{k,l}) \xi_{k,l} \right\} }\mid \Phi \right] \right),
    \end{aligned}
\end{equation}
where we have used the properties of the conditional expectation and the conditional independence condition at different times \ref{It:SnapshotDiffTimesIndepCondPHI}. Let us now compute the inner conditional expectations
\begin{equation}
\label{Eq:FirstCondExpectPHIproofSnapshot}
    \begin{aligned}
        E\left[  e^{ i \left\{ R_{t_{k}}\xi_{k,m_{k}+1}  + \sum_{l = 1}^{m_{k}} Q_{t_{k}}(A_{k,l}) \xi_{k,l} \right\} }\mid \Phi \right] &= E\left[  e^{ i \left\{ N \xi_{k,m_{k}+1}  + \sum_{l = 1}^{m_{k}} Q_{t_{k}}(A_{k,l}) ( \xi_{k,l} - \xi_{k,m_{k}+1}) \right\} }\mid \Phi \right] \\
 &=e^{iN \xi_{k,m_{k}+1} }E\left\{  e^{ i \sum_{l = 1}^{m_{k}} Q_{t_{k}}(A_{k,l}) ( \xi_{k,l} - \xi_{k,m_{k}+1})}\mid \Phi \right\} \\
 &= e^{iN \xi_{k,m_{k}+1} }\prod_{l=1}^{m_{k}} E\left\{  e^{ i Q_{t_{k}}(A_{k,l}) ( \xi_{k,l} - \xi_{k,m_{k}+1})}\mid \Phi \right\} \\
 &=  e^{iN \xi_{k,m_{k}+1} }\prod_{l=1}^{m_{k}}\left\{ p e^{i(\xi_{k,l} - \xi_{k,m_{k}+1})} + 1-p \right\}^{\Phi_{t_{k}}(A_{k,l})}.
    \end{aligned}
\end{equation}
Let us introduce the sets $A_{k,m_{k}+1} = \Rd \setminus \left( \bigcup_{l=1}^{m_{k}}A_{k,l} \right)$ for every $k = 1 , \ldots , n$. We have then
\begin{equation}
\label{Eq:NequalsSumProofSnapshot}
    N =  \Phi_{t_{k}}(A_{k,m_{k}+1}) + \sum_{l=1}^{m_{k}}\Phi_{t_{k}}(A_{k,l}).
\end{equation}
Using \eqref{Eq:NequalsSumProofSnapshot} in \eqref{Eq:FirstCondExpectPHIproofSnapshot}, we obtain
\small
\begin{equation}
    \begin{aligned}
         E\left[  e^{ i \left\{ R_{t_{k}}\xi_{k,m_{k}+1}  + \sum_{l = 1}^{m_{k}} Q_{t_{k}}(A_{k,l}) \xi_{k,l} \right\} } \mid  \Phi \right] &=  e^{i \Phi_{t_{k}}(A_{k,m_{k}+1}) \xi_{k,m_{k}+1} }\prod_{l=1}^{m_{k}}\left\{ p e^{i\xi_{k,l}} + (1-p)e^{i\xi_{k,m_{k}+1}} \right\}^{\Phi_{t_{k}}(A_{k,l})}. \\
         &= \prod_{l=1}^{m_{k}+1}\left\{ p e^{i\xi_{k,l}} + (1-p)e^{i\xi_{k,m_{k}+1}} \right\}^{\Phi_{t_{k}}(A_{k,l})}
    \end{aligned}
\end{equation}
\normalsize
It is then concluded that the characteristic function $\varphi_{\mathscr{Q}}$ has the form
\begin{equation}
\label{Eq:CharFuncQSnapshotAsProbGenFunctionPHI}
    \varphi_{\mathscr{Q}}(\mathscr{\xi} ) = E\left\{ \prod_{k=1}^{n}\prod_{l=1}^{m_{k}+1} z_{k,l}^{ \Phi_{t_{k}}(A_{k,l}) } \right\},
\end{equation}
with the complex numbers $(z_{k,l})_{k,l}$ being
\begin{equation}
    z_{k,l} = pe^{i\xi_{k,l}} + (1-p)e^{i\xi_{k,m_{k+1}}}.
\end{equation}
In any case $|z_{k,l}| \leq 1$. \eqref{Eq:CharFuncQSnapshotAsProbGenFunctionPHI} is nothing but the probability generating function of the random arrangement $\left( \Phi_{t_{k}}(A_{k,l})  \right)_{k\in \lbrace 1 , \ldots , n \rbrace,l \in \lbrace 1 , \ldots , m_{k}+1 \rbrace}$ evaluated at $\mathscr{z} = (z_{k,l})_{k,l}$. Since $A_{k,1} , \ldots , A_{k,m_{k+1}}$ form a partition of $\Rd$ for every $k$, from Lemma \ref{Lemma:PHIisECM} the arrangement $\left\{ \Phi_{t_{k}}(A_{k,l})  \right\}_{k\in \lbrace 1 , \ldots , n \rbrace,l \in \lbrace 1 , \ldots , m_{k}+1 \rbrace}$ follows an ECM distribution with full path probabilities  \eqref{Eq:PathProbPHI}. Using Lemma \ref{Lemma:GenFunctionQ}, we obtain

\begin{equation}
\label{Eq:CharFunctionQSnapshotBeggining}
    \varphi_{\mathscr{Q}}(\mathscr{\xi} ) = \left\{ \sum_{l_{1}=1}^{m_{1}+1} \cdots \sum_{l_{n}=1}^{m_{n}+1} \mu_{X(t_{1}) , \ldots , X(t_{n})}( A_{1,l_{1}}\times \cdots \times A_{n,l_{n}} ) \prod_{k=1}^{n} z_{k,l_{k}} \right\}^{N}.
\end{equation}
In order to prove our desired result, we need to verify that \eqref{Eq:CharFunctionQSnapshotBeggining} can be written in the form \eqref{Eq:CharFunctionQ}. For simplifying the notation, let us denote $\pi_{l_{1} , \ldots , l_{n}}^{(0)} = \mu_{X(t_{1}) , \ldots , X(t_{n})}( A_{1,l_{1}}\times \cdots \times A_{n,l_{n}} ) $. Then,

\begin{equation}
\label{Eq:CharFunctionQSnapshotDevelopment}
\begin{aligned}
    \sum_{l_{1}=1}^{m_{1}+1} \cdots \sum_{l_{n=1}}^{m_{n}+1} \pi_{l_{1} , \ldots , l_{n}}^{(0)} \prod_{k=1}^{n} z_{k,l_{k}} &= \sum_{l_{1}=1}^{m_{1}+1} \sum_{l_{2}=1}^{m_{2}+1} \cdots \sum_{l_{n}=1}^{m_{n}+1} \pi_{l_{1} , \ldots , l_{n}}^{(0)} \left\{pe^{i\xi_{1,l_{1}}} + (1-p)e^{i\xi_{1,m_{1}+1}} \right\} \prod_{k=2}^{n} z_{k,l_{k}} \\
    &= \sum_{l_{1}=1}^{m_{1}}\sum_{l_{2}=1}^{m_{2}+1} \cdots \sum_{l_{n}=1}^{m_{n}+1} \pi_{l_{1} , \ldots , l_{n}}^{(0)} \left\{pe^{i\xi_{1,l_{1}}} + (1-p)e^{i\xi_{1,m_{1}+1}} \right\} \prod_{k=2}^{n} z_{k,l_{k}} \\
    &\quad + \sum_{l_{2}=1}^{m_{2}+1} \cdots \sum_{l_{n}=1}^{m_{n}+1} \pi_{m_{1}+1 , l_{2} , \ldots , l_{n}}^{(0)} e^{i\xi_{1,m_{1}+1}} \prod_{k=2}^{n} z_{k,l_{k}} \\
    &= \sum_{l_{1}=1}^{m_{1}}\sum_{l_{2}=1}^{m_{2}+1} \cdots \sum_{l_{n}=1}^{m_{n}+1} e^{i\xi_{1,l_{1}}} p \pi_{l_{1} , \ldots , l_{n}}^{(0)} \prod_{k=2}^{n} z_{k,l_{k}} \\
    &\quad + \sum_{l_{2}=1}^{m_{2}+1} \cdots \sum_{l_{n}=1}^{m_{n}+1} e^{i\xi_{1,m_{1}+1}} (1-p)\left(\sum_{l_{1}=1}^{m_{1}}\pi_{l_{1} , \ldots , l_{n}}^{(0)} \right)  \prod_{k=2}^{n} z_{k,l_{k}} \\
    &\quad + \sum_{l_{2}=1}^{m_{2}+1} \cdots \sum_{l_{n}=1}^{m_{n}+1} e^{i\xi_{1,m_{1}+1}} \pi_{m_{1}+1 , l_{2} , \ldots , l_{n}}^{(0)} \prod_{k=2}^{n} z_{k,l_{k}} \\
    &= \sum_{l_{1}=1}^{m_{1}+1} \cdots \sum_{l_{n=1}}^{m_{n}+1} e^{i \xi_{1,l_{1}}} \pi_{l_{1} , \ldots , l_{n}}^{(1)} \prod_{k=2}^{n}z_{k,l_{k}},
    \end{aligned}
\end{equation}
where  the coefficients $ \pi_{l_{1} , \ldots , l_{n}}^{(1)} $ are
\begin{equation}
\label{Eq:ProbsPi(1)}
     \pi_{l_{1} , \ldots , l_{n}}^{(1)} = (1-\delta_{l_{1},m_{1}+1})p \pi_{l_{1} , \ldots , l_{n}}^{(0)} + \delta_{l_{1},m_{1}+1}\left\{ \pi_{m_{1}+1,l_{2} , \ldots , l_{n}}^{(0)} + (1-p)\sum_{l_{1}'=1}^{m_{1}}\pi_{l_{1}',l_{2} , \ldots , l_{n}}^{(0)} \right\}.
\end{equation}
Using the same splitting-sum arguments as shown in the sequence of equations \eqref{Eq:CharFunctionQSnapshotDevelopment}, using $z_{2,l_{2}} = pe^{i\xi_{2,l_{2}}} + (1-p)e^{i\xi_{2,m_{2}+1}}$, one can find coefficients $\pi_{l_{1} , \ldots , l_{n}}^{(2)}$ such that
\begin{equation}
\label{Eq:CharFunctionUsingpi(2)}
    \sum_{l_{1}=1}^{m_{1}+1} \cdots \sum_{l_{n=1}}^{m_{n}+1} \pi_{l_{1} , \ldots , l_{n}}^{(0)} \prod_{k=1}^{n} z_{k,l_{k}} = \sum_{l_{1}=1}^{m_{1}+1} \cdots \sum_{l_{n=1}}^{m_{n}+1} e^{i (\xi_{1,l_{1}} + \xi_{2,l_{2}})} \pi_{l_{1} , \ldots , l_{n}}^{(2)} \prod_{k=3}^{n}z_{k,l_{k}}.
\end{equation}
And repeating such arguments $n$ times one obtains
\begin{equation}
\label{Eq:CharFunctionUsingpi(n)}
     \sum_{l_{1}=1}^{m_{1}+1} \cdots \sum_{l_{n=1}}^{m_{n}+1} \pi_{l_{1} , \ldots , l_{n}}^{(0)} \prod_{k=1}^{n} z_{k,l_{k}} = \sum_{l_{1}=1}^{m_{1}+1} \cdots \sum_{l_{n=1}}^{m_{n}+1} e^{i (\xi_{1,l_{1}} + \cdots + \xi_{n,l_{n}})} \pi_{l_{1} , \ldots , l_{n}}^{(n)}.
\end{equation}
For sake of completeness, we provide the recursive formula for the coefficients $\pi_{l_{1} , \ldots , l_{n}}^{(n)}$:
\small
\begin{equation}
\label{Eq:Pi(n)Recursive}
    \pi_{l_{1} , \ldots , l_{n}}^{(k)} = (1 - \delta_{l_{k},m_{k}+1} ) p \pi_{l_{1} , \ldots , l_{n}}^{(k-1)} +  \delta_{l_{k},m_{k}+1} \left\{ \pi_{l_{1}, \ldots , l_{k-1} , m_{k}+1 , l_{k+1} , \ldots , l_{n}}^{(k-1)} + (1-p) \sum_{l_{k}'=1}^{m_{k}}\pi_{l_{1} , \ldots , l_{k-1} , l_{k}', l_{k+1} , \ldots,l_{n}}^{(k-1)} \right\},
\end{equation}
\normalsize
for $k \geq 1$. We conclude hence
\begin{equation}
\label{Eq:CharFunctionQSnapshotFinal}
    \varphi_{\mathscr{Q}}(\mathscr{\xi}) = \left\{ \sum_{l_{1}=1}^{m_{1}+1} \cdots \sum_{l_{n=1}}^{m_{n}+1} e^{i (\xi_{1,l_{1}} + \cdots + \xi_{n,l_{n}})} \pi_{l_{1} , \ldots , l_{n}}^{(n)} \right\}^{N},
\end{equation}
which is indeed the characteristic function of an ECM random arrangement, with full path probabilities given by the coefficients $\pi_{l_{1} , \ldots , l_{n}}^{(n)}$. $\blacksquare$

\subsection{Proof of Proposition \ref{Prop:MomentsSnapshot}}
\label{Proof:MomentsSnapshot}

For the expectation, we use the properties of the conditional expectation and we obtain
\begin{equation}
\begin{aligned}
        E\left\{ Q_{t}(A) \right\} &= E\left[ E\left\{ Q_{t}(A) \mid \Phi \right\}   \right] \\
        &= E\left\{ p \Phi_{t}(A)
  \right\} = p E\left\{  \Phi_{t}(A)
  \right\},
\end{aligned}
\end{equation}
where we have used that $Q_{t}(A)$, conditioned on $\Phi$, is a binomial random variable with size $\Phi_{t}(A)$ and success probability $p$ (Rule \ref{It:SnapshotDisjSetsIndepCondPHIBinomial}). The same argument can be used for the variance:
\begin{equation}
\begin{aligned}
        \Var\{ Q_{t}(A) \} &= E\left\{ Q_{t}^{2}(A)\right\} - E\left\{ Q_{t}(A)\right\}^{2} \\
        &= E\left[  E\left\{ Q_{t}^{2}(A) \mid \Phi \right\} \right] - p^{2}E\left\{ \Phi_{t}(A) \right\}^{2} \\
        &=E\left\{ p(1-p)\Phi_{t}(A) + p^{2}\Phi_{t}^{2}(A)  \right\} - p^{2}E\left\{ \Phi_{t}(A) \right\}^{2} \\
        &= p(1-p)E\left\{ \Phi_{t}(A) \right\} + p^{2}\Var\{ \Phi_{t}(A) \}.
\end{aligned}
\end{equation}
Now, if we consider $A,B$ and $t,s$ such that $A \cap B = \emptyset$ or such that $t \neq s$, we have
\begin{equation}
\begin{aligned}
      \Cov\{ Q_{t}(A) , Q_{s}(B) \} &= E\left\{ Q_{t}(A)Q_{s}(B) \right\} - E\left\{ Q_{t}(A) \right\} E\left\{ Q_{s}(B) \right\} \\
      &= E\left[ E\left\{ Q_{t}(A)Q_{s}(B) \mid \Phi \right\} \right] - p^{2}E\left\{ \Phi_{t}(A) \right\} E\left\{ \Phi_{s}(B) \right\} \\
      &= E\left[ E\left\{ Q_{t}(A)  \mid  \Phi \right) E\left( Q_{s}(B)  \mid  \Phi \right\}\right] - p^{2}E\left\{ \Phi_{t}(A) \right\} E\left\{ \Phi_{s}(B) \right\} \\
      &= E\left\{ p\Phi_{t}(A) p \Phi_{s}(B)\right\} - p^{2}E\left\{ \Phi_{t}(A) \right\} E\left\{ \Phi_{s}(B) \right\} \\
      &= p^{2}\Cov\{ \Phi_{t}(A) , \Phi_{s}(B) \},
\end{aligned}
\end{equation}
where we have used that $Q_{t}(A)$ and $Q_{s}(B)$ are independent binomial random variables conditioned on $\Phi$, according to rules \ref{It:SnapshotDiffTimesIndepCondPHI} and \ref{It:SnapshotDisjSetsIndepCondPHIBinomial}. This proves the desired result. $\blacksquare$

\subsection{Proof of Theorem \ref{Theo:DistributionTc}}

Suppose $\tilde{X}$ satisfies conditions \ref{It:FreeTrajProbDcNotNull} and \ref{It:ContinuityFreeTraj} and that there exists an extended real random variable $T_{c}$ satisfying the required axioms. Let $\mu_{T_{c}}$ be its probability law over $(t_{0},\infty]$. Let $t > t_{0}$ and $\Delta t > 0 $. Then, 
\begin{equation}
\label{Eq:ProbTcInSmallInterval}
\begin{aligned}
    \mu_{T_{c}}\left(  [t , t + \Delta t] \right) &= \mathbb{P}\left( T_{c} \in  [t , t + \Delta t]
  \right) \\
  &=\mathbb{P}\left\{  T_{c} \in [t , t + \Delta t] \ , \ \tilde{X}(t) \in D_{c}  \right\} + \mathbb{P}\left\{  T_{c} \in [t , t + \Delta t] \ , \ \tilde{X}(t) \notin D_{c} \right\}. 
\end{aligned}
\end{equation}
Let us study each term of this sum. For the first, we have
\small
\begin{equation}
\label{Eq:FactProbaTcSmallIntervalXtfreeInDc}
    \begin{aligned}
        \mathbb{P}\left\{  T_{c} \in [t , t + \Delta t]  \ , \ \tilde{X}(t) \in D_{c}  \right\} &= \mathbb{P}\left\{  T_{c} \in [t , t + \Delta t]  \ , \ T_{c} \geq t \ , \ \tilde{X}(t) \in D_{c}   \right\} \\
        &= \mathbb{P}\left\{  T_{c} \in [t , t + \Delta t] \mid T_{c} \geq t \ , \ \tilde{X}(t) \in D_{c}  \right\}\mathbb{P}\left\{ T_{c} \geq t \ , \ \tilde{X}(t) \in D_{c}  \right\}.
    \end{aligned}
\end{equation}
\normalsize
Axiom \ref{Axiom:StatRateGrowthCaptProb} implies in particular that $ \mathbb{P}\left\{  T_{c} \in [t , t + \Delta t] \mid T_{c} \geq t \ , \ \tilde{X}(t) \in D_{c}  \right\} \to 0$ as $\Delta t \to 0 $. Thus, expression \eqref{Eq:FactProbaTcSmallIntervalXtfreeInDc} also goes to $0$ as $\Delta t \to 0$. On the other hand we have
\small
\begin{equation}
\label{Eq:DevelProbTcSmallIntXfreetNotDc}
    \begin{aligned}
        \mathbb{P}\left\{  T_{c} \in [t , t + \Delta t] \ , \ \tilde{X}(t) \notin D_{c} \right\} &= \int_{[t , t + \Delta t]} \mathbb{P}\left\{ \tilde{X}(t) \in D_{c}^{c}  \mid  T_{c} = u   \right\} d\mu_{T_{c}}(u) \\
        &= \int_{[t , t + \Delta t]} \mu_{\tilde{X}(t) \mid \tilde{X}(u) \in D_{c} }(D_{c}^{c})d\mu_{T_{c}}(u) && \hbox{(axiom \ref{Axiom:CaptPosInfo})} \\
        &= \int_{(t , t + \Delta t]} \mu_{\tilde{X}(t) \mid \tilde{X}(u) \in D_{c} }(D_{c}^{c})d\mu_{T_{c}}(u) && (  \mu_{\tilde{X}(t) \mid \tilde{X}(t) \in D_{c} }(D_{c}^{c}) = 0 ) \\
        &\leq  \mu_{T_{c}}\left\{ (t , t + \Delta t] \right\}  \to 0 \ \hbox{as} \ \Delta t \to 0.
    \end{aligned}
\end{equation}
\normalsize
Since \eqref{Eq:FactProbaTcSmallIntervalXtfreeInDc} and \eqref{Eq:DevelProbTcSmallIntXfreetNotDc} go to $0$ as $\Delta t \to 0$, it is concluded that  $\mu_{T_{c}}\left( [t , t + \Delta t] \right) \to \mu_{T_{c}}\left( \lbrace t \rbrace \right) = 0$ for every $t > t_{0}$. Thus, $\mu_{T_{c}}$ has null discrete part and $T_{c}$ is indeed a continuous random variable. 

Let us now prove the existence of the limit $\lim_{\Delta t \to 0} \mu_{T_{c}}\left( [t,t+\Delta t] \right)/ \Delta t$. For this we divide by $\Delta t$ in equations \eqref{Eq:ProbTcInSmallInterval}, \eqref{Eq:FactProbaTcSmallIntervalXtfreeInDc}, and \eqref{Eq:DevelProbTcSmallIntXfreetNotDc} to obtain
\begin{equation}
\label{Eq:ProbTcSmallInterDivideddt}
    \begin{aligned}
        \frac{\mu_{T_{c}}\left( [t,t+\Delta t] \right)}{\Delta t} &= \frac{\mathbb{P}\left\{  T_{c} \in [t , t + \Delta t] \mid T_{c} \geq t \ , \ \tilde{X}(t) \in D_{c}  \right\} }{\Delta t}\mathbb{P}\left\{ T_{c} \geq t \ , \ \tilde{X}(t) \in D_{c}  \right\} \\
        &\quad+ \frac{1}{\Delta t} \int_{[t , t + \Delta t]} \mu_{\tilde{X}(t) \mid \tilde{X}(u) \in D_{c} }(D_{c}^{c})d\mu_{T_{c}}(u)
    \end{aligned}
\end{equation}
The first term in the sum in the right-side of \eqref{Eq:ProbTcSmallInterDivideddt} converges to $\alpha \mathbb{P}\left\{ T_{c} \geq t \ , \ \tilde{X}(t) \in D_{c}  \right\}$ due to axiom \ref{Axiom:StatRateGrowthCaptProb}. We claim that the second term converges to $0$. To verify this, we first consider that the positivity and continuity conditions  \ref{It:FreeTrajProbDcNotNull}, \ref{It:ContinuityFreeTraj} imply that the functions
\begin{equation}
     t \mapsto \mu_{\tilde{X}(t)}(D_{c}) \quad , \quad  \quad (t,u) \mapsto \mu_{\tilde{X}(t)\mid \tilde{X}(u) \in D_{c}}(D_{c}) \quad , \quad  (t,u) \mapsto \mu_{\tilde{X}(t)\mid \tilde{X}(u) \in D_{c}}(D_{c}^{c}),
\end{equation}
are all continuous over their respective open domains. Indeed, the function $t \mapsto \mu_{\tilde{X}(t)}(D_{c})$ is the restriction of the continuous function $(t,u) \mapsto \mu_{\tilde{X}(t),\tilde{X}(u)}(D_{c}\times D_{c})$ to the diagonal $\lbrace t = u \rbrace$ so it is continuous; the function $(t,u) \mapsto \mu_{\tilde{X}(t)\mid \tilde{X}(u) \in D_{c}}(D_{c}) $ is the division of two continuous functions with never-null denominator; and $ (t,u) \mapsto \mu_{\tilde{X}(t)\mid \tilde{X}(u) \in D_{c}}(D_{c}^{c}) = 1 -\mu_{\tilde{X}(t)\mid \tilde{X}(u) \in D_{c}}(D_{c}) $, so it is also continuous. Now, since $\mu_{\tilde{X}(t)\mid \tilde{X}(t) \in D_{c}}(D_{c}^{c}) = 0$, by continuity one has
\begin{equation}
\label{Eq:MuFreeConditionalNotInDcGoesTo0}
    \lim_{\Delta t \to 0 } \sup_{u \in [t,t+\Delta t]} \mu_{\tilde{X}(t)\mid \tilde{X}(u) \in D_{c}}(D_{c}^{c}) =   \mu_{\tilde{X}(t)\mid \tilde{X}(t) \in D_{c}}(D_{c}^{c}) = 0.
\end{equation}
Let $\epsilon > 0$ (smaller than $1$). Suppose $\Delta t$ is small enough so 
\small
\begin{equation}
\label{Eq:BoundincProbasWithEpsilon}
    \frac{\mathbb{P}\left\{  T_{c} \in [t , t + \Delta t] \mid T_{c} \geq t \ , \ \tilde{X}(t) \in D_{c}  \right\} }{\Delta t} \leq \alpha + \epsilon \quad \hbox{and} \quad \sup_{u \in [t,t+\Delta t]} \mu_{\tilde{X}(t)\mid \tilde{X}(u) \in D_{c}}(D_{c}^{c}) \leq \epsilon.
\end{equation}
\normalsize
Then, we can bound \eqref{Eq:ProbTcSmallInterDivideddt} through
\small
\begin{equation}
\begin{aligned}
    \frac{\mu_{T_{c}}\left( [t,t+\Delta t] \right)}{\Delta t} &\leq (\alpha + \epsilon)\mathbb{P}\left\{ T_{c} \geq t \ , \ \tilde{X}(t) \in D_{c}  \right\}  + \frac{\mu_{T_{c}}\left( [t,t+\Delta t] \right)}{\Delta t}\sup_{u \in [t,t+\Delta t]} \mu_{\tilde{X}(t)\mid \tilde{X}(u) \in D_{c}}(D_{c}^{c})  \\
    &\leq \alpha + \epsilon + \epsilon\frac{\mu_{T_{c}}\left( [t,t+\Delta t] \right)}{\Delta t},
    \end{aligned}
\end{equation}
\normalsize
and so
\begin{equation}
    \frac{\mu_{T_{c}}\left( [t,t+\Delta t] \right)}{\Delta t} \leq \frac{\alpha + \epsilon}{1- \epsilon}.
\end{equation}
$\mu_{T_{c}}\left( [t,t+\Delta t] \right)/ \Delta t$ is thus bounded as $\Delta t \to 0$. We obtain thus,
\small
\begin{equation}
    \frac{1}{\Delta t} \int_{[t , t + \Delta t]} \mu_{\tilde{X}(t) \mid \tilde{X}(u) \in D_{c} }(D_{c}^{c})d\mu_{T_{c}}(u) \leq \frac{\mu_{T_{c}}\left( [t,t+\Delta t] \right)}{\Delta t}\sup_{u \in [t,t+\Delta t]} \mu_{\tilde{X}(t)\mid \tilde{X}(u) \in D_{c}}(D_{c}^{c}) \to 0.
\end{equation}
\normalsize
Doing $\Delta t \to 0 $ in \eqref{Eq:ProbTcSmallInterDivideddt} we conclude
\begin{equation}
\label{Eq:LimitProbTcSmallIntDivdt}
    \lim_{\Delta t \to 0 } \frac{\mu_{T_{c}}\left( [t,t+\Delta t] \right)}{\Delta t} = \alpha \mathbb{P}\left\{ T_{c} \geq t \ , \ \tilde{X}(t) \in D_{c}  \right\}.
\end{equation}
This implies that the cumulative probability function of $T_{c}$ has a right-derivative at every $t > t_{0}$, which we call $f_{T_{c}}$:
\begin{equation}
\label{Eq:DeffTcAsRightDer}
    f_{T_{c}}(t) = \lim_{\Delta t \to 0 } \frac{\mu_{T_{c}}\left( [t,t+\Delta t] \right)}{\Delta t} = \alpha \mathbb{P}\left\{ T_{c} \geq t \ , \ \tilde{X}(t) \in D_{c}  \right\}.
\end{equation}
We can develop expression \eqref{Eq:DeffTcAsRightDer} using the axiom \ref{Axiom:CaptPosInfo} and basic probability computations
\begin{equation}
\label{Eq:fTcDevelopment}
\begin{aligned}
    f_{T_{c}}(t) &= \alpha \left[ \mathbb{P}\left\{ \tilde{X}(t) \in D_{c}  \right\} - \mathbb{P}\left\{ \tilde{X}(t) \in D_{c} , T_{c} < t \right\} \right]  \\
    &= \alpha \left[ \mu_{\tilde{X}(t)} (D_{c}) - \int_{(t_{0} , t)} \mathbb{P}\left\{ \tilde{X}(t) \in D_{c} \mid T_{c} = u  \right\}d\mu_{T_{c}}(u)  \right] \\
    &= \alpha \left[ \mu_{\tilde{X}(t)} (D_{c}) - \int_{(t_{0} , t)} \mu_{\tilde{X}(t) \mid \tilde{X}(u) \in D_{c}}(D_{c})d\mu_{T_{c}}(u)  \right].
\end{aligned}
\end{equation}
Now we claim that $f_{T_{c}}$ is continuous at every $t > t_{0}$. Since the function $t \mapsto \mu_{\tilde{X}(t)}(D_{c})$ is continuous, we only need to prove the continuity of $t \mapsto  \int_{(t_{0} , t)} \mu_{\tilde{X}(t) \mid \tilde{X}(u) \in D_{c}}(D_{c})d\mu_{T_{c}}(u)$. This can be done using a typical application of Lebesgue Dominated Convergence Theorem, which applies since the integrated is bounded and the measure is finite. Using the continuity of $(t,u) \mapsto  \mu_{\tilde{X}(t) \mid \tilde{X}(u) \in D_{c}}(D_{c})$, we have
\begin{equation}
    \lim_{t_{n} \to t^{-}} \int_{(t_{0} , t_{n})} \mu_{\tilde{X}(t_{n}) \mid \tilde{X}(u) \in D_{c}}(D_{c})d\mu_{T_{c}}(u) =\int_{(t_{0} , t)} \mu_{\tilde{X}(t) \mid \tilde{X}(u) \in D_{c}}(D_{c})d\mu_{T_{c}}(u), 
\end{equation}
which proves the left-continuity. On the other hand
\small
\begin{equation}
\begin{aligned}
  \lim_{t_{n} \to t^{+}} \int_{(t_{0} , t_{n})} \mu_{\tilde{X}(t_{n}) \mid \tilde{X}(u) \in D_{c}}(D_{c})d\mu_{T_{c}}(u) &= \int_{(t_{0} , t]} \mu_{\tilde{X}(t) \mid \tilde{X}(u) \in D_{c}}(D_{c})d\mu_{T_{c}}(u)   \\
  &= \int_{(t_{0} , t)} \mu_{\tilde{X}(t) \mid \tilde{X}(u) \in D_{c}}(D_{c})d\mu_{T_{c}}(u) + \underbrace{\mu_{\tilde{X}(t) \mid \tilde{X}(t) \in D_{c}}(D_{c})}_{=1}\underbrace{\mu_{T_{c}}(\lbrace t \rbrace ) }_{=0} \\
  &=  \int_{(t_{0} , t)} \mu_{\tilde{X}(t) \mid \tilde{X}(u) \in D_{c}}(D_{c})d\mu_{T_{c}}(u),
\end{aligned}
\end{equation}
\normalsize
where we have used that $\mu_{T_{c}}$ has no atoms. This proves the right-continuity of $f_{T_{c}}$ and thus $f_{T_{c}}$ is continuous over $(t_{0}, \infty)$.

Finally, since the cumulative distribution function of $T_{c}$, $F_{T_{c}}(t) = \mu_{T_{c}}\{(t_{0} , t ]\}$ is continuous ($\mu_{T_{c}}$ has no atoms) and has continuous right-derivative, it is differentiable\footnote{See for example the Math-Exchange discussion \url{https://math.stackexchange.com/questions/418737/continuous-right-derivative-implies-differentiability} on the subject, consulted for the last time on July the $16$th $2024$.}. The measure $\mu_{T_{c}}$ has thus a density over $(t_{0},\infty)$ which coincides with $f_{T_{c}} = F_{T_{c}}'$. From \eqref{Eq:fTcDevelopment}, $f_{T_{c}}$ satisfies the integral equation
\begin{equation}
\label{Eq:VolterrafTcInProof}
    f_{T_{c}}(t) =  \alpha \left\{ \mu_{\tilde{X}(t)} (D_{c}) - \int_{t_{0}}^{t} \mu_{\tilde{X}(t) \mid \tilde{X}(u) \in D_{c}}(D_{c})f_{T_{c}}(u)du  \right\}.
\end{equation}
The continuity of the functions $u \mapsto \mu_{\tilde{X}(u)}(D_{c})$ and $(t,u) \mapsto \mu_{\tilde{X}(t) \mid \tilde{X}(u) \in Dc}(D_{c})$ guarantees the existence and the uniqueness of a continuous function solution to  \eqref{Eq:VolterrafTcInProof} (well known result on Volterra integral equations, see \citep{volterra1913leccons,brunner2017volterra}). It follows that such solution must coincide with $f_{T_{c}}$ which is then the only possible density for the probability distribution of $T_{c}$ over $(t_{0},\infty)$. $\blacksquare$

\subsection{Proof of Proposition \ref{Prop:AlphaSmallTcPosInt}}
\label{Proof:AlphaSmallTcPosInt}

We express the problem as a fixed-point problem involving a contraction. Consider the following Lemma.

\begin{lemma}
\label{Lemma:FixedInSubset}
    Let $E$ be a real Banach space with norm $\| \cdot \|_{E}$. Let $\mathcal{L}: E \to E$ be a contraction mapping with constant $r < 1$. Let $\tilde{E} \subset E$ be a non-empty closed subset of $E$. Suppose in addition that $\mathcal{L}(\tilde{E}) \subset \tilde{E}$. Then, the unique solution of the fixed-point problem $\mathcal{L}(x) = x $ is in $\tilde{E}$.
\end{lemma}

\textbf{Proof of Lemma \ref{Lemma:FixedInSubset}:} This is trivial following the same principles as in Banach fixed-point Theorem \citep[Theorem 3.46]{burkill2002second}: take $x_{0} \in \tilde{E}$, and define the sequence $x_{n} = \mathcal{L}^{(n)}(x_{0})$. This sequence is contained in $\tilde{E}$ and it converges to the unique solution of the fixed-point problem $x = \mathcal{L}(x)$. Since $\tilde{E}$ is closed, $x \in \tilde
{E}$. $\blacksquare$

\textbf{Proof of Proposition \ref{Prop:AlphaSmallTcPosInt}}: Consider the Banach space of continuous functions over $[t_{0},t_{H}]$, $E = C([t_{0} , t_{H}])$ endowed with the supremum norm $\| \cdot \|_{\infty}$. Consider the operator $\mathcal{L} : E \to E $ defined by
\begin{equation}
\label{Eq:DefOperatorL}
    \mathcal{L}(f)(t) = \alpha \mu_{\tilde{X}(t)}(D_{c}) - \alpha \int_{t_{0}}^{t} \mu_{\tilde{X}(t) \mid \tilde{X}(u) \in D_{c}}(D_{c}) f(u)du.
\end{equation}
Continuity and positivity conditions \ref{It:FreeTrajProbDcNotNull}, \ref{It:ContinuityFreeTraj} allow to ensure that $\mathcal{L}(f)$ is indeed a continuous function over $[t_{0} , t_{H}]$. The Volterra equation \eqref{Eq:fTcVolterra} can be thus re-written as the fixed-point problem
\begin{equation}
\label{Eq:VolterraAsFixedPoint}
    f = \mathcal{L}(f).
\end{equation}
Since we know that Volterra equation \eqref{Eq:fTcVolterra} has a unique continuous solution, the only task is to verify if such solution is indeed positive with total integral over $[t_{0} , t_{H}]$ smaller than $1$. For that, we consider the subset of $E$:
\begin{equation}
\label{Eq:DefEtilde}
    \tilde{E} = \lbrace \ f \in C([t_{0},t_{H}]) \mid 0 \leq f \leq \alpha \mu_{\tilde{X}(\cdot)}(D_{c}) \rbrace.
\end{equation}
Since the function $t \mapsto \mu_{\tilde{X}(t)}(D_{c})$ is strictly positive over $(t_{0} , t_{H}]$, then $\tilde{E}$ is a non-empty set (the same function $\alpha\mu_{\tilde{X}(\cdot)}(D_{c})$ belongs to it). It is trivial to verify that $\tilde{E}$ is closed under the supremum norm.

Now, let us verify $\mathcal{L}$ is a contraction. Set $r = \alpha(t_{H}-t_{0}) < 1$.  Let $f_{1},f_{2} \in E$. Then,
\begin{equation}
\begin{aligned}
    |\mathcal{L}(f_{1} - f_{2})(t)| &= \left| -\alpha \int_{t_{0}}^{t} \underbrace{\mu_{\tilde{X}(t) \mid \tilde{X}(u) \in D_{c}}(D_{c})}_{\leq 1} (f_{1}-f_{2})(u) du  \right| \\
    &\leq \alpha(t-t_{0}) \| f_{1} - f_{2} \|_{\infty} \\
    &\leq r \| f \|_{\infty}, && \hbox{for all } t \in [t_{0},t_{H}].
\end{aligned}
\end{equation}
This proves $\mathcal{L}$ is indeed a contraction with constant $r < 1$. Now, let $f \in \tilde{E}$. Since $f \geq 0$, it is immediate that $\mathcal{L}(f)(t) \leq \alpha \mu_{\tilde{X}(t)}(D_{c})$ for every $t \in [t_{0},t_{H}]$. On the other hand,
\begin{equation}
\begin{aligned}
      \mathcal{L}(f)(t) &=  \alpha \mu_{\tilde{X}(t)}(D_{c}) - \alpha \int_{t_{0}}^{t} \mu_{\tilde{X}(t) \mid \tilde{X}(u) \in D_{c} }(D_{c}) f(u)du \\
      &\geq \alpha \mu_{\tilde{X}(t)}(D_{c}) -  \alpha \int_{t_{0}}^{t} \mu_{\tilde{X}(t) \mid \tilde{X}(u) \in D_{c} }(D_{c}) \alpha \mu_{\tilde{X}(u)}(D_{c})  du  \\
      &=  \alpha \mu_{\tilde{X}(t)}(D_{c}) -  \alpha^{2} \int_{t_{0}}^{t} \mu_{\tilde{X}(t), \tilde{X}(u) }(D_{c} \times D_{c})  du \\
      &\geq \alpha \mu_{\tilde{X}(t)}(D_{c}) - \alpha^{2} \mu_{\tilde{X}(t)}(D_{c})(t-t_{0}) \\
      &\geq \alpha \mu_{\tilde{X}(t)}(D_{c}) - \alpha  \mu_{\tilde{X}(t)}(D_{c})r \\
      &= \alpha \mu_{\tilde{X}(t)}(D_{c})( 1 - r ) \geq 0.
\end{aligned}
\end{equation}
Thus $\mathcal{L}(f) \geq 0$ for every $f \in \tilde{E}$. We conclude that $\mathcal{L}(\tilde{E}) \subset \tilde{E}$, and from Lemma \ref{Lemma:FixedInSubset}, the solution $f_{T_{c}}$ of the Volterra fixed-point problem \eqref{Eq:VolterraAsFixedPoint} must be in $\tilde{E}$. Finally, since $f_{T_{c}} \in \tilde{E}$, we have $f_{T_{c}} \geq 0 $ and also
\begin{equation}
    \int_{t_{0}}^{t_{h}}f_{T_{c}}(t)dt \leq \int_{t_{0}}^{t_{h}} \alpha
     \mu_{\tilde{X}(t)}(D_{c})dt \leq \alpha(t_{H} - t_{0}) = r < 1.\quad \blacksquare
\end{equation}

\subsection{Justification of Example \ref{Ex:TcExponential}}

In this scenario $\mu_{\tilde{X}(t)}(D_{c}) = 1 = \mu_{\tilde{X}(t) \mid \tilde{X}(u) \in D_{c}}(D_{c})$ for every $t,u \geq t_{0}$. Therefore, the Volterra equation \eqref{Eq:fTcVolterra} becomes
\begin{equation}
    f_{T_{c}}(t) = \alpha - \alpha \int_{t_{0}}^{t}f_{T_{c}}(u)du,
\end{equation}
from where it is clear that $f_{T_{c}}$ must be the exponential density \eqref{Eq:TcExponential}. $\blacksquare$

\subsection{Justification of Example \ref{Ex:XfreeEscaping}}

In this scenario the free trajectory is $\tilde{X}(t) = V(t-t_{0})$, where $V \sim \mathcal{N}(0 , I_{d})$, being $I_{d}$ the identity matrix of size $d$. Since $D_{c}$ is the closed ball of unit radius centred at the origin, we have for $t > t_{0}$
\begin{equation}
\begin{aligned}
        \mathbb{P}\left\{ \tilde{X}(t) \in D_{c} \right\} &= \mathbb{P}\left\{  \| V(t-t_{0}) \| \leq 1 \right\} \\
        &= \mathbb{P}\left\{  \| V \|^{2} \leq (t-t_{0})^{-2} \right\} \\
        &= F_{\chi_{d}^{2}}\left\{ (t-t_{0})^{-2} \right\},
\end{aligned}
\end{equation}
where we have used that $\| V \|^{2}$ follows a chi-squared distribution with $d$ degrees of freedom. Now, if we consider $t_{0} < u \leq t$, then $  \| V(u-t_{0}) \| \leq \|V(t-t_{0})\|$, therefore
\begin{equation}
\begin{aligned}
    \mathbb{P}\left\{  \tilde{X}(t) \in D_{c} , \tilde{X}(u) \in D_{c} \right\} &= \mathbb{P}\left\{ \| V(t-t_{0}) \| \leq 1 , \| V(u-t_{0}) \| \leq 1 \right\} \\
    &= \mathbb{P}\left\{ \| V(t-t_{0}) \| \leq 1 \right\} \\
    &= F_{\chi^{2}_{d}}\left\{ (t-t_{0})^{-2}  \right\},
    \end{aligned}
\end{equation}
and thus
\begin{equation}
    \mu_{\tilde{X}(t) \mid \tilde{X}(u) \in D_{c}}(D_{c}) = \frac{F_{\chi^{2}_{d}}\left\{ (t-t_{0})^{-2}  \right\}}{F_{\chi^{2}_{d}}\left\{ (u-t_{0})^{-2}  \right\}}, \quad t_{0} < u \leq t.
\end{equation}
The Volterra equation \eqref{Eq:fTcVolterra} becomes
\begin{equation}
\label{Eq:VolterraExample2}
    f_{T_{c}}(t) = \alpha F_{\chi^{2}_{d}}\left\{ (t-t_{0})^{-2}  \right\} - \alpha \int_{t_{0}}^{t} \frac{F_{\chi^{2}_{d}}\left\{ (t-t_{0})^{-2}  \right\}}{F_{\chi^{2}_{d}}\left\{ (u-t_{0})^{-2}  \right\}} f_{T_{c}}(u)du.
\end{equation}
If we divide equation \eqref{Eq:VolterraExample2} by $ F_{\chi^{2}_{d}}\left\{ (t-t_{0})^{-2}  \right\}$, and we define
\begin{equation}
    \phi_{T_{c}}(t) = \frac{f_{T_{c}}(t)}{ F_{\chi^{2}_{d}}\left\{ (t-t_{0})^{-2}  \right\}},
\end{equation}
we obtain
\begin{equation}
    \phi_{T_{c}}(t) = \alpha - \alpha\int_{t_{0}}^{t}\phi_{T_{c}}(u)du.
\end{equation}
Therefore, $\phi_{T_{c}}$ is an exponential density,
\begin{equation}
    \phi_{T_{c}}(t) = \alpha e^{-\alpha(t-t_{0})},
\end{equation}
from which we obtain
\begin{equation}
    f_{T_{c}}(t) =  \alpha e^{-\alpha(t-t_{0})}F_{\chi^{2}_{d}}\left\{ (t-t_{0})^{-2}  \right\}.
\end{equation}
For the claim that $\mathbb{P}(T_{c} = \infty ) > 0,$ we argue that
\begin{equation}
\begin{aligned}
       \mathbb{P}(T_{c} < \infty ) &=  \int_{t_{0}}^{\infty} f_{T_{c}}(t)dt \\
       &= \int_{t_{0}}^{\infty} \alpha e^{-\alpha(t-t_{0})} F_{\chi^{2}_{d}}\left\{ (t-t_{0})^{-2}  \right\} dt \\
       &= e^{-\alpha(t-t_{0})} F_{\chi^{2}_{d}}\left\{ (t-t_{0})^{-2}  \right\}\Big|^{t=t_{0}}_{t= \infty} + \int_{t_{0}}^{\infty} e^{-\alpha(t-t_{0})} f_{\chi^{2}_{d}}\left\{ (t-t_{0})^{-2}  \right\} \frac{(-2)}{(t-t_{0})^{3}}dt \\
       &= 1 - 2\int_{t_{0}}^{\infty} e^{-\alpha(t-t_{0})} f_{\chi^{2}_{d}}\left\{ (t-t_{0})^{-2}  \right\} (t-t_{0})^{-3}dt \\
       &< 1,
\end{aligned}
\end{equation}
where we have used integration by parts, and $f_{\chi^{2}_{d}}$ denotes the probability density function of a chi-squared random variable with $d$ degrees of freedom. It follows that $\mathbb{P}(T_{c} = \infty) > 0.$ $\blacksquare$

\subsection{Proof of Proposition \ref{Prop:DistributionTripletTc2tildeTc1Xfree}}

 We verify that the distribution of the triplet $\tilde{X},T_{c}^{(1)},\tilde{T}_{c}^{(2)}$ is correctly defined hierarchically. The distribution of $T_{c}^{(1)}$ is determined by rule \ref{Rule:DistTc1DistTc}. Rule \ref{Rule:DistTc2tildeGivenTc1} and the condition $\tilde{T}_{c}^{(2)} = \infty$ if $T_{c}^{(1)} > t_{L}$ fully determine the distribution of $\tilde{T}_{c}^{(2)}$ given $T_{c}^{(1)}$:
\begin{equation}
\label{Eq:MuTc2GivenTc1Proof}
    \mu_{\tilde{T}_{c}^{(2)} \mid T_{c}^{(1)} = u} = \mu_{T_{c} \mid T_{c} > u , \tilde{X}(u) \in D_{c}} \mathbbm{1}_{(t_{0} , t_{L}]}(u) + \delta_{\infty} \mathbbm{1}_{(t_{L}, \infty]}(u).
\end{equation}
Let $u \in (t_{0}, \infty) $ and let $B \subset (t_{0},t_{L}]$. Then,
\begin{equation}
\begin{aligned}
    \mu_{\tilde{T}_{c}^{(2)} \mid T_{c}^{(1)} = u}(B) &=  \mu_{T_{c} \mid T_{c} > u , \tilde{X}(u) \in D_{c}}(B)  \\
    &= \frac{\mathbb{P}\left\{ T_{c} \in B \cap (u,\infty) , \tilde{X}(u) \in D_{c} \right\} }{\mathbb{P}\left\{ T_{c} > u , \tilde{X}(u) \in D_{c} \right\} } \\
    &=  \frac{ \int_{B} \mathbbm{1}_{(u,\infty)}(v)   \mathbb{P}\left\{ \tilde{X}(u) \in D_{c} \mid T_{c} = v \right\} f_{T_{c}}(v)dv }{\mathbb{P}\left\{ \tilde{X}(u) \in D_{c} \right\} - \int_{t_{0}}^{u} \mathbb{P}\left\{ \tilde{X}(u) \in D_{c} \mid T_{c} = v \right\}   dv  } &&\hbox{(total probabilities)} \\
    &= \frac{ \int_{B} \mathbbm{1}_{(u,\infty)}(v)   \mu_{\tilde{X}(u)  \mid \tilde{X}(v) \in D_{c}}(D_{c}) f_{T_{c}}(v)dv }{ \underbrace{\mu_{\tilde{X}(u)}(D_{c})  - \int_{t_{0}}^{u} \mu_{\tilde{X}(u)  \mid \tilde{X}(v) \in D_{c}}(D_{c})dv }_{ = \frac{1}{\alpha}f_{T_{c}}(u) } } \\
    &= \int_{B} \mathbbm{1}_{(u,\infty)}(v)   \alpha \mu_{\tilde{X}(u)  \mid \tilde{X}(v) \in D_{c}}(D_{c}) \frac{f_{T_{c}}(v)}{f_{T_{c}}(u)} dv.
\end{aligned}
\end{equation}
It follows that the conditional distribution $\mu_{\tilde{T}_{c}^{(2)} \mid T_{c}^{(1)} = u}$ for $u \in (t_{0} , t_{L}]$ has the density over $(t_{0} , t_{L})$
\small
\begin{equation}
    f_{\tilde{T}_{c}^{(2)} \mid T_{c}^{(1)} = u}(u) = \alpha \mu_{\tilde{X}(u)  \mid \tilde{X}(v) \in D_{c}}(D_{c}) \frac{f_{T_{c}}(v)}{f_{T_{c}}(u)}\mathbbm{1}_{(u,\infty)}(v) = \alpha\mu_{\tilde{X}(v)  \mid \tilde{X}(u) \in D_{c}}(D_{c}) \frac{\phi_{T_{c}}(v) }{ \phi_{T_{c}}(u) }\mathbbm{1}_{(u,\infty)}(v),
\end{equation}
\normalsize
which proves expression \eqref{Eq:DensityTc2tildeGivenTc1}.

Concerning $\tilde{X}$ given $(T_{c}^{(1)} , \tilde{T}_{c}^{(2)}) = (u,v)$, the rule \ref{Rule:CaptPosInfoWithLiberation} equation \eqref{Eq:CapturePositionFirst&SecondCapture} covers the case $u \leq t_{L}, u < v < \infty$. The cases $u > t_{L}$ and $ u \leq t_{L}, v = \infty$ are still left to be made precise.

Let $t_{1} ,\ldots , t_{n} > t_{0}$ be arbitrary time-points. Consider the case $u > t_{L}$. Since $T_{c}^{(1)} > t_{L}$ implies $\tilde{T}_{c}^{(2)} = \infty$, one must have
\begin{equation}
\label{Eq:MuTildeXgiveTc1uTc2infty}
     \mu_{\tilde{X}(t_{1}) , \ldots , \tilde{X}(t_{n}) \mid T_{c}^{(1)} = u , \tilde{T}_{c}^{(2)} = \infty } = \mu_{\tilde{X}(t_{1}) , \ldots , \tilde{X}(t_{n}) \mid T_{c}^{(1)} = u }, \quad  u > t_{L}.
\end{equation}
Rule \ref{Rule:CaptPosInfoWithLiberation} Eq. \eqref{Eq:CapturePositionFirstCapture} provides the case $u < \infty$. For the case $u = \infty$ we assume $\mathbb{P}( T_{c}^{(1)} = \infty ) = \mu_{T_{c}}(\lbrace \infty \rbrace) > 0$, since otherwise the case $T_{c}^{(1)} = \infty$ cannot be considered in the conditioning. Then,
\begin{equation}
    \mu_{T_{c}}(\lbrace \infty \rbrace)\mu_{\tilde{X}(t_{1}) , \ldots , \tilde{X}(t_{n})\mid T_{c}^{(1)} = \infty } = \mu_{\tilde{X}(t_{1}) , \ldots , \tilde{X}(t_{n}) } -  \int_{t_{0}}^{\infty} \mu_{\tilde{X}(t_{1}) , \ldots , \tilde{X}(t_{n}) \mid T_{c}^{(1)} = u} f_{T_{c}}(u)du, 
\end{equation}
so we conclude
\begin{equation}
    \mu_{\tilde{X}(t_{1}) , \ldots , \tilde{X}(t_{n}) \mid T_{c}^{(1)} = u , \tilde{T}_{c}^{(2)} = \infty } = \begin{cases}
        \mu_{\tilde{X}(t_{1}) , \ldots , \tilde{X}(t_{n})\mid \tilde{X}(u) \in D_{c}  } & \hbox{ if }  u < \infty, \\
        \frac{\mu_{\tilde{X}(t_{1}) , \ldots , \tilde{X}(t_{n}) } - \int_{t_{0}}^{\infty} \mu_{\tilde{X}(t_{1}) , \ldots , \tilde{X}(t_{n}) \mid \tilde{X}(u) \in D_{c}  } f_{T_{c}}(u)du}{  \mu_{T_{c}}( \lbrace \infty \rbrace )   } & \hbox{ if } u = \infty.
    \end{cases} 
\end{equation}
Now, for the case $u\leq t_{L}$, $v = \infty$, consider $B \subset (t_{0},t_{L}]$. Then,
\small
\begin{equation}
\begin{aligned}
\label{Eq:MuXtildeT1T2tildeInfty1}
    \mu_{\tilde{X}(t_{1}) , \ldots , \tilde{X}(t_{n}), T_{c}^{(1)} , \tilde{T}_{c}^{(2)}}(\ \cdot \ \times B \times \lbrace \infty \rbrace )  &= \int_{B} \int_{\lbrace \infty \rbrace } \mu_{\tilde{X}(t_{1}) , \ldots , \tilde{X}(t_{n}) \mid T_{c}^{(1)} = u , \tilde{T}_{c}^{(2)} = v} d \mu_{\tilde{T}_{c}^{(2)} \mid T_{c}^{(1)} = u }(v) f_{T_{c}}(u)du \\
    &= \int_{B}  \mu_{\tilde{X}(t_{1}) , \ldots , \tilde{X}(t_{n}) \mid T_{c}^{(1)} = u , \tilde{T}_{c}^{(2)} = \infty} \mu_{\tilde{T}_{c}^{(2)} \mid T_{c}^{(1)} = u }( \lbrace \infty \rbrace) f_{T_{c}}(u)du. 
    \end{aligned}
\end{equation}
\normalsize
On the other hand, using complementary sets, we have
\small
\begin{equation}
\label{Eq:MuXtildeT1T2tildeInfty2}
    \begin{aligned}
         &\mu_{\tilde{X}(t_{1}) , \ldots , \tilde{X}(t_{n}), T_{c}^{(1)} , \tilde{T}_{c}^{(2)}}(\ \cdot \ \times B\times \lbrace \infty \rbrace ) \\
         &=  \mu_{\tilde{X}(t_{1}) , \ldots , \tilde{X}(t_{n}), T_{c}^{(1)}}(\ \cdot \ \times B ) -  \mu_{\tilde{X}(t_{1}) , \ldots , \tilde{X}(t_{n}), T_{c}^{(1)} , \tilde{T}_{c}^{(2)}}\{  \ \cdot \ \times B \times (t_{0} , \infty) \} \\
         &= \int_{B}\left\{   \mu_{\tilde{X}(t_{1}) , \ldots , \tilde{X}(t_{n})\mid T_{c}^{(1)}=u} - \int_{u}^{\infty}  \mu_{\tilde{X}(t_{1}) , \ldots , \tilde{X}(t_{n})\mid T_{c}^{(1)}=u , \tilde{T}_{c}^{(2)}=v}f_{\tilde{T}_{c}^{(2)}\mid T_{c}^{(1)}=u}(v)dv   \right\}f_{T_{c}(u)}du \\
         &= \int_{B}\left\{   \mu_{\tilde{X}(t_{1}) , \ldots , \tilde{X}(t_{n})\mid \tilde{X}(u) \in D_{c}} - \int_{u}^{\infty}  \mu_{\tilde{X}(t_{1}) , \ldots , \tilde{X}(t_{n})\mid\tilde{X}(u) \in D_{c} , \tilde{X}(v) \in D_{c}}f_{\tilde{T}_{c}^{(2)}\mid T_{c}^{(1)}=u}(v)dv   \right\}f_{T_{c}(u)}du.
    \end{aligned}
\end{equation}
\normalsize
Since \eqref{Eq:MuXtildeT1T2tildeInfty1} and \eqref{Eq:MuXtildeT1T2tildeInfty2} must hold for every $B \subset (t_{0},t_{L}]$, we obtain
\small
\begin{equation}
    \mu_{\tilde{X}(t_{1}) , \ldots , \tilde{X}(t_{n}) \mid T_{c}^{(1)} = u , \tilde{T}_{c}^{(2)} = \infty} = \frac{  \mu_{\tilde{X}(t_{1}) , \ldots , \tilde{X}(t_{n})\mid \tilde{X}(u) \in D_{c}} - \int_{u}^{\infty}  \mu_{\tilde{X}(t_{1}) , \ldots , \tilde{X}(t_{n})\mid\tilde{X}(u) \in D_{c} , \tilde{X}(v) \in D_{c}}f_{\tilde{T}_{c}^{(2)}\mid T_{c}^{(1)}=u}(v)dv  }{\mu_{\tilde{T}_{c}^{(2)} \mid T_{c}^{(1)} = u }( \lbrace \infty \rbrace)}
\end{equation}
\normalsize
for every $u \in (t_{0},t_{L}]$ such that $\mu_{\tilde{T}_{c}^{(2)} \mid T_{c}^{(1)} = u }( \lbrace \infty \rbrace) > 0$. This proves that the conditional distribution of $\tilde{X}$ given $T_{c}^{(1)}, \tilde{T}_{c}^{(2)}$ can be fully specified given the rules \ref{Rule:DistTc1DistTc},\ref{Rule:DistTc2tildeGivenTc1}, \ref{Rule:CaptPosInfoWithLiberation}. $\blacksquare$

\subsection{Proof of Proposition \ref{Prop:MomentsCapture}}

We consider just the case $N=1$, the other cases follow immediately by multiplying the mean and the covariance by $N$ considering the independence assumption on the trajectories (see Eq. \eqref{Eq:CovQ}). With $N=1$, the variable $Q_{t}(A)$ is a binary random variable which equals $1$ when the individual is captured in $A$ before the time $t$:
\begin{equation}
\label{Eq:Qt(A)=1t<=tL}
    \lbrace Q_{t}(A) = 1 \rbrace = \lbrace \tilde{X}(T_{c}^{(1)}) \in A , T_{c}^{(1)} \leq t \rbrace.
\end{equation}
 Thus, its expectation for $t \leq t_{L}$ is
\begin{equation}
\label{Eq:E(Qt(A))beforetL}
\begin{aligned}
    E\{Q_{t}(A)\} &= \mathbb{P}\{Q_{t}(A) = 1 \} \\
    &= \mathbb{P}\{ T_{c}^{(1)} \leq t_{L} , \tilde{X}(T_{c}^{(1)}) \in A \} \\
    &= \int_{t_{0}}^{t} \mathbb{P}\{ \tilde{X}(T_{c}^{(1)}) \in A  \mid T_{c}^{(1)} = u \} f_{T_{c}}(u)du \\
    &= \int_{t_{0}}^{t} \mu_{\tilde{X}(u) \mid \tilde{X}(u) \in D_{c}}(A) f_{T_{c}}(u)du \\
    &= \int_{t_{0}}^{t} \mu_{\tilde{X}(u)}(A) \phi_{T_{c}}(u)du, \quad t \leq t_{L}.
    \end{aligned}
\end{equation}
If $t > t_{L}$, $Q_{t}(A) = 1$ when the individual is either captured in $A$ at a first time $T_{c}^{(1)} \in (t_{L},t]$ (no capture before $t_{L}$), or if it has been captured for a second time at a time $T_{c}^{(2)} \in (t_{L},t]$ at a point in $A$. Equivalently,
\small
\begin{equation}
\label{Eq:Qt(A)=1t>tL}
    \lbrace Q_{t}(A) = 1  \rbrace = \lbrace \tilde{X}(\tilde{T}_{c}^{(2)}) \in A , T_{c}^{(1)} \leq t_{L}, \tilde{T}_{c}^{(2)} \in (T_{c}^{(1)} , T_{c}^{(1)} + t - t_{L} ] \rbrace \cup \lbrace \tilde{X}(T_{c}^{(1)}) \in A, T_{c}^{(1)} \in (t_{L},t] \rbrace.
\end{equation}
\normalsize
Hence,
\small
\begin{equation}
    \begin{aligned}
        E\{Q_{t}(A)\}  &= \mathbb{P}\left\{ \tilde{X}(\tilde{T}_{c}^{(2)}) \in A , T_{c}^{(1)} \leq t_{L}, \tilde{T}_{c}^{(2)} \in (T_{c}^{(1)} , T_{c}^{(1)} + t - t_{L} ]  \right\} + \mathbb{P}\left\{  \tilde{X}(T_{c}^{(1)}) \in A, T_{c}^{(1)} > t_{L}  \right\} \\
        &= \int_{t_{0}}^{t_{L}} \int_{u}^{u+ t - t_{L}} \mu_{\tilde{X}(v) \mid T_{c}^{(1)} = u , \tilde{T}_{c}^{(2)} = v }(A) f_{\tilde{T}_{c}^{(2)}\mid T_{c}^{(1)} = u}(v)dv f_{T_{c}}(u) du + \int_{t_{L}}^{t} \mu_{\tilde{X}(u) \mid  T_{c}^{(1)} = u}(A)f_{T_{c}}(u)du \\
        &= \int_{t_{0}}^{t_{L}} \int_{u}^{u+ t - t_{L}} \mu_{\tilde{X}(v) \mid \tilde{X}(u) \in D_{c}, \tilde{X}(v) \in D_{c} }(A) \alpha \mu_{\tilde{X}(v) \mid \tilde{X}(u) \in D_{c} }(D_{c}) \frac{\phi_{T_{c}}(v)}{\phi_{T_{c}}(u)} dv \mu_{\tilde{X}(u)}(D_{c})\phi_{T_{c}}(u) du  \\
        &\quad+ \int_{t_{L}}^{t} \mu_{\tilde{X}(u)\mid \tilde{X}(u) \in D_{c} }(A)f_{T_{c}}(u)du  \\
        &= \alpha\int_{t_{0}}^{t_{L}} \int_{u}^{u+ t - t_{L}} \mu_{\tilde{X}(u) , \tilde{X}(v) }(D_{c}\times A)  \phi_{T_{c}}(v) dv  du + \int_{t_{L}}^{t} \mu_{\tilde{X}(u)}(A) \phi_{T_{c}}(u)du, \quad t > t_{L}.
    \end{aligned}
\end{equation}
\normalsize
For two times $t,s$ and two subsets $A,B \subset D_{c}$, the product $Q_{t}(A)Q_{s}(B)$ is also a binary random variable which equals $1$ only if both $Q_{t}(A)$ and $Q_{s}(B)$ equal one. We apply thus the same idea following the equivalences \eqref{Eq:Qt(A)=1t<=tL} and \eqref{Eq:Qt(A)=1t>tL}. If $t,s \leq t_{L}$ then
\begin{equation}
\begin{aligned}
        E\left\{ Q_{t}(A)Q_{s}(B) \right\} &= \mathbb{P}\left\{Q_{t}(A)Q_{s}(B) = 1 \right\} \\
        &= \mathbb{P}\left\{ T_{c}^{(1)} \leq t \wedge s, \tilde{X}(T_{c}^{(1)}) \in A \cap B \right\} \\
        &= \int_{t_{0}}^{t \wedge s} \mu_{\tilde{X}(u)}(A \cap B ) \phi_{T_{c}}(u)du. 
\end{aligned}
\end{equation}
If $t,s > t_{L}$, we have
\small
\begin{equation}
\begin{aligned}
    E\left\{Q_{t}(A)Q_{s}(B) \right\} &= \mathbb{P}\left\{ \tilde{X}(\tilde{T}_{c}^{(2)}) \in A \cap B, T_{c}^{(1)} \leq t_{L} , \tilde{T}_{c}^{(2)} \in (T_{c}^{(1)} , T_{c}^{(1)} + t\wedge s - t_{L}]  \right\} \\
    &\quad +\mathbb{P}\left\{ \tilde{X}(T_{c}^{(1)}) \in A \cap B , T_{c}^{(1)} \in (t_{L} , t \wedge s] \right\}\\ 
    &= \int_{t_{0}}^{t_{L}} \int_{u}^{u + t\wedge s - t_{L}} \mu_{\tilde{X}(u),\tilde{X}(v)}\{D_{c}\times(A\cap B) \} \phi_{T_{c}}(v)dvdu + \int_{t_{L}}^{t\wedge s} \mu_{\tilde{X}(u)}(A\cap B) \phi_{T_{c}}(u)du.
\end{aligned}
\end{equation}
\normalsize
Finally, for $t \leq t_{L} < s$ we have
\begin{equation}
    \begin{aligned}
    E\left\{Q_{t}(A)Q_{s}(B) \right\} &= \mathbb{P}\left\{ \tilde{X}(T_{c}^{(1)}) \in A, T_{c}^{(1)} \leq t , \tilde{X}(\tilde{T}_{c}^{(2)}) \in B, \tilde{T}_{c}^{(2)} \in ( T_{c}^{(1)} , T_{c}^{(1)} + s - t_{L} ]  \right\} \\ 
    &= \int_{t_{0}}^{t} \int_{u}^{u + t\wedge s - t_{L}} \mu_{\tilde{X}(u),\tilde{X}(v)}( A \times B ) \phi_{T_{c}}(v)dvdu. \quad \blacksquare.
\end{aligned}
\end{equation}

\subsection{Proof of Proposition \ref{Prop:CaptTimetoCaptPosCond}}

For number \ref{It:CondDistCaptPosCapturedOnce}, we remark that the case $T_{c}^{(1)} = u > t_{L}$ implies $T_{c}^{(2)} = \tilde{T}_{c}^{(2)} = \infty > t_{H}$. Thus (see Eq. \eqref{Eq:MuTildeXgiveTc1uTc2infty} in the Proof of Proposition \ref{Prop:DistributionTripletTc2tildeTc1Xfree}),
\small
\begin{equation}
    \mu_{\tilde{X}(T_{c}^{(1)})\mid  T_{c}^{(1)} = u }(A) = \mu_{\tilde{X}(u)\mid T_{c}^{(1)} = u, \tilde{T}_{c}^{(2)} = \infty }(A) = \mu_{\tilde{X}(u)\mid\tilde{X}(u) \in D_{c} }(A) = \frac{\mu_{\tilde{X}(u)}(A)}{\mu_{\tilde{X}(u)}(D_{c})}, \quad u > t_{L}, A \subset D_{c}.
\end{equation}
\normalsize
Now, for the case $T_{c}^{(1)} = u \leq t_{L}$ but $T_{c}^{(2)} > t_{H}$, we need to consider the fact that the individual has not been captured twice within the time horizon. Let $A \subset D_{c}$. Then,

\begin{equation}
\begin{aligned}
&\mu_{\tilde{X}(T_{c}^{(1)}) \mid T_{c}^{(1)} = u , T_{c}^{(2)} > t_{H}}(A) = \mathbb{P}\left\{  \tilde{X}(T_{c}^{(1)}) \in A \mid T_{c}^{(1)} = u , \tilde{T}_{c}^{(2)} > u + t_{H} - t_{L}  \right\} \\
    &= \frac{ \mathbb{P}\left\{  \tilde{X}(T_{c}^{(1)}) \in A , \tilde{T}_{c}^{(2)} > u + t_{H} - t_{L}  \mid T_{c}^{(1)} = u  \right\} }{  \mathbb{P}\left\{ \tilde{T}_{c}^{(2)} > u + t_{H} - t_{L}  \mid T_{c}^{(1)} = u \right\} } \\
    &= \frac{   \mathbb{P}\left\{  \tilde{X}(T_{c}^{(1)}) \in A   \mid T_{c}^{(1)} = u  \right\}  - \mathbb{P}\left\{  \tilde{X}(T_{c}^{(1)}) \in A , \tilde{T}_{c}^{(2)} \leq u + t_{H} - t_{L}  \mid T_{c}^{(1)} = u  \right\} }{ 1 -  \mathbb{P}\left\{ \tilde{T}_{c}^{(2)} \leq u + t_{H} - t_{L}  \mid T_{c}^{(1)} = u \right\}  } \\
    &= \frac{ \mu_{\tilde{X}(u) \mid \tilde{X}(u)\in D_{c}}(A) - \int_{u}^{u + t_{H} - t_{L}} \mu_{\tilde{X}(u) \mid  T_{c}^{(1)} = u , \tilde{T}_{c}^{(2)} = v }(A) f_{\tilde{T}_{c}^{(2)} \mid T_{c}^{(1)} = u  }(v)dv }{ 1 - \int_{u}^{u + t_{H} - t_{L}}  f_{\tilde{T}_{c}^{(2)} \mid T_{c}^{(1)} = u  }(v)dv  } \\
    &= \frac{ \frac{\mu_{\tilde{X}(u)}(A)}{\mu_{\tilde{X}(u)}(D_{c})}  - \int_{u}^{u + t_{H} - t_{L}} \mu_{\tilde{X}(u) \mid \tilde{X}(u)\in D_{c}, \tilde{X}(v) \in D_{c}}(A) \alpha \mu_{\tilde{X}(v)\mid\tilde{X}(u)\in D_{c}}(D_{c}) \frac{\phi_{T_{c}}(v)}{\phi_{T_{c}}(u)}dv }{ 1 - \int_{u}^{u + t_{H} - t_{L}}  \alpha \mu_{\tilde{X}(v)\mid\tilde{X}(u)\in D_{c}}(D_{c}) \frac{\phi_{T_{c}}(v)}{\phi_{T_{c}}(u)}dv  } \\
    &= \frac{ \mu_{\tilde{X}(u)}(A)  - \frac{\alpha}{\phi_{T_{c}}(u)} \int_{u}^{u + t_{H} - t_{L}} \mu_{\tilde{X}(u), \tilde{X}(v)}(A \times D_{c})\phi_{T_{c}}(v)dv }{ \mu_{\tilde{X}(u)}(D_{c}) - \frac{\alpha}{\phi_{T_{c}}(u)} \int_{u}^{u + t_{H} - t_{L}}  \mu_{\tilde{X}(u),\tilde{X}(v)}(D_{c}\times D_{c})\phi_{T_{c}}(v)  dv  },
\end{aligned}
\end{equation}
which coincides with expression \eqref{Eq:CondDistCaptPosCapturedOnce} for $u \leq t_{L}$, for every $A \subset D_{c}$.

For number \ref{It:CondDistCaptPosCapturedTwice} we consider simply $T_{c}^{(1)} = u$, $\tilde{T}_{c}^{(2)} = v $ with $u \leq t_{L}, u < v \leq u + t_{H}-t_{L}$ and we obtain following rule \ref{Rule:CaptPosInfoWithLiberation}:
\begin{equation}
\begin{aligned}
 \mu_{\tilde{X}(T_{c}^{(1)}) , \tilde{X}(T_{c}^{(2)}) \mid \tilde{T}_{c}^{(2)} = v , T_{c}^{(1)} = u }(A \times B ) &=  \mu_{\tilde{X}(u) , \tilde{X}(v) \mid \tilde{T}_{c}^{(2)} = v , T_{c}^{(1)} = u }(A \times B ) \\
 &=  \mu_{\tilde{X}(u) , \tilde{X}(v) \mid \tilde{X}(u) \in D_{c} , \tilde{X}(v) \in D_{c} }(A \times B ) \\
 &= \frac{\mu_{\tilde{X}(u) , \tilde{X}(v)}(A\times B)}{\mu_{\tilde{X}(u) , \tilde{X}(v)}(D_{c}\times D_{c})},
\end{aligned}
\end{equation}
for every $A,B \subset D_{c}$. $\blacksquare$

\end{appendices}

\bibliography{mibib}

\begin{thebibliography}{}

\bibitem [\protect \citeauthoryear {%
Andrewartha%
\ \BBA {} Birch%
}{%
Andrewartha%
\ \BBA {} Birch%
}{%
{\protect \APACyear {1954}}%
}]{%
andrewartha1954distribution}
\APACinsertmetastar {%
andrewartha1954distribution}%
\begin{APACrefauthors}%
Andrewartha, H\BPBI G.%
\BCBT {}\ \BBA {} Birch, L\BPBI C.%
\end{APACrefauthors}%
\unskip\
\newblock
\APACrefYear{1954}.
\newblock
\APACrefbtitle {{The distribution and abundance of animals}} {{The distribution
  and abundance of animals}}\ (\BNUM\ Edn 1).
\PrintBackRefs{\CurrentBib}

\bibitem [\protect \citeauthoryear {%
Bevilacqua%
, Gaetan%
, Mateu%
\BCBL {}\ \BBA {} Porcu%
}{%
Bevilacqua%
\ \protect \BOthers {.}}{%
{\protect \APACyear {2012}}%
}]{%
bevilacqua2012estimating}
\APACinsertmetastar {%
bevilacqua2012estimating}%
\begin{APACrefauthors}%
Bevilacqua, M.%
, Gaetan, C.%
, Mateu, J.%
\BCBL {}\ \BBA {} Porcu, E.%
\end{APACrefauthors}%
\unskip\
\newblock
\APACrefYearMonthDay{2012}{}{}.
\newblock
{\BBOQ}\APACrefatitle {{Estimating space and space-time covariance functions
  for large data sets: a weighted composite likelihood approach}} {{Estimating
  space and space-time covariance functions for large data sets: a weighted
  composite likelihood approach}}.{\BBCQ}
\newblock
\APACjournalVolNumPages{Journal of the American Statistical
  Association}{107}{497}{268--280}.
\PrintBackRefs{\CurrentBib}

\bibitem [\protect \citeauthoryear {%
Brunner%
}{%
Brunner%
}{%
{\protect \APACyear {2017}}%
}]{%
brunner2017volterra}
\APACinsertmetastar {%
brunner2017volterra}%
\begin{APACrefauthors}%
Brunner, H.%
\end{APACrefauthors}%
\unskip\
\newblock
\APACrefYear{2017}.
\newblock
\APACrefbtitle {{Volterra integral equations: an introduction to theory and
  applications}} {{Volterra integral equations: an introduction to theory and
  applications}}\ (\BVOL~30).
\newblock
\APACaddressPublisher{}{Cambridge University Press}.
\PrintBackRefs{\CurrentBib}

\bibitem [\protect \citeauthoryear {%
Burkill%
\ \BBA {} Burkill%
}{%
Burkill%
\ \BBA {} Burkill%
}{%
{\protect \APACyear {2002}}%
}]{%
burkill2002second}
\APACinsertmetastar {%
burkill2002second}%
\begin{APACrefauthors}%
Burkill, J\BPBI C.%
\BCBT {}\ \BBA {} Burkill, H.%
\end{APACrefauthors}%
\unskip\
\newblock
\APACrefYear{2002}.
\newblock
\APACrefbtitle {{A second course in mathematical analysis}} {{A second course
  in mathematical analysis}}.
\newblock
\APACaddressPublisher{}{Cambridge University Press}.
\PrintBackRefs{\CurrentBib}

\bibitem [\protect \citeauthoryear {%
Cao%
, Genton%
, Keyes%
\BCBL {}\ \BBA {} Turkiyyah%
}{%
Cao%
\ \protect \BOthers {.}}{%
{\protect \APACyear {2022}}%
}]{%
cao2022tlrmvnmvt}
\APACinsertmetastar {%
cao2022tlrmvnmvt}%
\begin{APACrefauthors}%
Cao, J.%
, Genton, M\BPBI G.%
, Keyes, D\BPBI E.%
\BCBL {}\ \BBA {} Turkiyyah, G\BPBI M.%
\end{APACrefauthors}%
\unskip\
\newblock
\APACrefYearMonthDay{2022}{}{}.
\newblock
{\BBOQ}\APACrefatitle {tlrmvnmvt: Computing high-dimensional multivariate
  normal and student-t probabilities with low-rank methods in r} {tlrmvnmvt:
  Computing high-dimensional multivariate normal and student-t probabilities
  with low-rank methods in r}.{\BBCQ}
\newblock
\APACjournalVolNumPages{Journal of Statistical Software}{101}{}{1--25}.
\PrintBackRefs{\CurrentBib}

\bibitem [\protect \citeauthoryear {%
Chil{\`e}s%
\ \BBA {} Delfiner%
}{%
Chil{\`e}s%
\ \BBA {} Delfiner%
}{%
{\protect \APACyear {1999}}%
}]{%
chiles1999geostatistics}
\APACinsertmetastar {%
chiles1999geostatistics}%
\begin{APACrefauthors}%
Chil{\`e}s, J.%
\BCBT {}\ \BBA {} Delfiner, P.%
\end{APACrefauthors}%
\unskip\
\newblock
\APACrefYear{1999}.
\newblock
\APACrefbtitle {{Geostatistics: Modeling Spatial Uncertainty}} {{Geostatistics:
  Modeling Spatial Uncertainty}}.
\newblock
\APACaddressPublisher{}{John Wiley {\&} Sons}.
\PrintBackRefs{\CurrentBib}

\bibitem [\protect \citeauthoryear {%
Cressie%
\ \BBA {} Wikle%
}{%
Cressie%
\ \BBA {} Wikle%
}{%
{\protect \APACyear {2015}}%
}]{%
cressie2015statistics}
\APACinsertmetastar {%
cressie2015statistics}%
\begin{APACrefauthors}%
Cressie, N.%
\BCBT {}\ \BBA {} Wikle, C\BPBI K.%
\end{APACrefauthors}%
\unskip\
\newblock
\APACrefYear{2015}.
\newblock
\APACrefbtitle {Statistics for spatio-temporal data} {Statistics for
  spatio-temporal data}.
\newblock
\APACaddressPublisher{}{John Wiley \& Sons}.
\PrintBackRefs{\CurrentBib}

\bibitem [\protect \citeauthoryear {%
Daskalakis%
, Kamath%
\BCBL {}\ \BBA {} Tzamos%
}{%
Daskalakis%
\ \protect \BOthers {.}}{%
{\protect \APACyear {2015}}%
}]{%
daskalakis2015structure}
\APACinsertmetastar {%
daskalakis2015structure}%
\begin{APACrefauthors}%
Daskalakis, C.%
, Kamath, G.%
\BCBL {}\ \BBA {} Tzamos, C.%
\end{APACrefauthors}%
\unskip\
\newblock
\APACrefYearMonthDay{2015}{}{}.
\newblock
{\BBOQ}\APACrefatitle {On the structure, covering, and learning of poisson
  multinomial distributions} {On the structure, covering, and learning of
  poisson multinomial distributions}.{\BBCQ}
\newblock
\BIn{} \APACrefbtitle {2015 IEEE 56th annual symposium on foundations of
  computer science} {2015 ieee 56th annual symposium on foundations of computer
  science}\ (\BPGS\ 1203--1217).
\PrintBackRefs{\CurrentBib}

\bibitem [\protect \citeauthoryear {%
Diggle%
, Kaimi%
\BCBL {}\ \BBA {} Abellana%
}{%
Diggle%
\ \protect \BOthers {.}}{%
{\protect \APACyear {2010}}%
}]{%
diggle2010partial}
\APACinsertmetastar {%
diggle2010partial}%
\begin{APACrefauthors}%
Diggle, P\BPBI J.%
, Kaimi, I.%
\BCBL {}\ \BBA {} Abellana, R.%
\end{APACrefauthors}%
\unskip\
\newblock
\APACrefYearMonthDay{2010}{}{}.
\newblock
{\BBOQ}\APACrefatitle {{Partial-likelihood analysis of spatio-temporal
  point-process data}} {{Partial-likelihood analysis of spatio-temporal
  point-process data}}.{\BBCQ}
\newblock
\APACjournalVolNumPages{Biometrics}{66}{2}{347--354}.
\PrintBackRefs{\CurrentBib}

\bibitem [\protect \citeauthoryear {%
Doi%
}{%
Doi%
}{%
{\protect \APACyear {1976}}%
}]{%
doi1976stochastic}
\APACinsertmetastar {%
doi1976stochastic}%
\begin{APACrefauthors}%
Doi, M.%
\end{APACrefauthors}%
\unskip\
\newblock
\APACrefYearMonthDay{1976}{}{}.
\newblock
{\BBOQ}\APACrefatitle {Stochastic theory of diffusion-controlled reaction}
  {Stochastic theory of diffusion-controlled reaction}.{\BBCQ}
\newblock
\APACjournalVolNumPages{Journal of Physics A: Mathematical and
  General}{9}{9}{1479}.
\PrintBackRefs{\CurrentBib}

\bibitem [\protect \citeauthoryear {%
Edelsparre%
, Hefley%
, Rodr{\'\i}guez%
, Fitzpatrick%
\BCBL {}\ \BBA {} Sokolowski%
}{%
Edelsparre%
\ \protect \BOthers {.}}{%
{\protect \APACyear {2021}}%
}]{%
edelsparre2021scaling}
\APACinsertmetastar {%
edelsparre2021scaling}%
\begin{APACrefauthors}%
Edelsparre, A\BPBI H.%
, Hefley, T\BPBI J.%
, Rodr{\'\i}guez, M\BPBI A.%
, Fitzpatrick, M\BPBI J.%
\BCBL {}\ \BBA {} Sokolowski, M\BPBI B.%
\end{APACrefauthors}%
\unskip\
\newblock
\APACrefYearMonthDay{2021}{}{}.
\newblock
{\BBOQ}\APACrefatitle {{Scaling up: understanding movement from individual
  differences to population-level dispersal}} {{Scaling up: understanding
  movement from individual differences to population-level dispersal}}.{\BBCQ}
\newblock
\APACjournalVolNumPages{bioRxiv}{}{}{2021--01}.
\PrintBackRefs{\CurrentBib}

\bibitem [\protect \citeauthoryear {%
Eisaguirre%
, Booms%
, Barger%
, Goddard%
\BCBL {}\ \BBA {} Breed%
}{%
Eisaguirre%
\ \protect \BOthers {.}}{%
{\protect \APACyear {2021}}%
}]{%
eisaguirre2021multistate}
\APACinsertmetastar {%
eisaguirre2021multistate}%
\begin{APACrefauthors}%
Eisaguirre, J\BPBI M.%
, Booms, T\BPBI L.%
, Barger, C\BPBI P.%
, Goddard, S\BPBI D.%
\BCBL {}\ \BBA {} Breed, G\BPBI A.%
\end{APACrefauthors}%
\unskip\
\newblock
\APACrefYearMonthDay{2021}{}{}.
\newblock
{\BBOQ}\APACrefatitle {{Multistate Ornstein--Uhlenbeck approach for practical
  estimation of movement and resource selection around central places}}
  {{Multistate Ornstein--Uhlenbeck approach for practical estimation of
  movement and resource selection around central places}}.{\BBCQ}
\newblock
\APACjournalVolNumPages{Methods in Ecology and Evolution}{12}{3}{507--519}.
\PrintBackRefs{\CurrentBib}

\bibitem [\protect \citeauthoryear {%
Eisaguirre%
\ \protect \BOthers {.}}{%
Eisaguirre%
\ \protect \BOthers {.}}{%
{\protect \APACyear {2023}}%
}]{%
eisaguirre2023informing}
\APACinsertmetastar {%
eisaguirre2023informing}%
\begin{APACrefauthors}%
Eisaguirre, J\BPBI M.%
, Williams, P\BPBI J.%
, Lu, X.%
, Kissling, M\BPBI L.%
, Schuette, P\BPBI A.%
, Weitzman, B\BPBI P.%
\BDBL {}Hooten, M\BPBI B.%
\end{APACrefauthors}%
\unskip\
\newblock
\APACrefYearMonthDay{2023}{}{}.
\newblock
{\BBOQ}\APACrefatitle {Informing management of recovering predators and their
  prey with ecological diffusion models} {Informing management of recovering
  predators and their prey with ecological diffusion models}.{\BBCQ}
\newblock
\APACjournalVolNumPages{Frontiers in Ecology and the
  Environment}{21}{10}{479--488}.
\PrintBackRefs{\CurrentBib}

\bibitem [\protect \citeauthoryear {%
Fleming%
\ \protect \BOthers {.}}{%
Fleming%
\ \protect \BOthers {.}}{%
{\protect \APACyear {2014}}%
}]{%
fleming2014fine}
\APACinsertmetastar {%
fleming2014fine}%
\begin{APACrefauthors}%
Fleming, C\BPBI H.%
, Calabrese, J\BPBI M.%
, Mueller, T.%
, Olson, K\BPBI A.%
, Leimgruber, P.%
\BCBL {}\ \BBA {} Fagan, W\BPBI F.%
\end{APACrefauthors}%
\unskip\
\newblock
\APACrefYearMonthDay{2014}{}{}.
\newblock
{\BBOQ}\APACrefatitle {{From fine-scale foraging to home ranges: a semivariance
  approach to identifying movement modes across spatiotemporal scales}} {{From
  fine-scale foraging to home ranges: a semivariance approach to identifying
  movement modes across spatiotemporal scales}}.{\BBCQ}
\newblock
\APACjournalVolNumPages{The American Naturalist}{183}{5}{E154--E167}.
\PrintBackRefs{\CurrentBib}

\bibitem [\protect \citeauthoryear {%
Gardiner%
}{%
Gardiner%
}{%
{\protect \APACyear {2009}}%
}]{%
gardiner2009stochastic}
\APACinsertmetastar {%
gardiner2009stochastic}%
\begin{APACrefauthors}%
Gardiner, C.%
\end{APACrefauthors}%
\unskip\
\newblock
\APACrefYear{2009}.
\newblock
\APACrefbtitle {Stochastic methods} {Stochastic methods}\ (\BVOL~4).
\newblock
\APACaddressPublisher{}{Springer Berlin Heidelberg}.
\PrintBackRefs{\CurrentBib}

\bibitem [\protect \citeauthoryear {%
Gelfand%
\ \BBA {} Schliep%
}{%
Gelfand%
\ \BBA {} Schliep%
}{%
{\protect \APACyear {2018}}%
}]{%
gelfand2018bayesian}
\APACinsertmetastar {%
gelfand2018bayesian}%
\begin{APACrefauthors}%
Gelfand, A\BPBI E.%
\BCBT {}\ \BBA {} Schliep, E\BPBI M.%
\end{APACrefauthors}%
\unskip\
\newblock
\APACrefYearMonthDay{2018}{}{}.
\newblock
{\BBOQ}\APACrefatitle {Bayesian inference and computing for spatial point
  patterns} {Bayesian inference and computing for spatial point
  patterns}.{\BBCQ}
\newblock
\BIn{} \APACrefbtitle {NSF-CBMS regional conference series in probability and
  statistics} {Nsf-cbms regional conference series in probability and
  statistics}\ (\BVOL~10, \BPGS\ i--125).
\PrintBackRefs{\CurrentBib}

\bibitem [\protect \citeauthoryear {%
Glennie%
\ \protect \BOthers {.}}{%
Glennie%
\ \protect \BOthers {.}}{%
{\protect \APACyear {2021}}%
}]{%
glennie2021incorporating}
\APACinsertmetastar {%
glennie2021incorporating}%
\begin{APACrefauthors}%
Glennie, R.%
, Buckland, S\BPBI T.%
, Langrock, R.%
, Gerrodette, T.%
, Ballance, L.%
, Chivers, S.%
\BCBL {}\ \BBA {} Scott, M.%
\end{APACrefauthors}%
\unskip\
\newblock
\APACrefYearMonthDay{2021}{}{}.
\newblock
{\BBOQ}\APACrefatitle {Incorporating animal movement into distance sampling}
  {Incorporating animal movement into distance sampling}.{\BBCQ}
\newblock
\APACjournalVolNumPages{Journal of the American Statistical
  Association}{116}{533}{107--115}.
\PrintBackRefs{\CurrentBib}

\bibitem [\protect \citeauthoryear {%
Gurarie%
\ \protect \BOthers {.}}{%
Gurarie%
\ \protect \BOthers {.}}{%
{\protect \APACyear {2017}}%
}]{%
gurarie2017correlated}
\APACinsertmetastar {%
gurarie2017correlated}%
\begin{APACrefauthors}%
Gurarie, E.%
, Fleming, C\BPBI H.%
, Fagan, W\BPBI F.%
, Laidre, K\BPBI L.%
, Hern{\'a}ndez-Pliego, J.%
\BCBL {}\ \BBA {} Ovaskainen, O.%
\end{APACrefauthors}%
\unskip\
\newblock
\APACrefYearMonthDay{2017}{}{}.
\newblock
{\BBOQ}\APACrefatitle {{Correlated velocity models as a fundamental unit of
  animal movement: synthesis and applications}} {{Correlated velocity models as
  a fundamental unit of animal movement: synthesis and applications}}.{\BBCQ}
\newblock
\APACjournalVolNumPages{Movement Ecology}{5}{}{1--18}.
\PrintBackRefs{\CurrentBib}

\bibitem [\protect \citeauthoryear {%
Hefley%
\ \BBA {} Hooten%
}{%
Hefley%
\ \BBA {} Hooten%
}{%
{\protect \APACyear {2016}}%
}]{%
hefley2016hierarchical}
\APACinsertmetastar {%
hefley2016hierarchical}%
\begin{APACrefauthors}%
Hefley, T\BPBI J.%
\BCBT {}\ \BBA {} Hooten, M\BPBI B.%
\end{APACrefauthors}%
\unskip\
\newblock
\APACrefYearMonthDay{2016}{}{}.
\newblock
{\BBOQ}\APACrefatitle {Hierarchical species distribution models} {Hierarchical
  species distribution models}.{\BBCQ}
\newblock
\APACjournalVolNumPages{Current Landscape Ecology Reports}{1}{}{87--97}.
\PrintBackRefs{\CurrentBib}

\bibitem [\protect \citeauthoryear {%
Hefley%
, Hooten%
, Russell%
, Walsh%
\BCBL {}\ \BBA {} Powell%
}{%
Hefley%
\ \protect \BOthers {.}}{%
{\protect \APACyear {2017}}%
}]{%
hefley2017mechanism}
\APACinsertmetastar {%
hefley2017mechanism}%
\begin{APACrefauthors}%
Hefley, T\BPBI J.%
, Hooten, M\BPBI B.%
, Russell, R\BPBI E.%
, Walsh, D\BPBI P.%
\BCBL {}\ \BBA {} Powell, J\BPBI A.%
\end{APACrefauthors}%
\unskip\
\newblock
\APACrefYearMonthDay{2017}{}{}.
\newblock
{\BBOQ}\APACrefatitle {When mechanism matters: Bayesian forecasting using
  models of ecological diffusion} {When mechanism matters: Bayesian forecasting
  using models of ecological diffusion}.{\BBCQ}
\newblock
\APACjournalVolNumPages{Ecology Letters}{20}{5}{640--650}.
\PrintBackRefs{\CurrentBib}

\bibitem [\protect \citeauthoryear {%
Hooten%
, Garlick%
\BCBL {}\ \BBA {} Powell%
}{%
Hooten%
\ \protect \BOthers {.}}{%
{\protect \APACyear {2013}}%
}]{%
hooten2013computationally}
\APACinsertmetastar {%
hooten2013computationally}%
\begin{APACrefauthors}%
Hooten, M\BPBI B.%
, Garlick, M\BPBI J.%
\BCBL {}\ \BBA {} Powell, J\BPBI A.%
\end{APACrefauthors}%
\unskip\
\newblock
\APACrefYearMonthDay{2013}{}{}.
\newblock
{\BBOQ}\APACrefatitle {{Computationally efficient statistical differential
  equation modeling using homogenization}} {{Computationally efficient
  statistical differential equation modeling using homogenization}}.{\BBCQ}
\newblock
\APACjournalVolNumPages{Journal of agricultural, biological, and environmental
  statistics}{18}{}{405--428}.
\PrintBackRefs{\CurrentBib}

\bibitem [\protect \citeauthoryear {%
Hooten%
, Johnson%
, McClintock%
\BCBL {}\ \BBA {} Morales%
}{%
Hooten%
\ \protect \BOthers {.}}{%
{\protect \APACyear {2017}}%
}]{%
hooten2017animal}
\APACinsertmetastar {%
hooten2017animal}%
\begin{APACrefauthors}%
Hooten, M\BPBI B.%
, Johnson, D\BPBI S.%
, McClintock, B\BPBI T.%
\BCBL {}\ \BBA {} Morales, J\BPBI M.%
\end{APACrefauthors}%
\unskip\
\newblock
\APACrefYear{2017}.
\newblock
\APACrefbtitle {Animal movement: statistical models for telemetry data} {Animal
  movement: statistical models for telemetry data}.
\newblock
\APACaddressPublisher{}{CRC press}.
\PrintBackRefs{\CurrentBib}

\bibitem [\protect \citeauthoryear {%
Horne%
, Garton%
, Krone%
\BCBL {}\ \BBA {} Lewis%
}{%
Horne%
\ \protect \BOthers {.}}{%
{\protect \APACyear {2007}}%
}]{%
horne2007analyzing}
\APACinsertmetastar {%
horne2007analyzing}%
\begin{APACrefauthors}%
Horne, J\BPBI S.%
, Garton, E\BPBI O.%
, Krone, S\BPBI M.%
\BCBL {}\ \BBA {} Lewis, J\BPBI S.%
\end{APACrefauthors}%
\unskip\
\newblock
\APACrefYearMonthDay{2007}{}{}.
\newblock
{\BBOQ}\APACrefatitle {{Analyzing animal movements using Brownian bridges}}
  {{Analyzing animal movements using Brownian bridges}}.{\BBCQ}
\newblock
\APACjournalVolNumPages{Ecology}{88}{9}{2354--2363}.
\PrintBackRefs{\CurrentBib}

\bibitem [\protect \citeauthoryear {%
Hotelling%
}{%
Hotelling%
}{%
{\protect \APACyear {1927}}%
}]{%
hotelling1927differential}
\APACinsertmetastar {%
hotelling1927differential}%
\begin{APACrefauthors}%
Hotelling, H.%
\end{APACrefauthors}%
\unskip\
\newblock
\APACrefYearMonthDay{1927}{}{}.
\newblock
{\BBOQ}\APACrefatitle {Differential equations subject to error, and population
  estimates} {Differential equations subject to error, and population
  estimates}.{\BBCQ}
\newblock
\APACjournalVolNumPages{Journal of the American Statistical
  Association}{22}{159}{283--314}.
\PrintBackRefs{\CurrentBib}

\bibitem [\protect \citeauthoryear {%
K{\'e}ry%
, Royle%
\BCBL {}\ \protect \BOthers {.}}{%
K{\'e}ry%
\ \protect \BOthers {.}}{%
{\protect \APACyear {2016}}%
}]{%
kery2016modeling}
\APACinsertmetastar {%
kery2016modeling}%
\begin{APACrefauthors}%
K{\'e}ry, M.%
, Royle, J.%
\BCBL {}\ \BOthersPeriod {.}\end{APACrefauthors}%
\unskip\
\newblock
\APACrefYearMonthDay{2016}{}{}.
\newblock
{\BBOQ}\APACrefatitle {Modeling static occurrence and species distributions
  using site-occupancy models} {Modeling static occurrence and species
  distributions using site-occupancy models}.{\BBCQ}
\newblock
\APACjournalVolNumPages{Applied hierarchical modelling in
  ecology}{1}{}{551--630}.
\PrintBackRefs{\CurrentBib}

\bibitem [\protect \citeauthoryear {%
Kranstauber%
}{%
Kranstauber%
}{%
{\protect \APACyear {2019}}%
}]{%
kranstauber2019modelling}
\APACinsertmetastar {%
kranstauber2019modelling}%
\begin{APACrefauthors}%
Kranstauber, B.%
\end{APACrefauthors}%
\unskip\
\newblock
\APACrefYearMonthDay{2019}{}{}.
\newblock
{\BBOQ}\APACrefatitle {{Modelling animal movement as Brownian bridges with
  covariates}} {{Modelling animal movement as Brownian bridges with
  covariates}}.{\BBCQ}
\newblock
\APACjournalVolNumPages{Movement ecology}{7}{}{1--10}.
\PrintBackRefs{\CurrentBib}

\bibitem [\protect \citeauthoryear {%
Krebs%
}{%
Krebs%
}{%
{\protect \APACyear {2009}}%
}]{%
krebs2009ecology}
\APACinsertmetastar {%
krebs2009ecology}%
\begin{APACrefauthors}%
Krebs, C.%
\end{APACrefauthors}%
\unskip\
\newblock
\APACrefYearMonthDay{2009}{}{}.
\newblock
{\BBOQ}\APACrefatitle {{Ecology: the experimental analysis of distribution and
  abundance. University of British Columbia, Vancouver}} {{Ecology: the
  experimental analysis of distribution and abundance. University of British
  Columbia, Vancouver}}.{\BBCQ}
\newblock
\APACjournalVolNumPages{British Columbia}{}{}{655}.
\PrintBackRefs{\CurrentBib}

\bibitem [\protect \citeauthoryear {%
Lenzi%
, Bessac%
, Rudi%
\BCBL {}\ \BBA {} Stein%
}{%
Lenzi%
\ \protect \BOthers {.}}{%
{\protect \APACyear {2023}}%
}]{%
lenzi2023neural}
\APACinsertmetastar {%
lenzi2023neural}%
\begin{APACrefauthors}%
Lenzi, A.%
, Bessac, J.%
, Rudi, J.%
\BCBL {}\ \BBA {} Stein, M\BPBI L.%
\end{APACrefauthors}%
\unskip\
\newblock
\APACrefYearMonthDay{2023}{}{}.
\newblock
{\BBOQ}\APACrefatitle {{Neural networks for parameter estimation in intractable
  models}} {{Neural networks for parameter estimation in intractable
  models}}.{\BBCQ}
\newblock
\APACjournalVolNumPages{Computational Statistics \& Data
  Analysis}{185}{}{107762}.
\PrintBackRefs{\CurrentBib}

\bibitem [\protect \citeauthoryear {%
Lin%
, Wang%
\BCBL {}\ \BBA {} Hong%
}{%
Lin%
\ \protect \BOthers {.}}{%
{\protect \APACyear {2022}}%
}]{%
lin2022poisson}
\APACinsertmetastar {%
lin2022poisson}%
\begin{APACrefauthors}%
Lin, Z.%
, Wang, Y.%
\BCBL {}\ \BBA {} Hong, Y.%
\end{APACrefauthors}%
\unskip\
\newblock
\APACrefYearMonthDay{2022}{}{}.
\newblock
{\BBOQ}\APACrefatitle {{The Poisson multinomial distribution and its
  applications in voting theory, ecological inference, and machine learning}}
  {{The Poisson multinomial distribution and its applications in voting theory,
  ecological inference, and machine learning}}.{\BBCQ}
\newblock
\APACjournalVolNumPages{arXiv preprint arXiv:2201.04237}{}{}{}.
\PrintBackRefs{\CurrentBib}

\bibitem [\protect \citeauthoryear {%
Lindsay%
}{%
Lindsay%
}{%
{\protect \APACyear {1988}}%
}]{%
lindsay1988composite}
\APACinsertmetastar {%
lindsay1988composite}%
\begin{APACrefauthors}%
Lindsay, B\BPBI G.%
\end{APACrefauthors}%
\unskip\
\newblock
\APACrefYearMonthDay{1988}{}{}.
\newblock
{\BBOQ}\APACrefatitle {{Composite likelihood methods}} {{Composite likelihood
  methods}}.{\BBCQ}
\newblock
\APACjournalVolNumPages{Comtemporary Mathematics}{80}{1}{221--239}.
\PrintBackRefs{\CurrentBib}

\bibitem [\protect \citeauthoryear {%
Louvrier%
, Papaix%
, Duchamp%
\BCBL {}\ \BBA {} Gimenez%
}{%
Louvrier%
\ \protect \BOthers {.}}{%
{\protect \APACyear {2020}}%
}]{%
louvrier2020mechanistic}
\APACinsertmetastar {%
louvrier2020mechanistic}%
\begin{APACrefauthors}%
Louvrier, J.%
, Papaix, J.%
, Duchamp, C.%
\BCBL {}\ \BBA {} Gimenez, O.%
\end{APACrefauthors}%
\unskip\
\newblock
\APACrefYearMonthDay{2020}{}{}.
\newblock
{\BBOQ}\APACrefatitle {A mechanistic--statistical species distribution model to
  explain and forecast wolf (Canis lupus) colonization in South-Eastern France}
  {A mechanistic--statistical species distribution model to explain and
  forecast wolf (canis lupus) colonization in south-eastern france}.{\BBCQ}
\newblock
\APACjournalVolNumPages{Spatial Statistics}{36}{}{100428}.
\PrintBackRefs{\CurrentBib}

\bibitem [\protect \citeauthoryear {%
Lu%
\ \protect \BOthers {.}}{%
Lu%
\ \protect \BOthers {.}}{%
{\protect \APACyear {2020}}%
}]{%
lu2020nonlinear}
\APACinsertmetastar {%
lu2020nonlinear}%
\begin{APACrefauthors}%
Lu, X.%
, Williams, P\BPBI J.%
, Hooten, M\BPBI B.%
, Powell, J\BPBI A.%
, Womble, J\BPBI N.%
\BCBL {}\ \BBA {} Bower, M\BPBI R.%
\end{APACrefauthors}%
\unskip\
\newblock
\APACrefYearMonthDay{2020}{}{}.
\newblock
{\BBOQ}\APACrefatitle {Nonlinear reaction--diffusion process models improve
  inference for population dynamics} {Nonlinear reaction--diffusion process
  models improve inference for population dynamics}.{\BBCQ}
\newblock
\APACjournalVolNumPages{Environmetrics}{31}{3}{e2604}.
\PrintBackRefs{\CurrentBib}

\bibitem [\protect \citeauthoryear {%
Malchow%
\ \protect \BOthers {.}}{%
Malchow%
\ \protect \BOthers {.}}{%
{\protect \APACyear {2024}}%
}]{%
malchow2024fitting}
\APACinsertmetastar {%
malchow2024fitting}%
\begin{APACrefauthors}%
Malchow, A\BHBI K.%
, Fandos, G.%
, Kormann, U\BPBI G.%
, Gr{\"u}ebler, M\BPBI U.%
, K{\'e}ry, M.%
, Hartig, F.%
\BCBL {}\ \BBA {} Zurell, D.%
\end{APACrefauthors}%
\unskip\
\newblock
\APACrefYearMonthDay{2024}{}{}.
\newblock
{\BBOQ}\APACrefatitle {Fitting individual-based models of spatial population
  dynamics to long-term monitoring data} {Fitting individual-based models of
  spatial population dynamics to long-term monitoring data}.{\BBCQ}
\newblock
\APACjournalVolNumPages{Ecological Applications}{34}{4}{e2966}.
\PrintBackRefs{\CurrentBib}

\bibitem [\protect \citeauthoryear {%
May%
}{%
May%
}{%
{\protect \APACyear {1999}}%
}]{%
may1999unanswered}
\APACinsertmetastar {%
may1999unanswered}%
\begin{APACrefauthors}%
May, R.%
\end{APACrefauthors}%
\unskip\
\newblock
\APACrefYearMonthDay{1999}{}{}.
\newblock
{\BBOQ}\APACrefatitle {Unanswered questions in ecology} {Unanswered questions
  in ecology}.{\BBCQ}
\newblock
\APACjournalVolNumPages{Philosophical Transactions of the Royal Society of
  London. Series B: Biological Sciences}{354}{1392}{1951--1959}.
\PrintBackRefs{\CurrentBib}

\bibitem [\protect \citeauthoryear {%
Michelot%
, Gloaguen%
, Blackwell%
\BCBL {}\ \BBA {} {\'E}tienne%
}{%
Michelot%
\ \protect \BOthers {.}}{%
{\protect \APACyear {2019}}%
}]{%
michelot2019langevin}
\APACinsertmetastar {%
michelot2019langevin}%
\begin{APACrefauthors}%
Michelot, T.%
, Gloaguen, P.%
, Blackwell, P\BPBI G.%
\BCBL {}\ \BBA {} {\'E}tienne, M\BHBI P.%
\end{APACrefauthors}%
\unskip\
\newblock
\APACrefYearMonthDay{2019}{}{}.
\newblock
{\BBOQ}\APACrefatitle {{The Langevin diffusion as a continuous-time model of
  animal movement and habitat selection}} {{The Langevin diffusion as a
  continuous-time model of animal movement and habitat selection}}.{\BBCQ}
\newblock
\APACjournalVolNumPages{Methods in ecology and evolution}{10}{11}{1894--1907}.
\PrintBackRefs{\CurrentBib}

\bibitem [\protect \citeauthoryear {%
Newman%
\ \protect \BOthers {.}}{%
Newman%
\ \protect \BOthers {.}}{%
{\protect \APACyear {2014}}%
}]{%
newman2014modelling}
\APACinsertmetastar {%
newman2014modelling}%
\begin{APACrefauthors}%
Newman, K.%
, Buckland, S.%
, Morgan, B\BPBI J.%
, King, R.%
, Borchers, D.%
, Cole, D\BPBI J.%
\BDBL {}Thomas, L.%
\end{APACrefauthors}%
\unskip\
\newblock
\APACrefYearMonthDay{2014}{}{}.
\newblock
{\BBOQ}\APACrefatitle {Modelling population dynamics} {Modelling population
  dynamics}.{\BBCQ}
\newblock
\APACjournalVolNumPages{Modelling Population Dynamics: Model Formulation,
  Fitting and Assessment using State-Space Methods. Springer New York, New
  York, USA}{}{}{169--195}.
\PrintBackRefs{\CurrentBib}

\bibitem [\protect \citeauthoryear {%
Okubo%
, Levin%
\BCBL {}\ \protect \BOthers {.}}{%
Okubo%
\ \protect \BOthers {.}}{%
{\protect \APACyear {2001}}%
}]{%
okubo2001diffusion}
\APACinsertmetastar {%
okubo2001diffusion}%
\begin{APACrefauthors}%
Okubo, A.%
, Levin, S\BPBI A.%
\BCBL {}\ \BOthersPeriod {.}\end{APACrefauthors}%
\unskip\
\newblock
\APACrefYear{2001}.
\newblock
\APACrefbtitle {{Diffusion and ecological problems: modern perspectives}}
  {{Diffusion and ecological problems: modern perspectives}}\ (\BVOL~14).
\newblock
\APACaddressPublisher{}{Springer}.
\PrintBackRefs{\CurrentBib}

\bibitem [\protect \citeauthoryear {%
Padoan%
, Ribatet%
\BCBL {}\ \BBA {} Sisson%
}{%
Padoan%
\ \protect \BOthers {.}}{%
{\protect \APACyear {2010}}%
}]{%
padoan2010likelihood}
\APACinsertmetastar {%
padoan2010likelihood}%
\begin{APACrefauthors}%
Padoan, S\BPBI A.%
, Ribatet, M.%
\BCBL {}\ \BBA {} Sisson, S\BPBI A.%
\end{APACrefauthors}%
\unskip\
\newblock
\APACrefYearMonthDay{2010}{}{}.
\newblock
{\BBOQ}\APACrefatitle {{Likelihood-based inference for max-stable processes}}
  {{Likelihood-based inference for max-stable processes}}.{\BBCQ}
\newblock
\APACjournalVolNumPages{Journal of the American Statistical
  Association}{105}{489}{263--277}.
\PrintBackRefs{\CurrentBib}

\bibitem [\protect \citeauthoryear {%
Potts%
\ \BBA {} B{\"o}rger%
}{%
Potts%
\ \BBA {} B{\"o}rger%
}{%
{\protect \APACyear {2023}}%
}]{%
potts2023scale}
\APACinsertmetastar {%
potts2023scale}%
\begin{APACrefauthors}%
Potts, J\BPBI R.%
\BCBT {}\ \BBA {} B{\"o}rger, L.%
\end{APACrefauthors}%
\unskip\
\newblock
\APACrefYearMonthDay{2023}{}{}.
\newblock
{\BBOQ}\APACrefatitle {{How to scale up from animal movement decisions to
  spatiotemporal patterns: An approach via step selection}} {{How to scale up
  from animal movement decisions to spatiotemporal patterns: An approach via
  step selection}}.{\BBCQ}
\newblock
\APACjournalVolNumPages{Journal of Animal Ecology}{92}{1}{16--29}.
\PrintBackRefs{\CurrentBib}

\bibitem [\protect \citeauthoryear {%
Roques%
}{%
Roques%
}{%
{\protect \APACyear {2013}}%
}]{%
roques2013modeles}
\APACinsertmetastar {%
roques2013modeles}%
\begin{APACrefauthors}%
Roques, L.%
\end{APACrefauthors}%
\unskip\
\newblock
\APACrefYear{2013}.
\newblock
\APACrefbtitle {Mod{\`e}les de r{\'e}action-diffusion pour l'{\'e}cologie
  spatiale: Avec exercices dirig{\'e}s} {Mod{\`e}les de r{\'e}action-diffusion
  pour l'{\'e}cologie spatiale: Avec exercices dirig{\'e}s}.
\newblock
\APACaddressPublisher{}{Editions Quae}.
\PrintBackRefs{\CurrentBib}

\bibitem [\protect \citeauthoryear {%
Roques%
, Allard%
\BCBL {}\ \BBA {} Soubeyrand%
}{%
Roques%
\ \protect \BOthers {.}}{%
{\protect \APACyear {2022}}%
}]{%
roques2022spatial}
\APACinsertmetastar {%
roques2022spatial}%
\begin{APACrefauthors}%
Roques, L.%
, Allard, D.%
\BCBL {}\ \BBA {} Soubeyrand, S.%
\end{APACrefauthors}%
\unskip\
\newblock
\APACrefYearMonthDay{2022}{}{}.
\newblock
{\BBOQ}\APACrefatitle {{Spatial statistics and stochastic partial differential
  equations: A mechanistic viewpoint}} {{Spatial statistics and stochastic
  partial differential equations: A mechanistic viewpoint}}.{\BBCQ}
\newblock
\APACjournalVolNumPages{Spatial Statistics}{50}{}{100591}.
\PrintBackRefs{\CurrentBib}

\bibitem [\protect \citeauthoryear {%
Royle%
\ \BBA {} Dorazio%
}{%
Royle%
\ \BBA {} Dorazio%
}{%
{\protect \APACyear {2008}}%
}]{%
royle2008hierarchical}
\APACinsertmetastar {%
royle2008hierarchical}%
\begin{APACrefauthors}%
Royle, J\BPBI A.%
\BCBT {}\ \BBA {} Dorazio, R\BPBI M.%
\end{APACrefauthors}%
\unskip\
\newblock
\APACrefYear{2008}.
\newblock
\APACrefbtitle {Hierarchical modeling and inference in ecology: the analysis of
  data from populations, metapopulations and communities} {Hierarchical
  modeling and inference in ecology: the analysis of data from populations,
  metapopulations and communities}.
\newblock
\APACaddressPublisher{}{Elsevier}.
\PrintBackRefs{\CurrentBib}

\bibitem [\protect \citeauthoryear {%
Schoenberg%
, Brillinger%
\BCBL {}\ \BBA {} Guttorp%
}{%
Schoenberg%
\ \protect \BOthers {.}}{%
{\protect \APACyear {2002}}%
}]{%
schoenberg2002point}
\APACinsertmetastar {%
schoenberg2002point}%
\begin{APACrefauthors}%
Schoenberg, F\BPBI P.%
, Brillinger, D\BPBI R.%
\BCBL {}\ \BBA {} Guttorp, P.%
\end{APACrefauthors}%
\unskip\
\newblock
\APACrefYearMonthDay{2002}{}{}.
\newblock
{\BBOQ}\APACrefatitle {{Point processes, spatial-temporal}} {{Point processes,
  spatial-temporal}}.{\BBCQ}
\newblock
\APACjournalVolNumPages{Encyclopedia of environmetrics}{3}{}{1573--1577}.
\PrintBackRefs{\CurrentBib}

\bibitem [\protect \citeauthoryear {%
G\BPBI A.~Seber%
}{%
G\BPBI A.~Seber%
}{%
{\protect \APACyear {1986}}%
}]{%
seber1986review}
\APACinsertmetastar {%
seber1986review}%
\begin{APACrefauthors}%
Seber, G\BPBI A.%
\end{APACrefauthors}%
\unskip\
\newblock
\APACrefYearMonthDay{1986}{}{}.
\newblock
{\BBOQ}\APACrefatitle {A review of estimating animal abundance} {A review of
  estimating animal abundance}.{\BBCQ}
\newblock
\APACjournalVolNumPages{Biometrics}{}{}{267--292}.
\PrintBackRefs{\CurrentBib}

\bibitem [\protect \citeauthoryear {%
G\BPBI A.~Seber%
\ \BBA {} Schofield%
}{%
G\BPBI A.~Seber%
\ \BBA {} Schofield%
}{%
{\protect \APACyear {2023}}%
}]{%
seber2023estimating}
\APACinsertmetastar {%
seber2023estimating}%
\begin{APACrefauthors}%
Seber, G\BPBI A.%
\BCBT {}\ \BBA {} Schofield, M\BPBI R.%
\end{APACrefauthors}%
\unskip\
\newblock
\APACrefYear{2023}.
\newblock
\APACrefbtitle {{Estimating Presence and Abundance of Closed Populations}}
  {{Estimating Presence and Abundance of Closed Populations}}.
\newblock
\APACaddressPublisher{}{Springer}.
\PrintBackRefs{\CurrentBib}

\bibitem [\protect \citeauthoryear {%
G\BPBI A\BPBI F.~Seber%
}{%
G\BPBI A\BPBI F.~Seber%
}{%
{\protect \APACyear {1982}}%
}]{%
seber1982estimation}
\APACinsertmetastar {%
seber1982estimation}%
\begin{APACrefauthors}%
Seber, G\BPBI A\BPBI F.%
\end{APACrefauthors}%
\unskip\
\newblock
\APACrefYear{1982}.
\newblock
\APACrefbtitle {{The estimation of animal abundance and related parameters}}
  {{The estimation of animal abundance and related parameters}}.
\PrintBackRefs{\CurrentBib}

\bibitem [\protect \citeauthoryear {%
Sisson%
, Fan%
\BCBL {}\ \BBA {} Beaumont%
}{%
Sisson%
\ \protect \BOthers {.}}{%
{\protect \APACyear {2018}}%
}]{%
sisson2018handbook}
\APACinsertmetastar {%
sisson2018handbook}%
\begin{APACrefauthors}%
Sisson, S\BPBI A.%
, Fan, Y.%
\BCBL {}\ \BBA {} Beaumont, M.%
\end{APACrefauthors}%
\unskip\
\newblock
\APACrefYear{2018}.
\newblock
\APACrefbtitle {Handbook of approximate Bayesian computation} {Handbook of
  approximate bayesian computation}.
\newblock
\APACaddressPublisher{}{CRC press}.
\PrintBackRefs{\CurrentBib}

\bibitem [\protect \citeauthoryear {%
Skellam%
}{%
Skellam%
}{%
{\protect \APACyear {1951}}%
}]{%
skellam1951random}
\APACinsertmetastar {%
skellam1951random}%
\begin{APACrefauthors}%
Skellam, J\BPBI G.%
\end{APACrefauthors}%
\unskip\
\newblock
\APACrefYearMonthDay{1951}{}{}.
\newblock
{\BBOQ}\APACrefatitle {Random dispersal in theoretical populations} {Random
  dispersal in theoretical populations}.{\BBCQ}
\newblock
\APACjournalVolNumPages{Biometrika}{38}{1/2}{196--218}.
\PrintBackRefs{\CurrentBib}

\bibitem [\protect \citeauthoryear {%
Soubeyrand%
\ \BBA {} Roques%
}{%
Soubeyrand%
\ \BBA {} Roques%
}{%
{\protect \APACyear {2014}}%
}]{%
soubeyrand2014parameter}
\APACinsertmetastar {%
soubeyrand2014parameter}%
\begin{APACrefauthors}%
Soubeyrand, S.%
\BCBT {}\ \BBA {} Roques, L.%
\end{APACrefauthors}%
\unskip\
\newblock
\APACrefYearMonthDay{2014}{}{}.
\newblock
{\BBOQ}\APACrefatitle {Parameter estimation for reaction-diffusion models of
  biological invasions} {Parameter estimation for reaction-diffusion models of
  biological invasions}.{\BBCQ}
\newblock
\APACjournalVolNumPages{Population ecology}{56}{}{427--434}.
\PrintBackRefs{\CurrentBib}

\bibitem [\protect \citeauthoryear {%
Sullivan%
\ \protect \BOthers {.}}{%
Sullivan%
\ \protect \BOthers {.}}{%
{\protect \APACyear {2009}}%
}]{%
sullivan2009ebird}
\APACinsertmetastar {%
sullivan2009ebird}%
\begin{APACrefauthors}%
Sullivan, B\BPBI L.%
, Wood, C\BPBI L.%
, Iliff, M\BPBI J.%
, Bonney, R\BPBI E.%
, Fink, D.%
\BCBL {}\ \BBA {} Kelling, S.%
\end{APACrefauthors}%
\unskip\
\newblock
\APACrefYearMonthDay{2009}{}{}.
\newblock
{\BBOQ}\APACrefatitle {{eBird: A citizen-based bird observation network in the
  biological sciences}} {{eBird: A citizen-based bird observation network in
  the biological sciences}}.{\BBCQ}
\newblock
\APACjournalVolNumPages{Biological conservation}{142}{10}{2282--2292}.
\PrintBackRefs{\CurrentBib}

\bibitem [\protect \citeauthoryear {%
Turchin%
}{%
Turchin%
}{%
{\protect \APACyear {1998}}%
}]{%
turchin1998quantitative}
\APACinsertmetastar {%
turchin1998quantitative}%
\begin{APACrefauthors}%
Turchin, P.%
\end{APACrefauthors}%
\unskip\
\newblock
\APACrefYearMonthDay{1998}{}{}.
\newblock
{\BBOQ}\APACrefatitle {{Quantitative analysis of movement: measuring and
  modeling population redistribution in animals and plants}} {{Quantitative
  analysis of movement: measuring and modeling population redistribution in
  animals and plants}}.{\BBCQ}
\newblock

\PrintBackRefs{\CurrentBib}

\bibitem [\protect \citeauthoryear {%
Volterra%
}{%
Volterra%
}{%
{\protect \APACyear {1913}}%
}]{%
volterra1913leccons}
\APACinsertmetastar {%
volterra1913leccons}%
\begin{APACrefauthors}%
Volterra, V.%
\end{APACrefauthors}%
\unskip\
\newblock
\APACrefYear{1913}.
\newblock
\APACrefbtitle {Le{\c{c}}ons sur les {\'e}quations int{\'e}grales et les
  {\'e}quations int{\'e}gro-diff{\'e}rentielles: Le{\c{c}}ons profess{\'e}es
  {\`a} la Facult{\'e} des sciences de Rome en 1910} {Le{\c{c}}ons sur les
  {\'e}quations int{\'e}grales et les {\'e}quations
  int{\'e}gro-diff{\'e}rentielles: Le{\c{c}}ons profess{\'e}es {\`a} la
  facult{\'e} des sciences de rome en 1910}.
\newblock
\APACaddressPublisher{}{Gauthier-Villars}.
\PrintBackRefs{\CurrentBib}

\bibitem [\protect \citeauthoryear {%
Wackernagel%
}{%
Wackernagel%
}{%
{\protect \APACyear {2003}}%
}]{%
wackernagel2003multivariate}
\APACinsertmetastar {%
wackernagel2003multivariate}%
\begin{APACrefauthors}%
Wackernagel, H.%
\end{APACrefauthors}%
\unskip\
\newblock
\APACrefYear{2003}.
\newblock
\APACrefbtitle {{Multivariate Geostatistics: an introduction with
  applications}} {{Multivariate Geostatistics: an introduction with
  applications}}\ (\PrintOrdinal{3}\ \BEd).
\newblock
\APACaddressPublisher{}{Springer}.
\PrintBackRefs{\CurrentBib}

\bibitem [\protect \citeauthoryear {%
Wikle%
}{%
Wikle%
}{%
{\protect \APACyear {2003}}%
}]{%
wikle2003hierarchical}
\APACinsertmetastar {%
wikle2003hierarchical}%
\begin{APACrefauthors}%
Wikle, C\BPBI K.%
\end{APACrefauthors}%
\unskip\
\newblock
\APACrefYearMonthDay{2003}{}{}.
\newblock
{\BBOQ}\APACrefatitle {Hierarchical Bayesian models for predicting the spread
  of ecological processes} {Hierarchical bayesian models for predicting the
  spread of ecological processes}.{\BBCQ}
\newblock
\APACjournalVolNumPages{Ecology}{84}{6}{1382--1394}.
\PrintBackRefs{\CurrentBib}

\bibitem [\protect \citeauthoryear {%
B\BPBI K.~Williams%
, Nichols%
\BCBL {}\ \BBA {} Conroy%
}{%
B\BPBI K.~Williams%
\ \protect \BOthers {.}}{%
{\protect \APACyear {2002}}%
}]{%
williams2002analysis}
\APACinsertmetastar {%
williams2002analysis}%
\begin{APACrefauthors}%
Williams, B\BPBI K.%
, Nichols, J\BPBI D.%
\BCBL {}\ \BBA {} Conroy, M\BPBI J.%
\end{APACrefauthors}%
\unskip\
\newblock
\APACrefYear{2002}.
\newblock
\APACrefbtitle {{Analysis and management of animal populations}} {{Analysis and
  management of animal populations}}.
\newblock
\APACaddressPublisher{}{Academic press}.
\PrintBackRefs{\CurrentBib}

\bibitem [\protect \citeauthoryear {%
P\BPBI J.~Williams%
, Hooten%
, Womble%
\BCBL {}\ \BBA {} Bower%
}{%
P\BPBI J.~Williams%
\ \protect \BOthers {.}}{%
{\protect \APACyear {2017}}%
}]{%
williams2017estimating}
\APACinsertmetastar {%
williams2017estimating}%
\begin{APACrefauthors}%
Williams, P\BPBI J.%
, Hooten, M\BPBI B.%
, Womble, J\BPBI N.%
\BCBL {}\ \BBA {} Bower, M\BPBI R.%
\end{APACrefauthors}%
\unskip\
\newblock
\APACrefYearMonthDay{2017}{}{}.
\newblock
{\BBOQ}\APACrefatitle {{Estimating occupancy and abundance using aerial images
  with imperfect detection}} {{Estimating occupancy and abundance using aerial
  images with imperfect detection}}.{\BBCQ}
\newblock
\APACjournalVolNumPages{Methods in Ecology and Evolution}{8}{12}{1679--1689}.
\PrintBackRefs{\CurrentBib}

\bibitem [\protect \citeauthoryear {%
Zamberletti%
, Papa{\"\i}x%
, Gabriel%
\BCBL {}\ \BBA {} Opitz%
}{%
Zamberletti%
\ \protect \BOthers {.}}{%
{\protect \APACyear {2022}}%
}]{%
zamberletti2022understanding}
\APACinsertmetastar {%
zamberletti2022understanding}%
\begin{APACrefauthors}%
Zamberletti, P.%
, Papa{\"\i}x, J.%
, Gabriel, E.%
\BCBL {}\ \BBA {} Opitz, T.%
\end{APACrefauthors}%
\unskip\
\newblock
\APACrefYearMonthDay{2022}{}{}.
\newblock
{\BBOQ}\APACrefatitle {Understanding complex spatial dynamics from mechanistic
  models through spatio-temporal point processes} {Understanding complex
  spatial dynamics from mechanistic models through spatio-temporal point
  processes}.{\BBCQ}
\newblock
\APACjournalVolNumPages{Ecography}{2022}{5}{e05956}.
\PrintBackRefs{\CurrentBib}

\bibitem [\protect \citeauthoryear {%
Zammit-Mangion%
, Sainsbury-Dale%
\BCBL {}\ \BBA {} Huser%
}{%
Zammit-Mangion%
\ \protect \BOthers {.}}{%
{\protect \APACyear {2024}}%
}]{%
zammit2024neural}
\APACinsertmetastar {%
zammit2024neural}%
\begin{APACrefauthors}%
Zammit-Mangion, A.%
, Sainsbury-Dale, M.%
\BCBL {}\ \BBA {} Huser, R.%
\end{APACrefauthors}%
\unskip\
\newblock
\APACrefYearMonthDay{2024}{}{}.
\newblock
{\BBOQ}\APACrefatitle {{Neural Methods for Amortised Parameter Inference}}
  {{Neural Methods for Amortised Parameter Inference}}.{\BBCQ}
\newblock
\APACjournalVolNumPages{arXiv preprint arXiv:2404.12484}{}{}{}.
\PrintBackRefs{\CurrentBib}

\end{thebibliography}
\bibliographystyle{apacite}

\end{document}